\documentclass[aps,prf,preprint,groupedaddress, amsmath, amssymb]{revtex4-2}

\usepackage{graphicx}
\usepackage{amsmath}
\usepackage{amssymb}
\usepackage{newtxtext}
\usepackage{newtxmath}
\usepackage{natbib}
\usepackage{comment}
\usepackage{hyperref}
\usepackage{bbm}
\usepackage{multirow}

\usepackage[mathlines]{lineno}

\hypersetup{
    colorlinks = true,
    urlcolor   = blue,
    citecolor  = black,
}

\newcommand{\RomanNumeralCaps}[1]

\newcommand{\q}{q}

\newcommand{\xx}{\underline{x}}
\newcommand{\uu}{\underline{u}}

\begin{document}

\title{On the analysis of Rayleigh-Bénard convection using Latent Dirichlet Allocation} 



\author{B. Podvin},
\author{L. Soucasse},
\affiliation{{EM2C, Centralesupélec, CNRS, Université Paris-Saclay}}
 \author{F. Yvon}
 \affiliation{LISN, CNRS, Université Paris-Saclay }

\begin{abstract}
We apply a probabilistic clustering method,  Latent Dirichlet Allocation (LDA),
to characterize the large-scale dynamics of Rayleigh-Bénard convection.
The method, introduced in Frihat et al. 2021, is applied to a collection of
snapshots in the vertical mid-planes of a cubic cell
for Rayleigh numbers in the range $[10^6, 10^8]$.
For the convective heat flux, temperature and kinetic energy, 
the decomposition identifies latent factors, called motifs, which consist
of connex regions of fluid. 
Each snapshot is modelled with a sparse combination of motifs, the coefficients of which
are called the weights.
The spatial extent of the motifs  varies across the cell and with the Rayleigh number.
We show that the method is able to provide a compact representation of the heat flux
and displays good generative properties.
At all Rayleigh numbers the dominant heat flux motifs consist of elongated structures
located mostly within the vertical boundary layer, at a quarter of the cavity height.
Their weights depend on the orientation of the large-scale circulation (LSC).
A simple model relating the conditionally averaged weight of the motifs  to
 the relative strength  of the corner rolls and of the large-scale circulation, is found
to predict well the average LSC reorientation rate.
Application of LDA to the temperature fluctuations
shows that temperature motifs are well correlated with heat flux motifs
in space as well as in time, and to some lesser extent with kinetic energy motifs.
The abrupt decrease of the reorientation rate observed at $10^8$ is associated with 
a strong concentration of plumes impinging onto the corners of the cell, which decrease 
the temperature difference within the corner structures. It is also associated with a
reinforcement of the longitudinal wind  through formation and entrainment of new plumes.
\end{abstract}

\maketitle


\section{Introduction}
\label{sec:headings}

Rayleigh-Bénard convection, in which a fluid is heated from below and cooled from above,
represents an idealized configuration to study thermal convection phenomena. These
characterize  a variety of applications ranging from industrial processes such as heat exchangers to geophysical flows in the atmosphere or the ocean.
A central question is to determine how the heat transfer depends
on nondimensional parameters such as the Prandtl number $Pr= \nu/ \kappa$ where $\nu$ is the kinematic viscosity 
and $\kappa$ the thermal diffusitivity,  and the Rayleigh number 
\begin{equation}
    Ra=\frac{g\beta \Delta T H^3}{\nu a},
\end{equation}
where $g$ is the gravity,  $\beta$ is the thermal expansion coefficient, $\Delta T$ the temperature difference and $H$ the cell dimension.
The \citet{grossmann00} theory constitutes a unified approach to address  this question. 
It is based on a local description of the physics:  the contributions from the bulk averaged thermal and kinetic dissipation rate 
are split into two subsets, one 
corresponding to the boundary layers, and one corresponding to the bulk. 
This theory was further refined in~\citet{kn:grossmannlohse04}, where the thermal dissipation rate  was split into
a contribution from the plumes and a contribution from the turbulent background.
Through the action of buoyancy, the thermal boundary layers generate plumes which 
 create a large-scale circulation, as evidenced by \citet{kn:xilamxia04}, also called "wind" \citep{castaing89}.
 The distribution of temperature fluctuations depends on plume clustering effects
 \citep{kn:wang22}, but it is also affected by interaction with turbulent fluctuations in the bulk, resulting in fragmentation \citep{kn:bosbach12}.

 \citet{shang03} showed that plume-dominated regions were located near the sidewalls and the conducting surfaces and that
thermal plumes carry most of the convective heat flux, which  contributes to the production of both kinetic and  thermal fluctuations. 
The morphology of plumes and its effect on the heat transfer have  been given careful attention.
The plumes have a sheet-like structure near the boundary layer and progressively become mushroom-like as they move into 
the bulk region~\citep{kn:zhou07}.  \citet{kn:shishkinawagner08} found that very high values of the local heat flux were observed in 
 regions where the sheet-like plumes merged, constituting "stems" for the mushroom-like plumes developing
 in the bulk.
 The relative contributions of the plumes and turbulent background vary with the Rayleigh number:
\citet{kn:emranschumacher12}  have shown  that the fraction of plume-dominated regions decreases with the
Rayleigh number, while that of background-dominated regions increases.

The identification of local coherent structures such as plumes is therefore an essential  step for the 
understanding of thermal convection flows.   Several definitions have been used:
some of the first criteria were based on the skewness of the temperature derivative \citep{kn:belmontelibchaber96}  or  the temperature difference \citep{kn:zhouxia10}.
\citet{kn:ching04} have proposed to use simultaneous measurements of the temperature and the velocity to define the velocity of the plumes using conditional averaging.
Following \citet{kn:huangprl13}, \citet{vdpoel15} identified plumes from both a temperature anomaly and an excess of convective heat flux.
\cite{kn:zhou16} relied on cliff-ramp-like structures in the temperature signals to determine the spatial characteristics of plumes. 
\citet{kn:emranschumacher12} and  \citet{kn:vishnu22} separated the plume from the background regions based on a threshold  on the convective heat flux.
\citet{kn:shevkar22} have recently proposed a dynamic criterion based on the 2-D velocity divergence to separate plumes from boundary layers.

As pointed out by \citet{kn:chillaschumacher12}, this multiplicity of criteria illustrates the difficulty of identifying coherent 
structures in a consistent and objective manner, which is a long-running question in various types of turbulent flows.  
To this end, Proper Orthogonal Decomposition (POD) \citep{kn:lumleyPOD} has proven a useful tool 
to analyze large-scale fluctuations in Rayleigh-Bénard convection. It has been used in particular to study reorientations of 
the large-scale circulation \citep{bailon-cuba10,foroozani17,podvin15,podvin17,soucasse19}.
Through spectral decomposition of the autocorrelation tensor, POD  provides  a basis of spatial modes, 
also called empirical modes, since they originate from the data.
The modes are energetically optimal to reconstruct the fluctuations. 
The POD modes typically have a global support, which is well suited to 
capture the large-scale organization of the flow. However, this can make physical interpretation difficult as there is no straightforward connection between a mode 
and a local coherent structure as a local structure is represented with a superposition of many POD modes, a situation also observed in Fourier analysis. 
\citet{soucasse21} have used POD to study the dynamics of the large-scale circulation for Rayleigh numbers
in the range $[10^6, 10^8]$.  They found that although the reorientation rate varied with the Rayleigh number,
the dominant structures remained similar across that range, albeit with some variations in their energy.
A new dissipation-based POD decomposition, proposed by  \citet{olesen23} and applied to Rayleigh-Bénard convection
\citep{olesen23b},  highlighted the importance of boundary layers for the dynamics, which points to the need for local descriptions.

As an alternative, \citet{kn:frihat21} have recently adapted a probabilistic method that can extract localized latent factors in turbulent flow measurements.
This method, Latent Dirichlet Allocation or LDA \citep{kn:griffiths02,kn:blei03}, was originally developed 
in the context of natural language processing, where it aims 
to extract topics from a collection of documents. In this framework, documents are represented by a non-ordered set of words taken from a fixed vocabulary.
A word count matrix can be built for the collection, where  
each column corresponds to a document, each line corresponds
to a vocabulary word and the matrix entry represents the number of times the word appears in the document.
LDA  provides a  probabilistic decomposition of the word count matrix, based on latent factors called {\it topics}.
Topics are defined by two distributions:  the distribution of topics within
each document (each document is associated with a mixture of topics, the coefficients of the mixture sum up to one)
and the distribution of vocabulary words with each topic (each topic is represented by a combination of words, the coefficients
of which also sum up to one).

The method has been adapted  for turbulent flows as follows:
we consider a collection of snapshots of a scalar field taken over a 2D domain discretized into cells. The equivalent of a document is therefore a snapshot, and
the cells (or snapshot pixels) constitute the vocabulary.
The digitized values of the scalar field over the  cells in a snapshot are gathered into a vector which is formally analogous to a column of the word count matrix. 
The "topics" produced by the decomposition, called {\it motifs},  correspond to fixed (in the Eulerian sense), 
spatially coherent regions of the flow.
The method was found to be well suited for the representation of intermittent data (\citet{kn:frihat21}, \citet{kn:fery22}).
It was succesfully applied to the analysis of the turbulent Reynolds stress in wall turbulence \citep{kn:frihat21}.
Moreoever, the method  provides a local description that is insensitive to the existence of global symmetries.
It proved a useful tool to identify synoptic objects  in weather data \citep{kn:fery22}.

In this paper, we apply this method to the analysis of  fluctuations
in a cubic Rayleigh-Bénard cell in the range of Rayleigh number $[10^6, 10^8]$.
The goal is to  track the local signature of the large-scale dynamics of the flow, and to determine
whether changes can be identified as the Rayleigh number increases.
To this end,  the technique is applied to 2D sections of a cubic Rayleigh-Bénard cell in the range of Rayleigh number $[10^6, 10^8]$.
The numerical configuration and the data set are described in Section~2. 
We first present the method for  the convective heat flux, using a comparison with POD
to highlight the similarities and differences of the approach. 
Proper Orthogonal Decomposition (POD) and Latent
Dirichlet Allocation (LDA) are respectively presented in Section~3 and~4.
We examine in Section~5 how LDA compares with POD and the extent to which it is able to capture the general features of the heat flux.
The characteristics of heat flux motifs and their connection with the reorientations of the large-scale circulation are discussed in Section~6. 
The analysis is then extended  to temperature fluctuations and to the kinetic energy in Section-7 
in order to provide further insight into the physics.
A conclusion is given in section 8.

\section{Numerical setting}

\subsection{Set-up}

Numerical setup and associated datasets are the same as used in~\citet{soucasse19, soucasse21}. The configuration studied is a cubic Rayleigh-Bénard cell filled with air, with isothermal horizontal walls and adiabatic side walls. The air is assumed to be transparent and thermal radiation effects are disregarded.
Direct numerical simulations have been performed at various values of the Rayleigh number. The Prandtl number is set to 0.707. All physical quantities are made dimensionless using the cell size $H$, the reference time $H^2/(a\sqrt{Ra})$ and the reduced temperature $\theta=(T-T_0)/\Delta T$, $T_0$ being the mean temperature between hot and cold walls. Spatial coordinates are denoted $x$, $y$, $z$ ($z$ being the vertical direction) and the origin is placed at a bottom corner of the cube.

Navier–Stokes equations under Boussinesq approximation are solved using a Chebyshev collocation method~\citep{xin02,xin-PCFD08}. Computations are made parallel using domain decomposition along the vertical direction. Time integration is performed through a second-order semi-implicit scheme. The velocity divergence-free condition is enforced using a projection method. Numerical parameters are given in Table~\ref{tab:dataset} for the four considered Rayleigh numbers $Ra=\{ 10^6 ; 3~10^6 ; 10^7 ; 10^8\}$. We have checked that the number of collocation points is sufficient to accurately discretize the boundary layers according to the criterion proposed by~\cite{shishkina10}. A number of 1000 snapshots have been extracted from the simulations for each Rayleigh number at a sampling period of 10 (at $Ra=\{ 10^6 ; 3~10^6 ; 10^7 \}$) or 5 (at $Ra=10^8$), in dimensionless time units. It is worth noting that the time separation between the snapshots is sufficient to describe the evolution of the  large-scale circulation but is not suited for a fine description of the plume emission or of the reorientation process.
For each Rayleigh number, a dataset satisfying the statistical  symmetries of the flow was then constructed from these 1000 snapshots, as  will be described in the next section.

\begin{table}
\centering
\begin{tabular}{|l|l|l|l|l|}
\hline
$Ra$ & $(N_x, N_y, N_z)$ & $N_S$ & $\Delta t$ & $\delta_{BL}$ \\ \hline
$10^6$ & (81,81,81) & 1000 & 10 & 0.056 \\
$3 $ $ 10^6$ & (81,81,81) & 1000 & 10 & 0.042 \\
$10^7$ & (81,81,81) & 1000 & 10 & 0.0297 \\
$10^8$ & (161,161,161) & 1000 & 5 & 0.0167 \\ \hline
\end{tabular}
\caption{Characteristics of the datasets at various Rayleigh numbers: spatial resolution $N_x$, $N_y$, $N_z$ in each direction of space, number of snapshot $N_S$, snapshot sampling period $\Delta t$ and thermal boundary layer thickness $\delta_{BL}$.}
\label{tab:dataset}
\end{table}

\subsection{Construction of the data set}

\begin{figure}
\centerline{ \includegraphics[height=4cm]{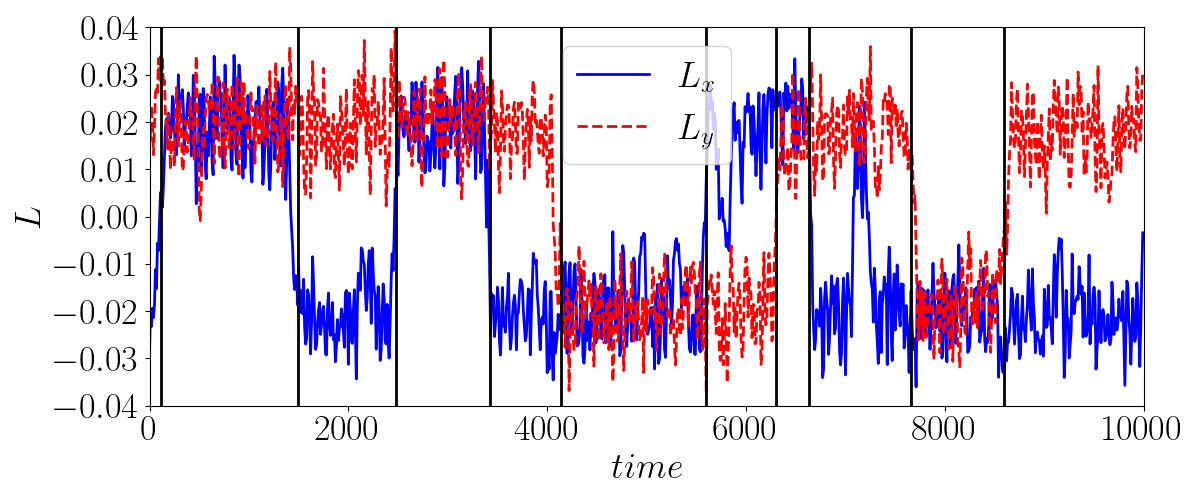}  } 
  \caption{ Evolution of the horizontal components of the angular momentum  at $Ra=10^7$. The vertical black lines correspond to reorientations of the large-scale circulation. }
\label{fig:Lxy}
\end{figure}

At each Rayleigh number, the data set consisted of
a collection of $N_S=1000$  snapshots $q(\xx,t_k)$, $k=1, \ldots N_S$.
Results will be presented first for the convective heat flux $q = w \theta$, then for the temperature
fluctuations $q=\theta'=\theta-\langle\theta\rangle$ ($\langle\theta\rangle$ being the time-averaged temperature) and for the kinetic energy $q= k = \frac{1}{2} (u^2 + v^2 + w^2)$, {$u$, $v$ and $w$ being the velocity components}.   
 We note that due to the velocity reference scale, the non-dimensional heat flux varies like $Nu Ra^{-1/2}$.
As in \citet{soucasse19}, the data set was first enriched by making use of the 
statistical symmetries of the flow~\citep{puigjaner08}. In the cubic Rayleigh-Bénard cell, four quasi-stable states are available for the flow for this Rayleigh number range: the large-scale circulation settles in one of the two diagonal planes of the cube with clockwise or counterclockwise motion. 
The evolution of the large-scale circulation can be tracked through that of the $x$ and $y$ components of the angular momentum of the cell $\underline{L}=\int (\xx-\underline{x_0}) \times\uu d\xx$ with respect to the cell center $\underline{x_0}$.
As Figure~\ref{fig:Lxy} shows at $Ra=10^7$, the angular momentum along each horizontal direction oscillates near a quasi-steady position for long periods of times  - several hundreds of convective time scales, before experiencing a rapid switch ($\mathcal{O}(10)$ convective time scales) to the opposite value, which corresponds
to a reorientation.
 On each plane we can define an indicator function $I$, which takes the value $sgn(L) 1$ 
 where  $L$ is the angular momentum component normal to the plane. 

Reorientations from one state to another occur during the time sequence but each state is not necessarily equally visited. In order to counteract this bias, we have built enlarged snapshot sets, obtained by the action of the symmetry group of the problem on the original snapshot sets.
The symmetries  are based on four independent symmetries $S_x$, $S_y$, $S_z$ and $S_d$ with respect to the planes $x=0.5$, $y=0.5$, $z=0.5$ and $x=y$. This generates a group of 16 symmetries for the cube, which should lead to a 16-fold in the number of snapshots. However, since we will exclusively consider the vertical mid-planes $x=0.5$ and $y=0.5$, which are invariant planes for respectively $S_x$ and $S_y$, the increase is reduced. The data set aggregates 1000 snapshots on each of the  planes $x=0.5$ and $y=0.5$, each of which undergoes a vertical flip, a horizontal flip and a combination of the two, yielding a total of $N_s=8000$ snapshots.

The LDA technique requires  transforming the data into a non-negative, integer field.
The signal defined on a grid of $\tilde{N_C}$ cells was digitized using a rescaling factor $s$.  
If the field was not of constant sign ({temperature, heat flux}),  
 positive and negative values were split onto two distinct grids, leading to a field defined on $ N_C = 2 \tilde{N}_C$ cells. 
{For the heat flux, this gives}
\begin{eqnarray}
 \q(\xx_j)=\q(\xx_j, t_m)& = & \mbox{Max}\left[ \mbox{Int}[ s \,w(\xx_j, t_m) \theta(\xx_j, t_m),0 \right]    \label{eq:digit1} \\
 \q(\xx_{j+\tilde{N}_C})=\q(\xx_{j+\tilde{N}_C}, t_m)& = & -\mbox{Max}\left[- \mbox{Int}[ s \,w(\xx_j, t_m) \theta(\xx_j, t_m),0 \right]  \label{eq:digit2}
\end{eqnarray}
where $s >0$,  $m \in [1, N_S]$ and $j \in [1, \tilde{N}_C]$
and $\xx_j$ represents the $j^\mathrm{th}$ cell location on the mid-planes
$x=0.5$ or $y=0.5$. 
We note that throughout the paper, the total field will directly be represented on the physical grid of size $\tilde{N}$ from the renormalized  difference
$[\q(\xx_j, t_m)-\q(\xx_{j+\tilde{N}_C}, t_m)]/s$.

\section{POD analysis}

\subsection{Method}

Proper Orthogonal Decomposition~\citep{kn:HLBR}
makes it possible to write  a collection of $N_S$ spatial fields $\q(\xx_j,t_m) $
defined on $N_C$ grid points,
as a superposition of spatial modes $\phi_n(\xx)$, the amplitude of which varies in time:
\begin{equation} 
\q(\xx_j, t_m) = \sum_{n=1}^{N_S} a_n(t_m) \phi_n(\xx_j),
\label{eq:pod}
\end{equation}
with $m \in [1, N_S]$ and $j \in [1, N_C]$. The amplitudes $a_n(t_m)$ are solution of the eigenvalue problem
\begin{equation}
C_{mp} a_n(t_p) = \lambda_n a_n(t_m),
\end{equation}
where $C$ is the temporal autocorrelation matrix 
\begin{equation}
C_{mp} = \frac{1}{N_S} \sum_{j=1}^{N_C} \q(\xx_j,t_m) \q(\xx_j,t_m) 
\end{equation}
The eigenvalues $\lambda_n$, such that  $\lambda_1 > \lambda_2 > \lambda_3 > \ldots$,  represent the respective contribution of the modes to the total variance. If we consider the $p$ most energetic modes,
the reconstruction based on $p$ modes minimizes the 
$L_2$-norm error between the set of snapshots and the projection of the set of snapshots onto a basis of size $p$.

\subsection{Application to the convective heat flux}

\begin{figure}
\begin{tabular}{cc}

 $n=1$ & 
 \multirow{2}{*} {\includegraphics[height=4cm]{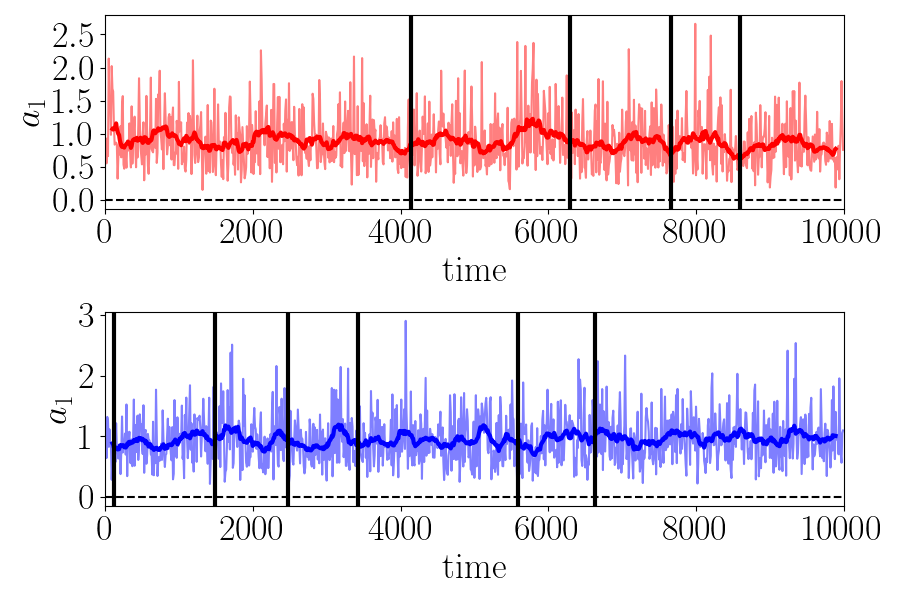} }\\
   \includegraphics[height=3cm]{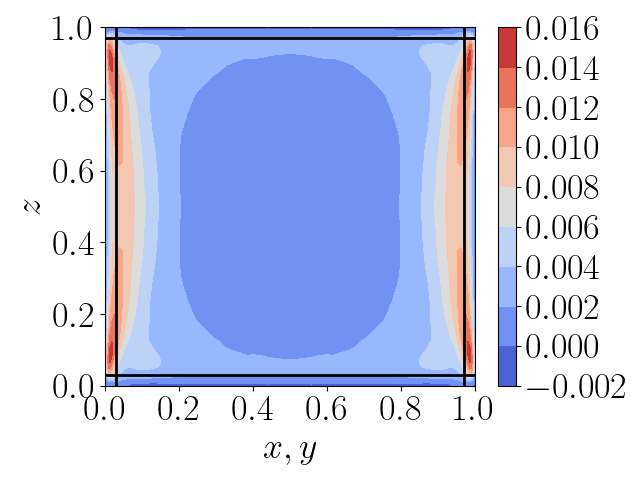} & \\

$n=2$ & 
 \multirow{2}{*} {\includegraphics[height=4cm]{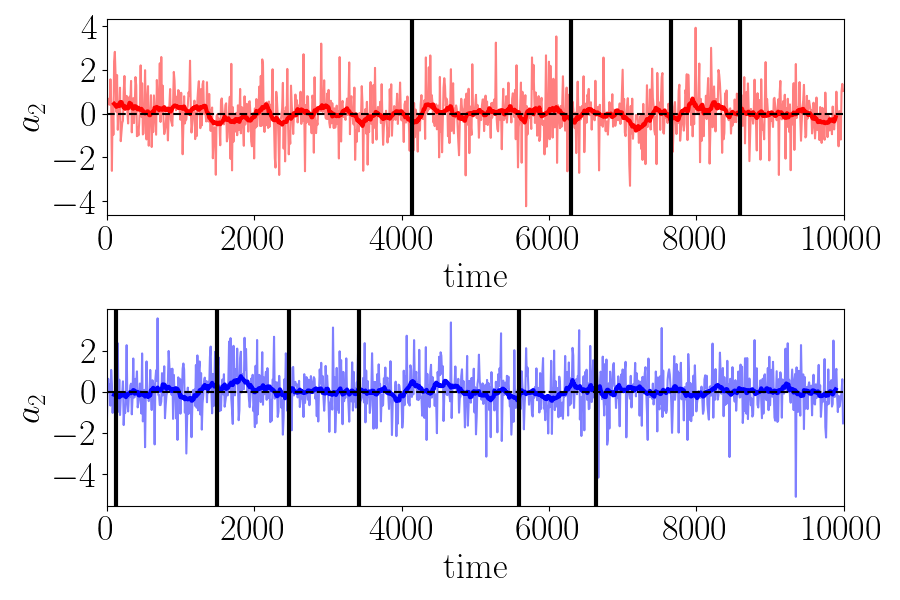} }\\
   \includegraphics[height=3cm]{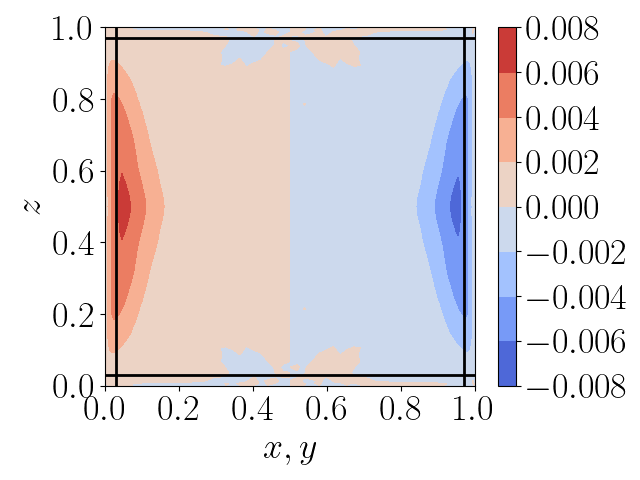} & \\

$n=3$ & 
 \multirow{2}{*} {\includegraphics[height=4cm]{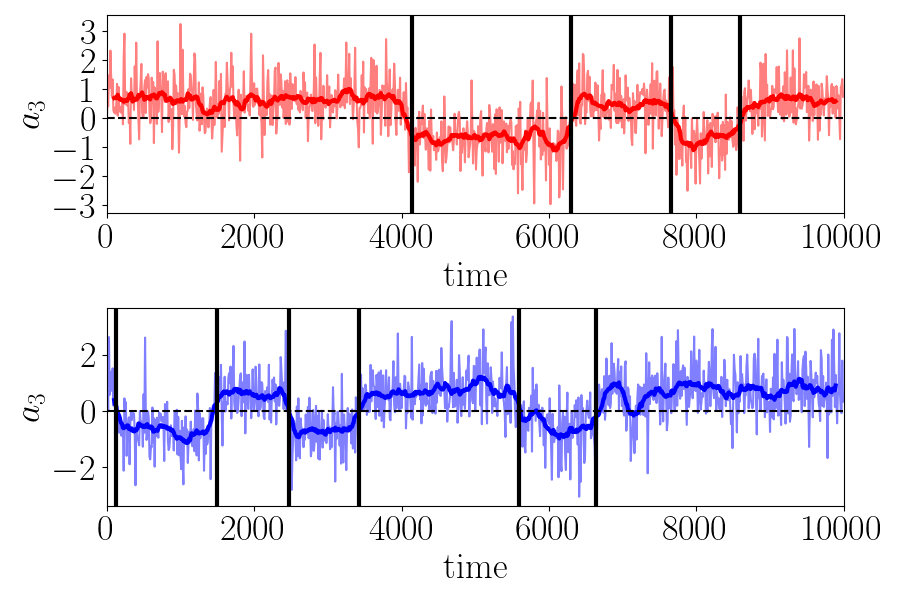} }\\
   \includegraphics[height=3cm]{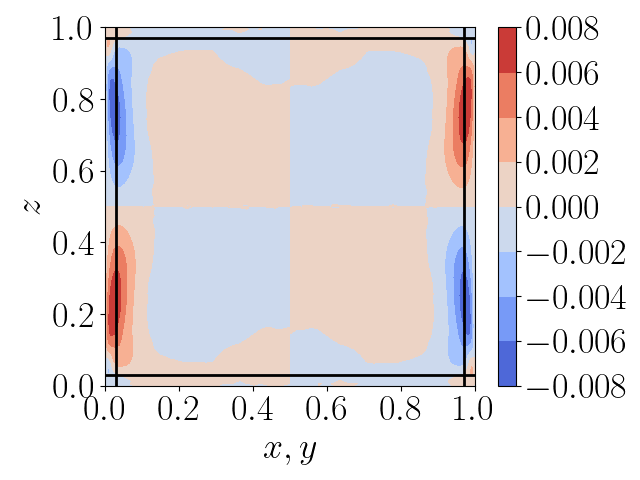} & \\

\end{tabular}
  \caption{POD dominant modes and amplitudes in the vertical mid-plane at $Ra=10^7$. Left : POD modes $\phi_n$ , Right : POD amplitudes $a_n$ associated with plane $x=0.5$ (in blue) and plane $y=0.5$ (in red). The vertical black lines correspond to changes in the component of the angular momentum. The darker line corresponds to a moving average over 200 convective units (4 $T_c$).}
\label{fig:vert:podmodes}
\end{figure}

POD is applied to the  digitized  heat flux signal $\q=\Phi$ defined in equations~\eqref{eq:digit1} and
\eqref{eq:digit2}. 
The first three POD  modes and POD coefficients are shown in Figure~\ref{fig:vert:podmodes} for $Ra=10^7$, where 
black vertical and horizontal lines indicate the thickness of the boundary layers. 
We checked that the first mode corresponds to the mean flow. 
The mode is most important in a region close to the wall, with a maximum within the vertical boundary layer at 
a height of about $z \sim 0.1$.   
The second mode corresponds to a dissymetry between the vertical sides and is most important at mid-height in 
the region outside the boundary layers.
The third mode is both antisymmetric in the vertical and in the horizontal direction.
It is maximum at the edge of the vertical boundary layers, at  a vertical distance of about $0.25$ from the horizontal surfaces. 
The pattern it is associated with corresponds to a more intense flux along a diagonal (bottom of one side
and top of the opposite side) and a less intense flux along the opposite diagonal.
As evidenced by application of a moving average performed over 
200 convective time units (about 4 recirculation times $T_c$, as was determined in~\citet{soucasse19}),
the evolution of the amplitude at large time scales matches that of the 
 horizontal angular momentum components $L_x$ and $L_y$ (compare with Figure~\ref{fig:Lxy}), unlike the two dominant modes. 
 This mode therefore appears to be the signature of the large-scale circulation, where
 the flux is more intense in the lower corner of the cell as hot plumes rise on one side
and in the upper corner of the opposite side of the cell as cold plumes go down.

\section{Latent Dirichlet Allocation \label{sec:lda}}

\subsection{Principles}

We briefly review the principles of Latent Dirichlet Allocation and refer the reader to~\citet{kn:frihat21} for more details. 
LDA is an inference approach to identify latent factors in a collection of observed data, which relies on Dirichlet distributions as priors. 
We first recall the definition of a Dirichlet distribution, which 
is a multivariate probability distribution over the space
of multinomial distributions. It is parameterized by a vector of positive-valued
parameters
$\underline{\alpha}=(\alpha_1, \alpha_2, \ldots, \alpha_N)$ as follows
\begin{equation}
p (\theta_1, . . . , \theta_N; \alpha_1, . . . , \alpha_N) = \frac{1}{B(\underline{\alpha})}\prod_{n=1}^N \theta_n^{\alpha_n-1},
\end{equation}
where $B$ is a normalizing factor, which can be expressed in terms of the Gamma
function~$\Gamma$:
\begin{equation}
B(\underline{\alpha})=\frac{\prod_{n=1}^{N} \Gamma(\alpha_n)}{\Gamma(\sum_{n=1}^N\alpha_n)}.
\end{equation}
The components $\{\alpha_n,n=1 \dots N\}$ of $\underline{\alpha}$ control the sparsity of the
distribution: values of $\alpha_n$ larger than unity correspond to evenly dense distributions,
while values lower than unity correspond to sparse distributions.
Here $\theta$ will represent either the motif-cell distribution
or the snapshot-motif distribution.

As mentioned above, 
the data to which  LDA is applied consists of  a collection of non-negative, integer fields that are defined in equations~\eqref{eq:digit1} and \eqref{eq:digit2}.
For each snapshot $m$,
the integer value $\q(\xx_j)$ measured at cell $j$ is interpreted as 
an integer count of  the cell $j$. The key is to interpret this integer count as the  number of times cell $j$ appears in the composition of snapshot $m$.
A snapshot $m$ is therefore defined as a list of tuples of the form $(\xx_j, \q(\xx_j))$.

The main assumptions of LDA are the following:
\begin{enumerate}
\item Each snapshot consists of a mixture of $N_T$ latent factors called motifs.
$N_T$ is a user-defined parameter (analogous to a number of clusters).

\item Each motif $n$ is associated with a  multinomial distribution over the grid cells $\psi_n$
so that the probability to observe the $j^\mathrm{th}$ grid cell located at $\xx_j$ given the motif $n$ is $\psi_n(x_j)$.
The distribution $\psi_n$ is modelled with a Dirichlet prior parameterized with an $N_C$-dimensional vector $\underline{\eta}$. Low values of $\eta_l$ mean that the motif is distributed over a small number of cells.

\item Each snapshot $m$ is associated with a distribution $b^m$ over motifs such that
the probability that motif $n$ is present in snapshot $m$ will be denoted $b_n(t_m)$. This
distribution is modelled with a $N_T$-dimensional Dirichlet distribution of parameter 
$\underline{\alpha}$. The magnitude of $\alpha$ characterizes the sparsity of the distribution. 
Low values of $\alpha_l$ mean that relatively few motifs are observed in each snapshot.
\end{enumerate}

\subsection{Implementation}

The snapshot–motif distribution $b_n$ and the motif-cell distribution $\psi_n$ are determined
from the observed snapshots $\q(\xx)$ and constitute $N_T$ and $N_C$-dimensional categorical
distributions. Finding the distributions $b^m$  and $\psi_n$ that are most compatible with the
observations constitutes an inference problem. The problem can be solved either with 
 a Markov chain Monte-Carlo (known as MCMC) algorithm such as
Gibbs sampling \citep{kn:griffiths02}, or  by a variational approach
\citep{kn:blei03}, which aims to minimize the Kullback–Leibler divergence between the true posterior and its variational approximation.
In both cases, the computational complexity of the problem of the order of $(N_C N_S N_T )$.

The solution {\it a priori} depends on the number of motifs $N_T$ as well as on the values of the Dirichlet parameters $\underline{\alpha}$  and $\underline{\eta}$.
Special attention was therefore given to establish the robustness of the results reported here.
Non-informative default values were used for the Dirichlet parameters i.e. the prior distributions  were taken with symmetric parameters equal 
to $\forall n, \alpha_n=1/N_T$  and $\forall l, \eta_l=1/N_T$.   
Practical implementation was performed in Python using gensim \citep{rehurek2011gensim}. 
No significant change was observed in the results 
when the value of the quantization $s$ was  high enough (however it had
to be kept reasonably  low in order to limit the computational time).
Although multiple tests were carried out for varying values of $s \in [40,600]$, all results reported in this paper were obtained with $s = 600$ for the heat flux.
Values of $s=40$  and $s=50$  were respectively used for the temperature fluctuations and for the kinetic energy. 
Analyses were also performed for varying numbers of motifs $N_T$, ranging from 50 to 400.


\subsection{LDA as a generative process}

The standard generative process performed by LDA with $N_T$ motifs is the following.
\begin{itemize}
\item[$(i)$]
 For each motif $n$, a $N_C$-dimensional cell–motif distribution $\psi_n$ is drawn from the Dirichlet
distribution of parameter $\underline{\eta}$.
\item[$(ii)$]
 To generate snapshot $m$:
 \begin{itemize}
 \item[$(a)$]
 a  $N_T$-dimensional snapshot–motif distribution $b^m$ is drawn according to a Dirichlet distribution parameterized by $\underline{\alpha}$
  \item[$(b)$]
  a total integer count $\q(t_m)$  is drawn.
  This number corresponds to the total number of cell integer counts associated with snapshot $m$ i.e. $\q(t_m)= \sum_j \q(\xx_j,t_m)$. $\q_T(t_m)$ is typically 
sampled from a Poisson distribution that matches the statistics of the original database.
\item[$(c)$]
 for each $i=1, \ldots, \q_T(t_m)$:
\begin{itemize}
\item a motif $n$ is selected from $b(t_m)$ (since $b_n(t_m)$ represents the probability
that motif $n$ is present in the $m^\mathrm{th}$ snapshot)
\item once this motif $n$ is chosen, a cell $j$ is selected from $\psi_n$ (since $\psi_n(\xx_j)$ represents the probability that cell $j$ is present in motif $n$)
\end{itemize}
\end{itemize}
\end{itemize}
The snapshot $m$ then represents the set of $\q_T$ cells $j$ that have been drawn and can be reorganized as a list of $N_C$ cells with integer counts
$\q(\xx_j,t_m)$.

\begin{figure}
          \centerline{\includegraphics[height=6cm]{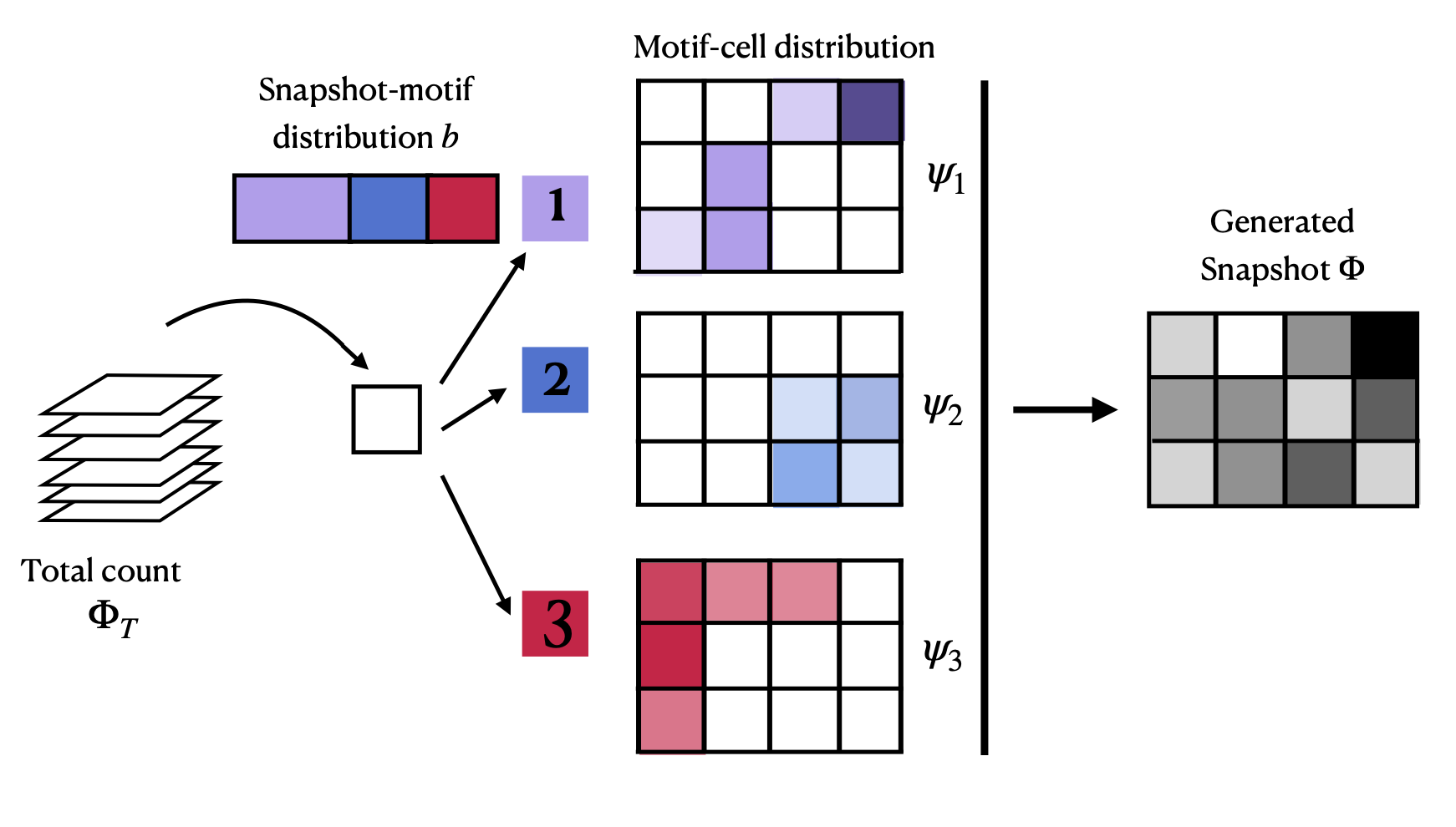}}
          \caption{ Schematics of the LDA generative model illustrated here for a field defined on 12 cells and generated with 3 motifs (corresponding to the blue, red and purple colours).  A snapshot $k$ is represented as a set of integer values defined on  an array of cells.
         Let $\Phi_T$ be the sum of  the integer values over the cells, which can be represented as a stack of $\Phi_T$ tokens of unit value.
         Each token is assigned in turn to a cell as follows: 
         a motif is selected by sampling the snapshot-motif distribution $b^k$ corresponding to this snapshot. Once the motif $n$ is selected, a cell is selected by sampling the motif-cell distribution $\psi_n$. At the end of the process, the  number of tokens at cell $j$ yields the value of the field $\Phi_k(\xx_j)$.
         As an alternative, the field can be  generated by summing directly the contributions of all distributions $\psi_n$ weighted by the corresponding component $b_n$   
         (see also text).}
        \label{fig:lda}
        \end{figure}

In Fluid Mechanics applications (\citep{kn:frihat21, kn:fery22}), sampling from the motif-cell distribution (step c)) can be replaced with a faster step, where the contribution of  each motif $n$ to snapshot $m$ is directly obtained from  the motif-cell distribution $\psi_n$ and the distribution
$b_n(t_m)$ and expressed as  $\q_T(t_m) b_n(t_m) \psi_n(\xx_j)$. The reconstructed field
is then the sum of the motif contributions. 
Figure~\ref{fig:lda} illustrates the LDA generative process on a $4 \times 3 $ grid for three topics.
 
\subsection{Interpretation and evaluation criteria}

By construction, the decomposition identifies {\it fixed} regions of space over which the intensity of the  scalar field is likely to be important at the same time. 
The connection between temperature motifs and plumes should be examined with caution
since plumes are  Lagrangian structures travelling and  possibly changing in shape and orientation through the shell. 
LDA motifs only aim to detect the Eulerian signature of  structures. 

Each  motif $n$ can be characterized in space through the motif-cell distribution $\psi_n^q$
(which integrates to 1 over the cells) and which will sometimes 
referred to as the motif in the absence of ambiguity. 
Each distribution has a maximum value $\psi_n^{max}$ and a maximum  location $\xx_n^{max}$ such that 
$\psi_n(\xx_n^{max})=\psi^{max}$.
One can also define   a characteristic area $\Sigma_n$ using
\begin{equation} \Sigma_n = \int_{\Omega}  1_{\{ \psi_n \ge   \psi_n^{max}/e \}} d\Omega
\label{eq:area}
\end{equation}
where $\Omega$ represents the plane of analysis and  the factor $1/e \sim 0.606$ is an arbitrary factor chosen by analogy with a Gaussian distribution.
Characteristic dimensions $l_i$ for the motif $n$ in the direction  $i$  can also be defined using
$ l_i^n = \left[\int  \psi_n (x_{n,i} - x_{n,i}^{max})^2 dx_i \right]^{1/2}$.
Each motif can also be characterized in time through the snapshot-motif distribution $b_n$, 
that will be  called the motif {\it weight} throughout the paper. 
The motifs can be ordered by their time-averaged weight, also called {\it prevalence},  
defined as  $ \langle b_n\rangle=\frac{1}{N_S} \sum_{m=1}^{N_S} b_n(t_m) $
where $\langle \cdot \rangle$ represents a time average.

LDA decompositions were carried out independently for the heat flux 
$\Phi= w \theta $, temperature fluctuations $\theta $ and the total kinetic energy 
$k = \frac{1}{2} (u^2+v^2+w^2)$.  To differentiate between these quantities, the motif topics and weights will be denoted respectively as  $\psi_n$, $\psi_n^\theta$ and $\psi_n^k$ and $b_n$, $b_n^\theta$ and $b_n^k$.
A useful tool for comparing the motifs associated with two different quantities is 
to compute the correlation coefficient matrix between the corresponding motif weights
(for instance if we compare the heat flux and the temperature motifs, each $nn'$ entry of the matrix will correspond to the correlation coefficient between $b_n$ and $b_{n'}^\theta)$).

As noted above, a reconstruction of the  field can be obtained  by using the inferred motif-cell distribution and snapshot-motif distribution to provide what we will call the LDA-Reconstructed field, defined as
\begin{equation}
 \q_R(\xx_j, t_m) =   \q_T(t_m)  \sum_{n=1}^{N_T} b_n^q(t_m) \psi_n^q(\xx_j)
\label{eq:lda}
\end{equation}
where $\q_T$ represents the sum  of the field values over the cells.
To evaluate the relevance of the decomposition, one can compute for each snapshot $m$ the instantaneous spatial correlation coefficient $C_m$ between a given field $\q$ and
its reconstruction $\q_R$ defined as
\begin{equation}
C_m(\q, \q_R)=\frac{\int (\tilde{q}(\xx,t_m) \tilde{q}_R(\xx,t_m) d\xx}
{\left(\int \tilde{q}^2(\xx,t_m) d\xx \int \tilde{q}_R^2(\xx,t_m) d\xx \right)^{1/2}}
\end{equation}
where $\tilde{q}$ represents the fluctuation 
$q(\xx,t_m)=q(\xx,t_m)-1/ |\Omega| \int_{\Omega} q(\xx,t_m) d\xx $
A global measure of the reconstruction is then given by 
$\langle C\rangle=\frac{1}{N_S}\sum_{m=1}^{N_S} C_m$, the average value of $C$ over all snapshots.

\section{Evaluation  of LDA for reconstruction and generation of the heat flux}

\subsection{Reconstruction}

\begin{figure}
\centering
	\begin{tabular}{ll}  
    
    \includegraphics[height=4cm]{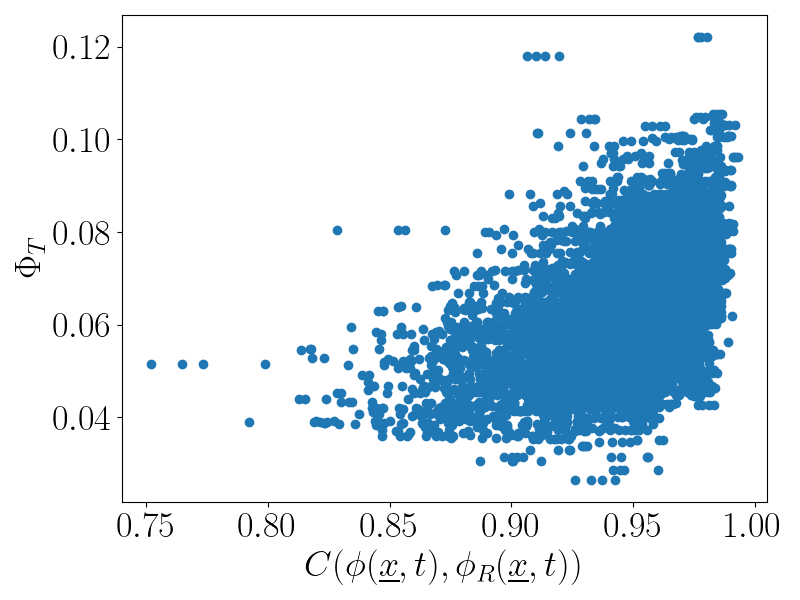}  & 
    \includegraphics[height=4cm]{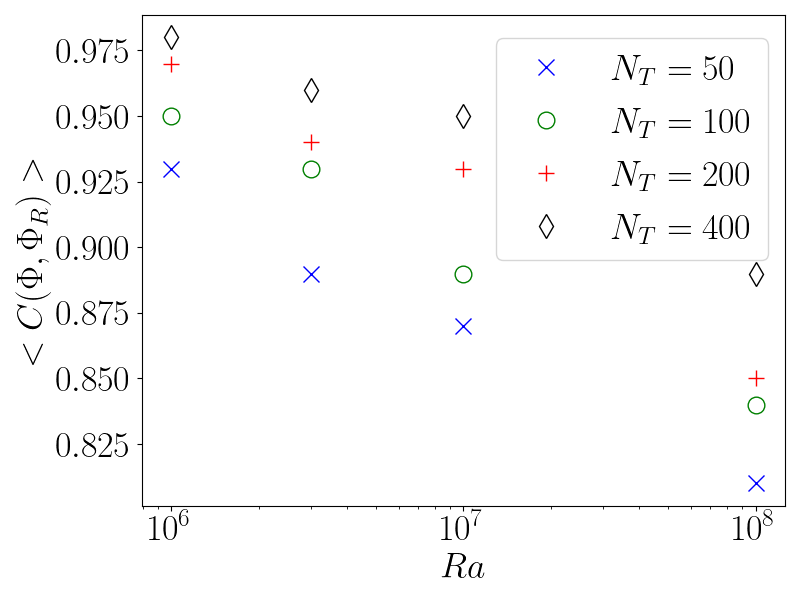}     \\  
	\end{tabular}
	  \caption{ 
   left) Instantaneous correlation coefficient between the projected and the true field as a function of the integral
convective heat flux for $N_T=100$ and $Ra=10^7$;
right) Average correlation coefficient $<C(\Phi,\Phi_R)>$ as a function of the Rayleigh number
and of the number of topics considered for both mid-planes. 
} 
   \label{correl}
\end{figure}

We first evaluate to which extent the LDA decomposition  provides an adequate reconstruction of the heat flux $\Phi$.
Figure~\ref{correl} (left) shows how the instantaneous value of the correlation coefficient $C_m(\Phi,\Phi_R)$ depends on the discrete integral of the field $q_T(t_m)=\sum \Phi(\xx_j,t_m)$. The Rayleigh considered is  $Ra=10^7$  and the number of topics is $N_T=100$, but the same trend was observed for all other Rayleigh numbers as well as all other values of $N_T$. 
Lower values of the correlation were associated with lower values of the total integrated heat flux, which illustrates
that the LDA representation is suited to capture extreme events.

Figure~\ref{correl} (right) presents the  mean correlation coefficient $\langle C_m(\Phi, \Phi_R)\rangle$ for different number of motifs and
different Rayleigh numbers on the vertical planes. 
Unsurprisingly, the correlation increases with the number of topics. 
It also decreases with the Rayleigh number, which is consistent with an increase in the complexity of the flow. However, the minimum value for the lower number of topics and the highest Rayleigh number was 0.8, which shows the relevance 
of the decomposition.

\begin{figure}
	\begin{tabular}{cccc}
         instantaneous field $\phi$ &   100 motifs  &   20 dominant motifs &   20-mode POD   \\   
	\includegraphics[trim=0cm 0 3cm 0, clip, height=3cm]{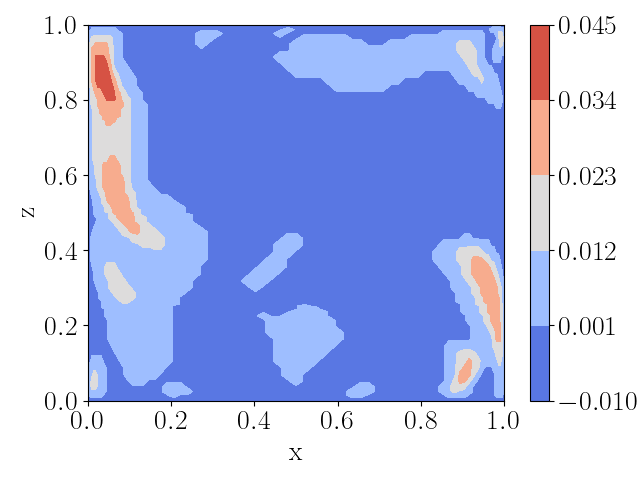} & 
	\includegraphics[trim=0cm 0 3cm 0, clip, height=3cm]{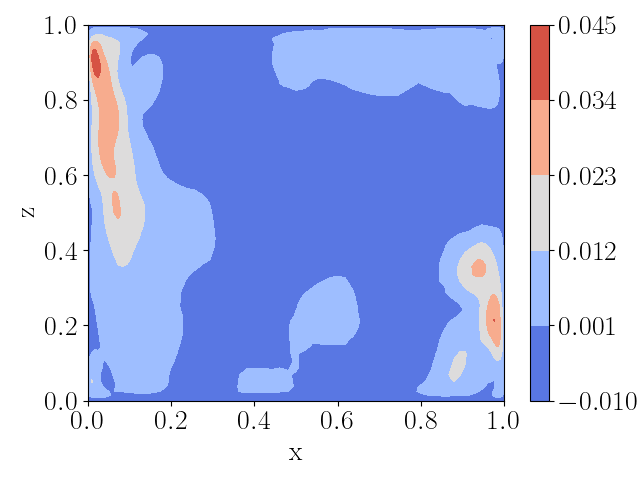} & 
	\includegraphics[trim=0cm 0 3cm 0, clip, height=3cm]{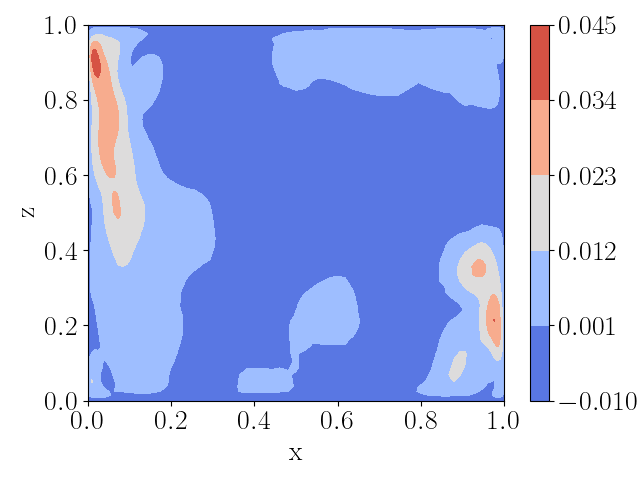} & 
	\includegraphics[trim=0cm 0 0cm 0, clip, height=3cm]{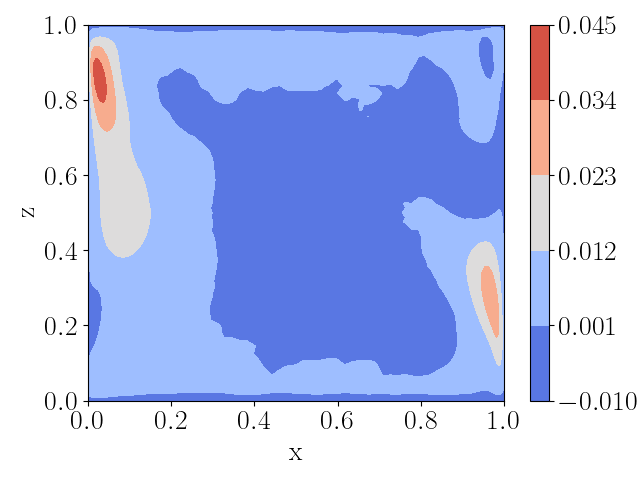}  \\ 

	\end{tabular}
	  \caption{ Example of an instantaneous snapshot 
   and its reconstructions at $Ra=10^7$. From left to right: original field, LDA-reconstructed field
   using $N_T=100$ motifs, LDA-reconstructed field using the 20 (instantaneously) most prevalent motifs, POD-reconstructed field using the  20 (on average) most energetic modes.
   }

	\label{fig:recons}
	\end{figure}

Figure \ref{fig:recons} 
compares an original  snapshot  at $Ra=10^7$ 
(based on the digitized signal) with different reconstructions:
i) the LDA-reconstruction based on $N_T=100$ motifs, ii) the reconstruction limited to the 20 most prevalent
topics (for this particular snapshot), iii) the  POD-based reconstruction based on the first 20 modes.
By construction, POD provides the best approximation of the field for a given number of modes. 
Since the distribution of the heat flux is intermittent in space and time, only a limited number of motifs is necesssary
to reconstruct the flow.
We note that  little difference was observed between the full LDA reconstruction and the reconstruction limited to the 20 most 
prevalent motifs, which highlights the intermittent nature of the field.
The relative error between the original and the reconstructed field is 29\% for the full LDA reconstruction, 34 \%
when the 20 most prevalent modes are retained in the reconstruction.
In contrast, limiting the POD to 20 global modes slightly lowers the  quality of the reconstruction, with a global error of 38\%. 
It should be noted that the 20 dominant POD modes correspond to an average over all snapshots, while the 20 most prevalent LDA modes are selected for that specific snapshot.  
On average,  the reconstructed field based on keeping the 20 most prevalent motifs differed by less than 10\%
from the full 100-mode reconstruction and the average correlation coefficient $C=<C_m(\Phi, \Phi_R)>$ decreased from 0.89 to 0.83. This shows that LDA can provide a compact representation of the local heat flux that compares reasonably well with POD.

\subsection{Generation}

The ability to generate  statistically relevant synthetic fields is of interest for a number of applications, such
as accelerating computations or developing multi-physics models. 
As a generative model, LDA makes it possible to produce such a set of  fields, the statistics  of which can be compared with those of the original fields used to extract the motifs, as well as with those of the corresponding
LDA-Reconstructed fields. 
It would also be useful to compare the generated LDA data set with 
one generated using POD.
To this end, we  generated two sets of 4000 new fields using both LDA and POD. The same number $N_T=100$ of POD modes and LDA motifs were used to generate the datasets. The plane in which the data is generated is assumed to be the
$y=0.5$ plane.
The different fields to be compared are therefore the following:
\begin{enumerate}
	\item the original (digitized) field $\Phi$ defined in section 3.2  with equations~\eqref{eq:digit1} and \eqref{eq:digit2}
	\item the LDA-Reconstructed field (LDA-R) as defined in equation~\eqref{eq:lda}
	\item the LDA-Generated field (LDA-G): the field is constructed by sampling weights $\tilde{b}_n(t_k)$ 
from snapshot-motif distributions and then reconstructing 
\begin{equation}
  \Phi^{LDA-G}(\xx_j, t_k) =   \Phi_T \sum_{n=1}^{N_T} \tilde{b}_n(t_k) \psi_n(\xx_j), \label{eq:ldag}
\end{equation}
where $\Phi_T$ is a random variable obtained by sampling a Poisson distribution with the same statistics as the original database.
\item the POD-Generated field (POD-G): the field is constructed by independently sampling $N_T$ POD mode amplitudes $\tilde{a}_n$   from the POD  amplitudes of the original database
  \begin{equation}
    \Phi^{POD-G}(\xx_j, t_k) =  \sum_{n=1}^{N_T} \tilde{a}_n(t_k) \phi_n(\xx_j)
    \label{eq:podg}
  \end{equation}
\end{enumerate}

\begin{figure}
	\begin{tabular}{cccc}
	original field & LDA-projected field & POD-generated field &  LDA-generated field \\
	 \includegraphics[trim={0.5cm 0 3.5cm 0},clip,height=3cm]{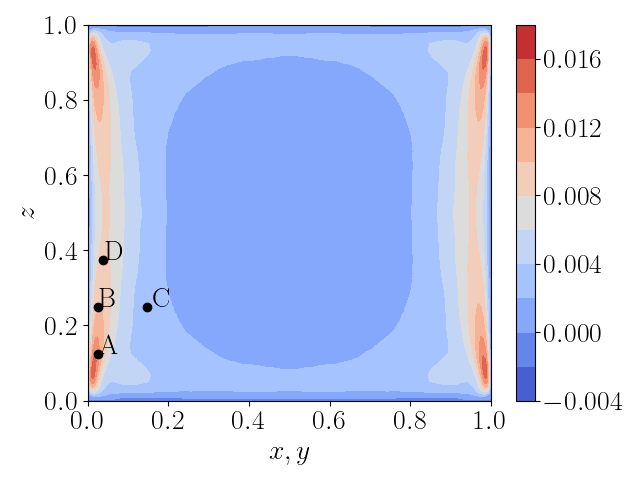} &
	 \includegraphics[trim={0.5cm 0 3.5cm 0},clip,height=3cm]{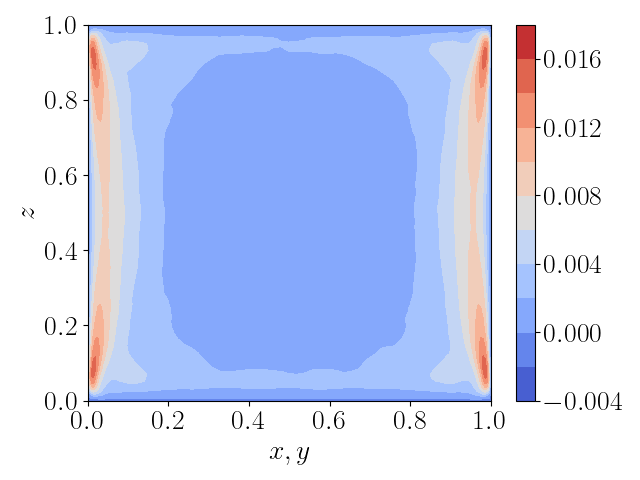} &
	 \includegraphics[trim={0.5cm 0 3.5cm 0},clip,height=3cm]{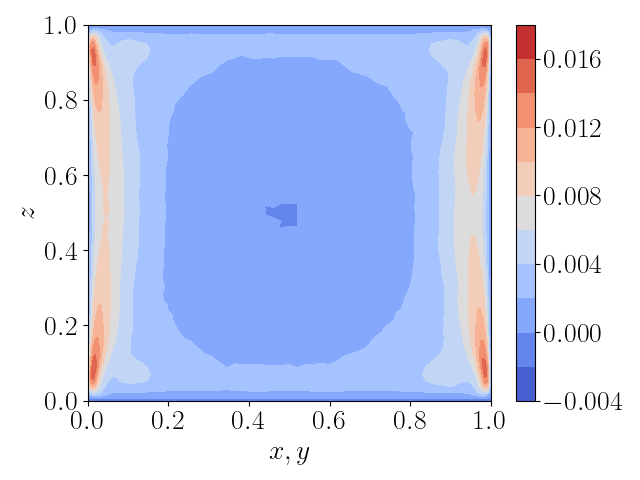} &
	 \includegraphics[trim={0.5cm 0 0cm 0},clip,height=3cm]{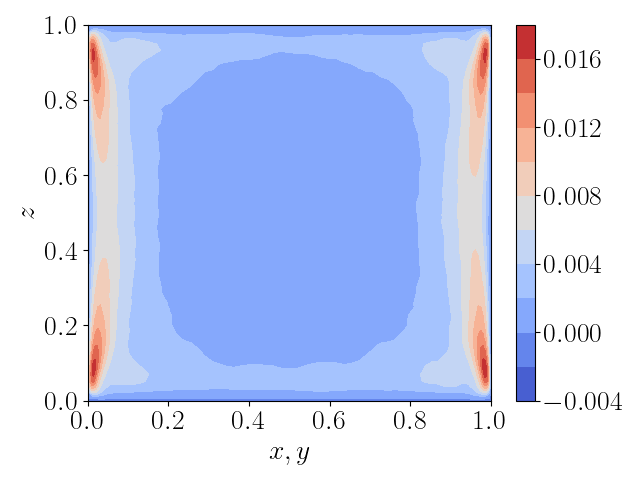} \\ 

	\end{tabular}
	  \caption{ Time-averaged value of the convective heat flux for different databases at $Ra=10^7$.
   From left to right: original fields, LDA-reconstructed (LDA-R) fields using $N_T=100$ motifs, POD-generated (POD-G) fields using 100 modes, LDA-generated (LDA-G) fields using $N_T=100$ modes.
   }
	\label{fig:stats}
	\end{figure}

The time-averaged fields corresponding to the different databases are compared in Figure~\ref{fig:stats}. 
A good agreement is observed for all datasets, with global errors
of 4\%, 8\% and 3\% for respectively the LDA-reconstructed, the
LDA-generated and the POD-generated datasets.
Although it provides the lowest error (as  could be expected), 
the POD-generated data set overestimates negative values in the core of the cell. 
       
\begin{figure}
	\begin{tabular}{cccc}
  A & B & C & D \\
	 \includegraphics[height=2.5cm]{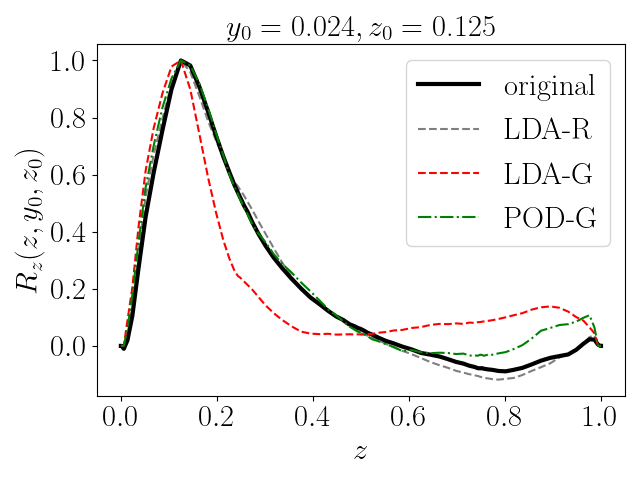} &
  	 \includegraphics[height=2.5cm]{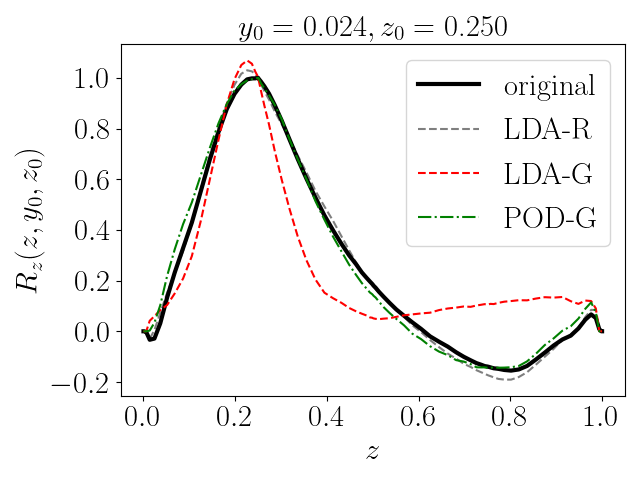} &
    	 \includegraphics[height=2.5cm]{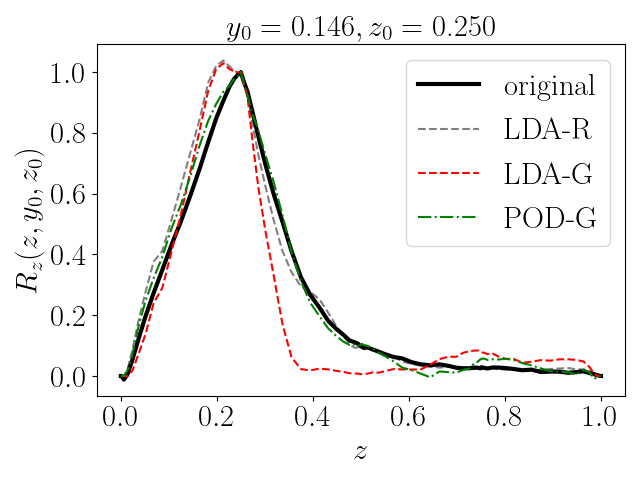} &
	 \includegraphics[height=2.5cm]{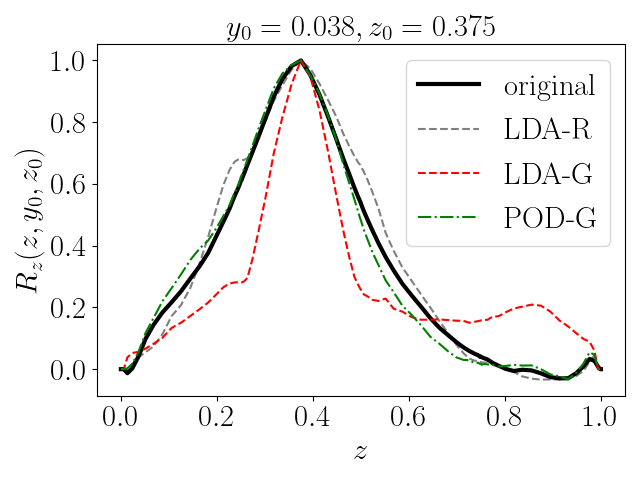} \\ 
	 \includegraphics[height=2.5cm]{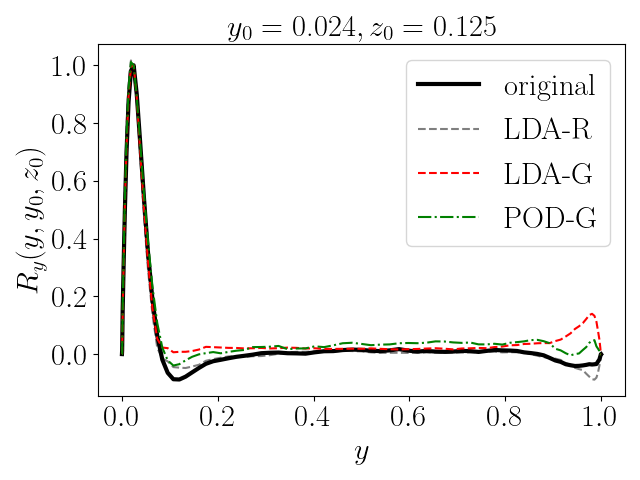} & 
   \includegraphics[height=2.5cm]{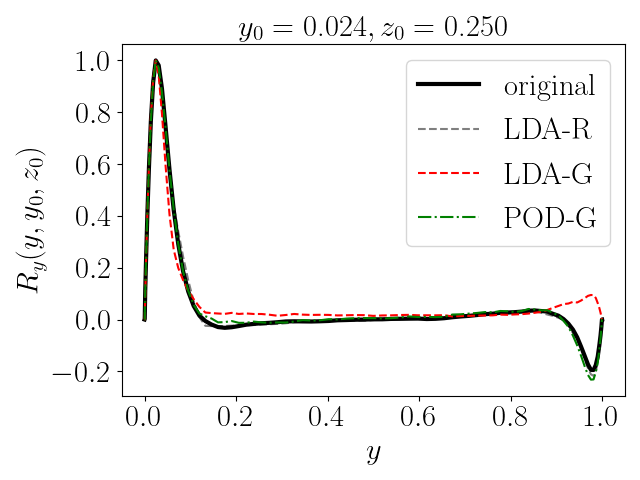} &
    	 \includegraphics[height=2.5cm]{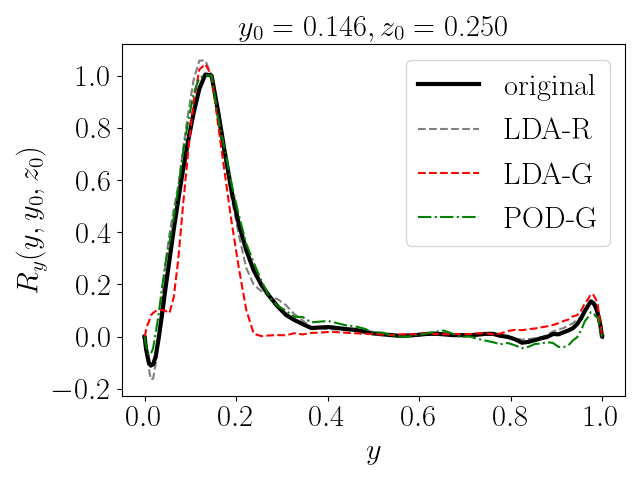} &
	 \includegraphics[height=2.5cm]{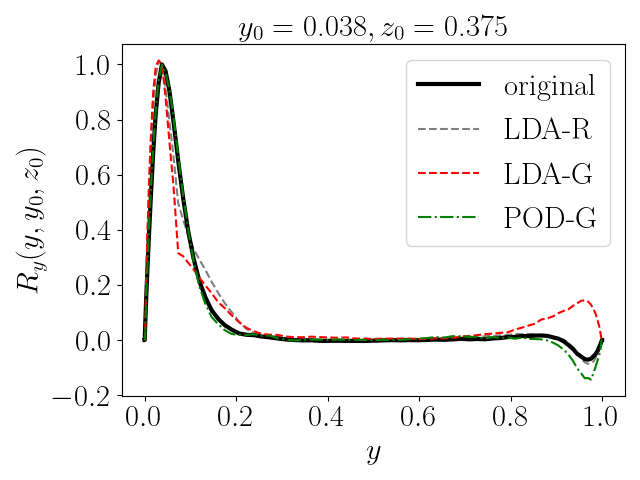}  \\ 
	\end{tabular}
	  \caption{ Autocorrelation of the  convective heat flux at selected locations (see Figure~\ref{fig:stats}). Top row: $R_y$ (horizontal direction); 
Bottom row: $R_z$ (vertical direction). See text (equations~\eqref{eq:Ry} and \eqref{eq:Rz}) for definition. }
	\label{fig:vert:autocorrel}
	\end{figure}

For a given location $(y_0, z_0)$, we  defined spatial autocorrelation functions in the horizontal and vertical directions as: 
\begin{equation} R_y(y,y_0,z_0)= \frac{ \langle \Phi(y,z_0,t) \Phi(y_0,z_0,t) \rangle}{ \langle \Phi(y_0,z_0,t)^2 \rangle}
\label{eq:Ry}
\end{equation} 
\begin{equation} 
 R_z(z,y_0,z_0)= \frac{ \langle \Phi(y_0,z,t) \Phi(y_0,z_0,t) \rangle}{ \langle \Phi(y_0,z_0,t)^2 \rangle} 
\label{eq:Rz}
\end{equation} 
The autocorrelation functions are  displayed  
in Figure~\ref{fig:vert:autocorrel} for the selected locations indicated in Figure~\ref{fig:stats}, which correspond
to regions of  high heat flux.
We can see that  that in all cases, the flux remains correlated over much longer
vertical extents than in the horizontal direction. 
Both  the LDA-reconstructed and the POD-generated autocorrelations approximate the original data well 
 - again, by construction, POD-based fields are optimal to reconstruct second-order statistics. 
The LDA-generated autocorrelation is not as close to the original one, but still manages to capture the characteristic spatial
scale over which the fields are correlated.

\begin{figure}
	\begin{tabular}{cccc}
 A & B & C & D \\ 
	 \includegraphics[height=2.5cm]{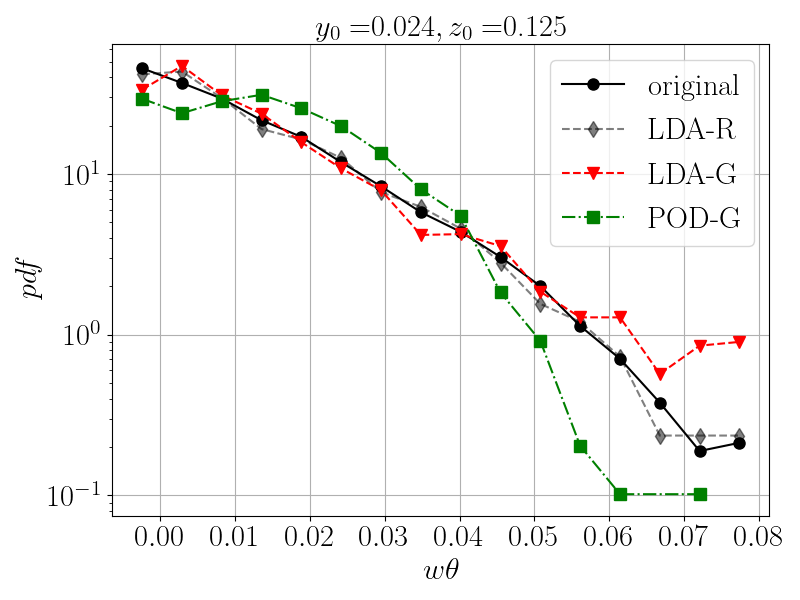} &
	 \includegraphics[height=2.5cm]{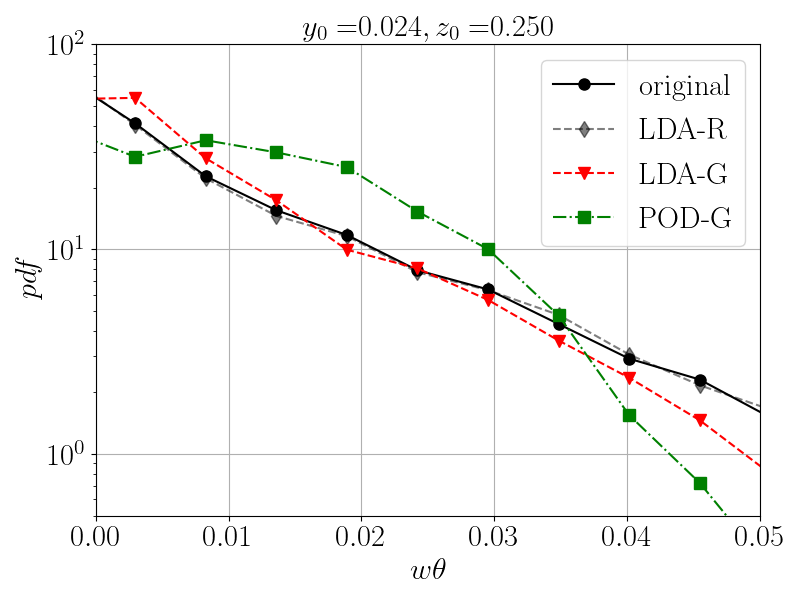} &
	 \includegraphics[height=2.5cm]{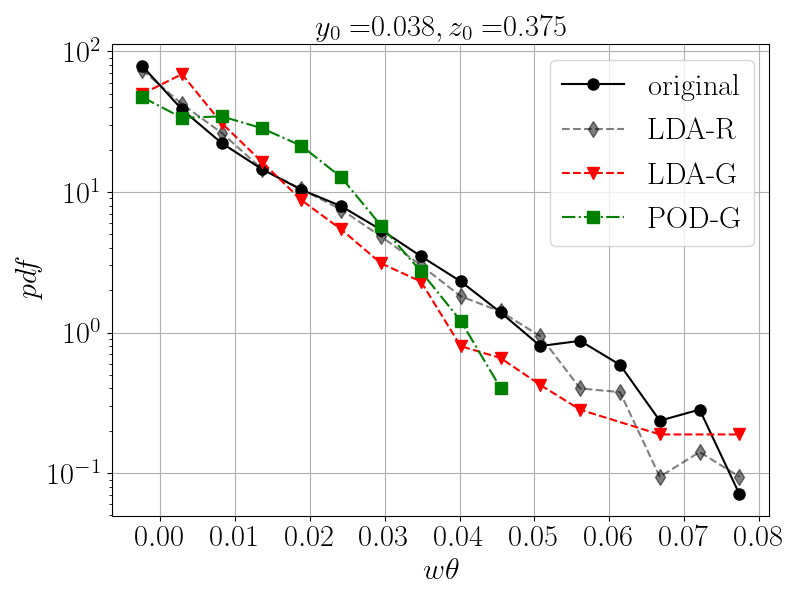} &
	 \includegraphics[height=2.5cm]{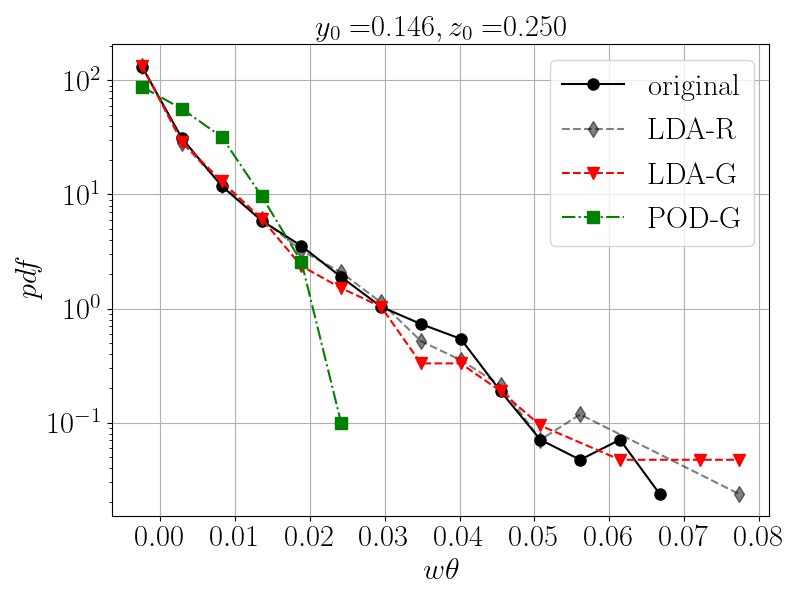} \\ 
	\end{tabular}
	  \caption{ Probability density function of the convective heat flux at the selected locations indicated in Figure~\ref{fig:stats}.}
	\label{fig:vert:hist}
	\end{figure}

One-point pdfs of the  flux $\Phi$ are represented in Figure~\ref{fig:vert:hist} for the same selected locations (again, indicated in Figure~\ref{fig:stats}).
POD-generated fields tend to overpredict lower values  and
underpredict higher values, which means that they do not capture well the intermittent features of the heat flux.
The LDA-generated fields display a better agreement with the original fields and are in particular able to reproduce the exponential tails of the distributions.

\section{Heat flux motifs}
\subsection{Spatial organization}

\begin{figure}
    \centering
	\begin{tabular}{ccc}  
 $Ra=10^6$ &  $Ra=10^7$ & $Ra=10^8$ \\
	\includegraphics[trim=0cm 0 0cm 0, clip, height=3.5cm]{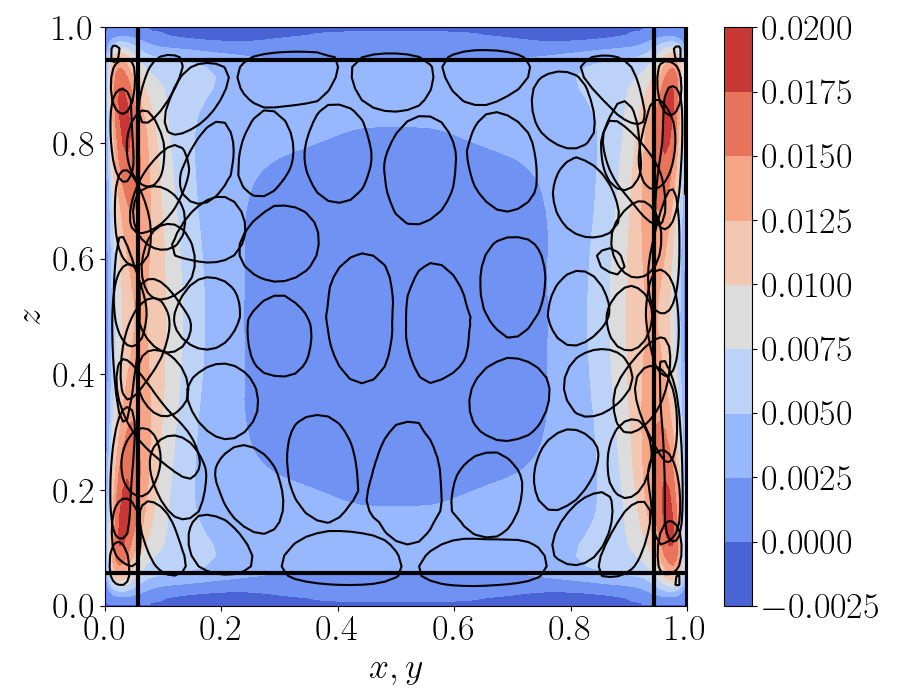} & 
        \includegraphics[trim=0cm 0 0cm 0, clip, height=3.5cm]{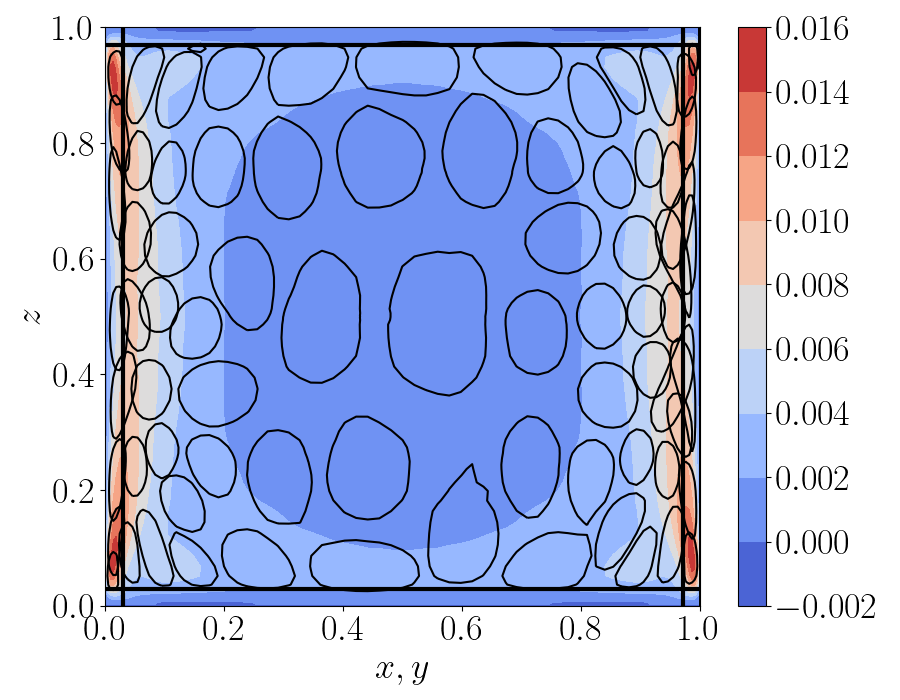} & 
	\includegraphics[trim=0cm 0 0cm 0, clip, height=3.5cm]{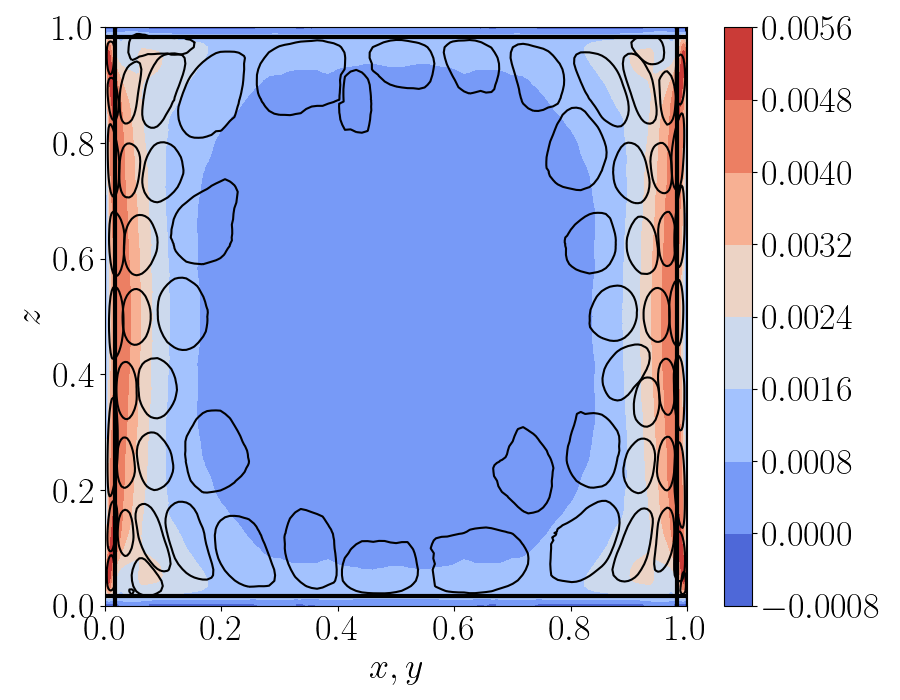}  \\ 
	\end{tabular}
  \caption{ 
Spatial distribution of the positive-flux motifs  in the vertical mid-plane for 
$N_T =100$. The motifs are materialized by a black line corresponding to a probability contour of $0.606\,\psi_n^{max}$.
The vertical lines correspond to the boundary layer thickness.
The time-averaged convective heat flux 
is represented in the background. 
}
	\label{fig:alltopics}
	\end{figure}

We now describe the spatial organization of the motifs through the motif-cell distribution $\psi_n$.
The general trends reported below held for all values of $N_T$ considered, which ranged
from 50 to 400. 
For all Rayleigh numbers, most LDA motifs were found to be associated with a positive flux (i.e were associated with the first $\tilde{N}$ cells in the decomposition).
A few negative (counter-gradient) motifs were also identified, but their average weight was generally very small (at most
10 \% of that of the dominant motif). 
We therefore chose to focus only on the  motifs with a positive contribution to the heat flux.
Figure~\ref{fig:alltopics} (left) displays these motifs for three different Rayleigh numbers for $N_T=100$. 
{ The case $Ra=3$ $10^6$ was omitted as it 
did not show significant differences with the case  $Ra=10^6$.}
The  motif-cell distribution is materialized by a black line corresponding to the iso-probability contour of $0.606 \psi_n^{max}$, which can be compared with the
average value of the heat flux at this location.
For all Rayleigh numbers, the motifs  are clustered in the regions of high heat flux, close to the vertical walls.
Within the vertical boundary layers, motifs are elongated in shape.
Outside the vertical boundary layers, the motifs are more isotropic and tend to increase in shape as one moves away from the walls. 
Outside the horizontal boundary layers, the motif-cell distributions are elongated in the direction of the wind, with a horizontal orientation
in the center of the cell, and a gradual vertical shift closer to
the walls.
Large motifs are found in the bulk at $Ra=10^6$ and $Ra=10^7$ (it was also the case at $Ra=3$ $10^6$). In contrast, fewer, smaller motifs are found in the bulk at $Ra=10^8$ in the central region
$x/y \in [0.2, 0.8]$, signalling a loss of spatial coherence in the bulk at this Rayleigh number.

\begin{figure}
    \centering
	\begin{tabular}{cccc}
	  $Ra=10^6$ & $Ra=3$ $10^6$  &  $Ra=10^7$ &  $Ra= 10^8$  \\  
	 \includegraphics[trim=0 0 0cm 0, clip,height=3cm]{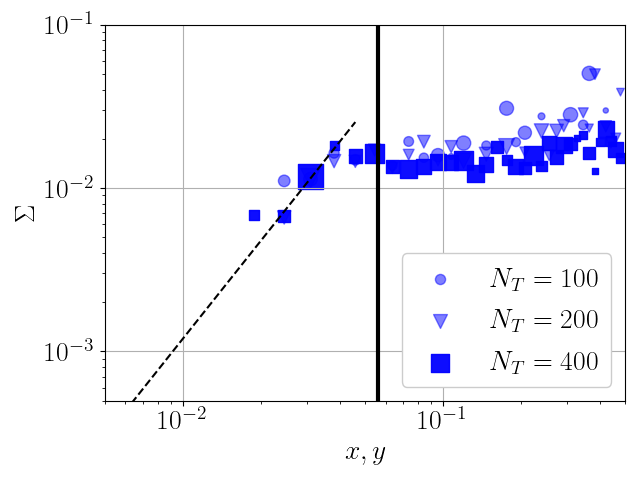}  &
	 \includegraphics[trim=2.5cm 0 0cm 0, clip,height=3cm]{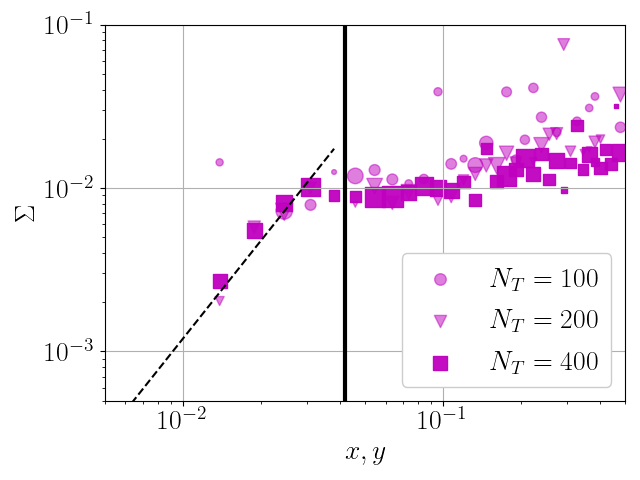} & 
	 \includegraphics[trim=2.5cm 0 0cm 0, clip,height=3cm]{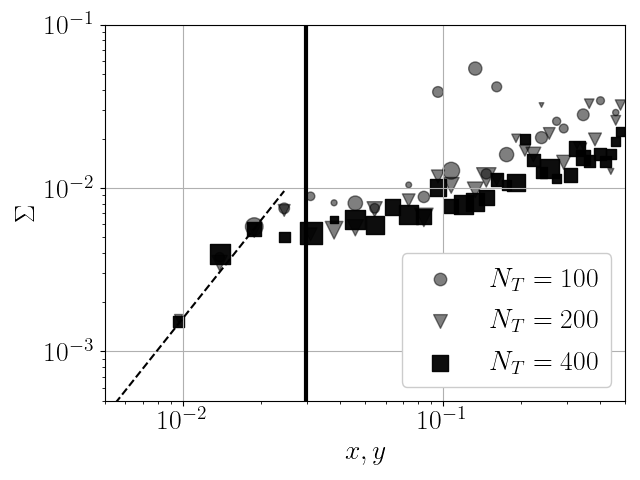} &
	 \includegraphics[trim=2.5cm 0 0cm 0, clip,height=3cm]{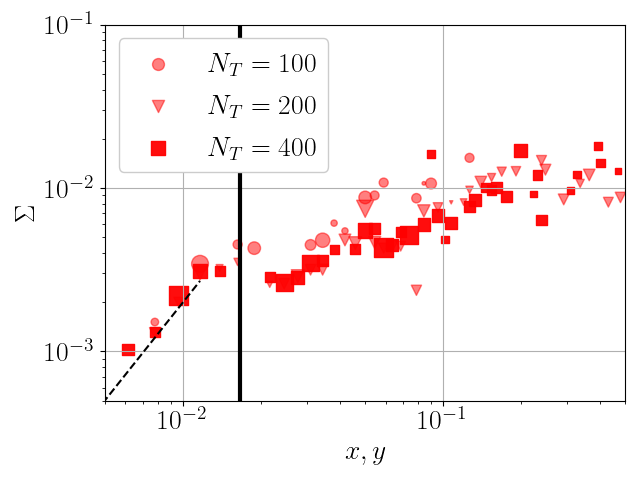} \\ 
	\end{tabular}
	  \caption{ Distribution of motif areas (see definition in equation~\eqref{eq:area}) in the vertical mid-plane with the distance from the lateral walls at varying Rayleigh numbers. The size of the symbols shown in the picture is proportional to the fraction of motifs over which the average was performed.
     The black solid lines indicate the boundary layer thickness. The dashed lines
     have slope 2.}
	\label{fig:vert:area}
	\end{figure}

In general, the motif size seems to decrease with the Rayleigh number. This is confirmed by
Figure~\ref{fig:vert:area}, which represents  the average motif area as a function of their distance from the
 vertical walls.
In order to avoid the influence of the horizontal plates, we only considered the motifs located at a vertical distance larger than $0.07$  from the horizontal walls (i.e. outside the horizontal boundary layer).  
The size of the symbols shown in the picture is proportional to the fraction of motifs over which the  average was performed.  
Results were relatively robust with respect to the number of topics $N_T$, although some dependence on $N_T$ is observed in the center of the cell.
Within the boundary layer, the motif area grows quadratically, which means that the characteristic size of the motif is proportional to the wall distance.
We note that a similar scaling was found for turbulent eddies in pressure-gradient 
driven turbulence such as channel flow \citep{kn:frihat21}. 
Further away from the vertical wall, after a short plateau at the edge of the boundary layer, a slower increase in the motif size was observed with a rate that increased
with the Rayleigh number, so that the motif area was about the same  (on the order of $0.02 $) for all Rayleigh numbers in the center of the cell. This suggests the presence of a double scaling for the motifs:
one based on the boundary layer thickness, and one based on the cell size.
The decrease in size with the Rayleigh number appears consistent with a dependence on the boundary layer thickness but also with an increase 
of the fragmentation by the bulk turbulent fluctuations, in agreement
with the literature \citep{kn:bosbach12, vdpoel15}.
The difference observed at the highest Rayleigh number also signals 
that the flow  is still evolving and has not reached an asymptotic state.

\subsection{Dominant motifs}

\subsubsection{Spatial description}

Owing to the symmetry of the database, the motifs in the vertical plane
$(x,z)$ (resp. $(y,z)$) should approximate the
symmetry $x \rightarrow 1-x$  (resp. $y \rightarrow 1-y$), and  $z \rightarrow 1-z$
(complete symmetry cannot be expected owing to the stochastic nature of the decomposition).   

\begin{figure}
\begin{tabular}{ccccc}
\includegraphics[trim=3cm 0 3cm 0, clip, height=4cm]{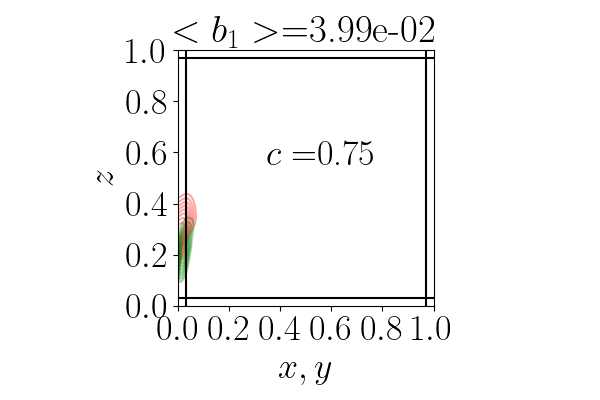}   &
\includegraphics[trim=3cm 0 3cm 0, clip, height=4cm]{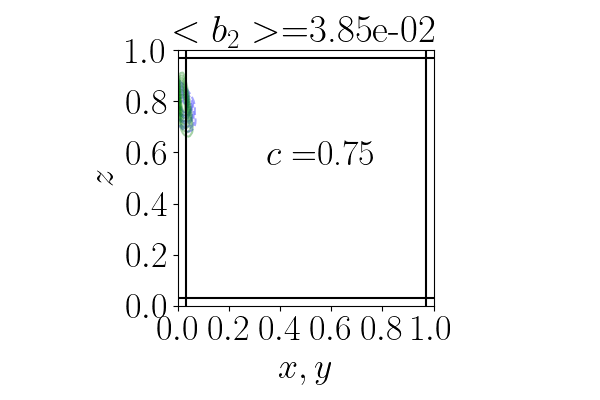}   &
\includegraphics[trim=3cm 0 3cm 0, clip, height=4cm]{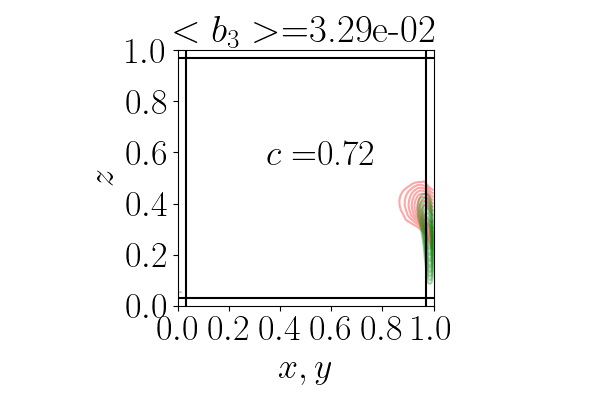}   &
\includegraphics[trim=3cm 0 3cm 0, clip, height=4cm]{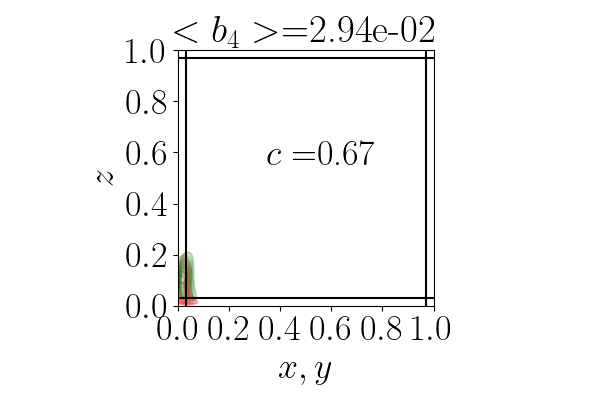}   & 
\\
\includegraphics[trim=3cm 0 3cm 0, clip, height=4cm]{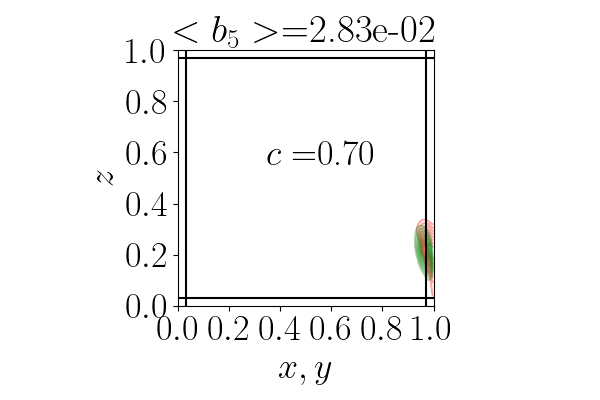}   &
\includegraphics[trim=3cm 0 3cm 0, clip, height=4cm]{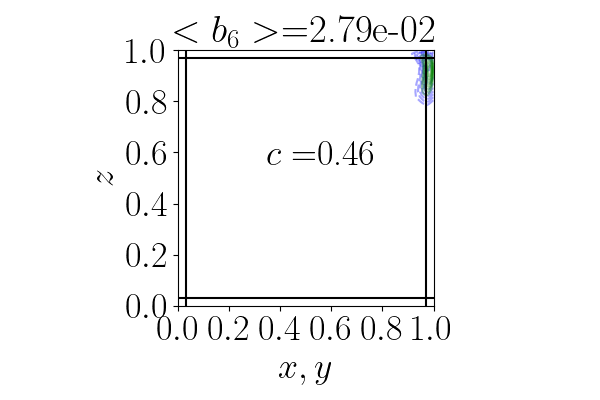}   &
\includegraphics[trim=3cm 0 3cm 0, clip, height=4cm]{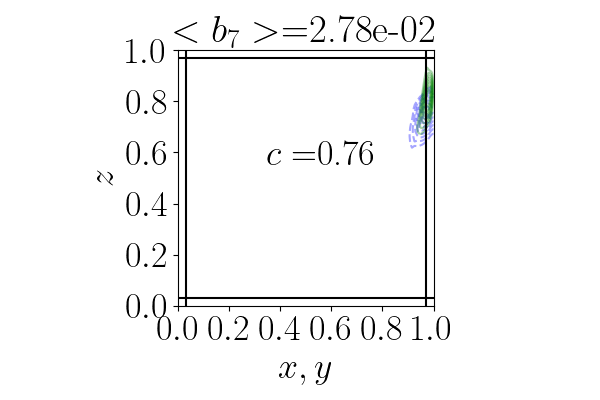}   &
\includegraphics[trim=3cm 0 3cm 0, clip, height=4cm]{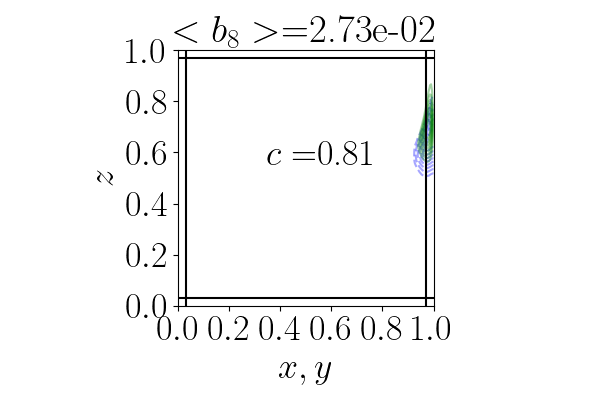}   & \\
\end{tabular}
\caption{ Dominant heat flux motif (green lines) ordered by prevalence 
and associated  temperature motifs (blue for negative and red for positive fluctuations)  at $Ra=10^7$. 
Contour levels go from $0.2$ to $0.9$ $\psi_n^{max}$ with increments of 0.1 $\psi_n^{max}$. 
$c$ is the maximum correlation coefficient between the heat flux and temperature motif weights.}
\label{fig:comptopicwttra1E7}
\end{figure}

{ To help interpret the heat flux motifs, we compare them with LDA motifs
corresponding to temperature fluctuations.
 The eight most prevalent heat motifs are represented in figure \ref{fig:comptopicwttra1E7}  (green lines). 
The prevalence of each motif is indicated at the top of each plot.
Most motifs have similar sizes and are located close to the 
side walls at about a similar  height, except for motifs 4 and 6, which have a smaller extent and are located closer
to the horizontal wall. 
The same value of $N_T=100$ was used for both heat flux and temperature.

For a heat flux motif $n$ with weight $b_n$,
we identified the temperature motif  $j$    
that maximized the correlation coefficient between the heat flux and the temperature motif weights
$C(b_n, b_j^\theta)$. 

The maximal value of this coefficient, denoted $c$, is represented on each plot and is generally very high (about 0.7)  
- especially in view of the intermittent nature of the weights.  
The best correlated heat flux and temperature motifs are close to each other in space, with 
a larger spread for temperature motifs.   
In all cases, flux motifs in the lower (resp. higher) portion of the side walls correspond to positive (resp. negative)  
fluctuations. 
Dominant heat flux motifs can be therefore interpreted as the wall imprint of hot plumes rising in
the boundary layer (resp. cold plumes descending in the boundary layer).
The same observations were made at all other Rayleigh numbers. 
}

\begin{figure}
          \centering
          \begin{tabular}{cc}
	  \includegraphics[trim=0 0 0cm 0, clip, height=4cm]{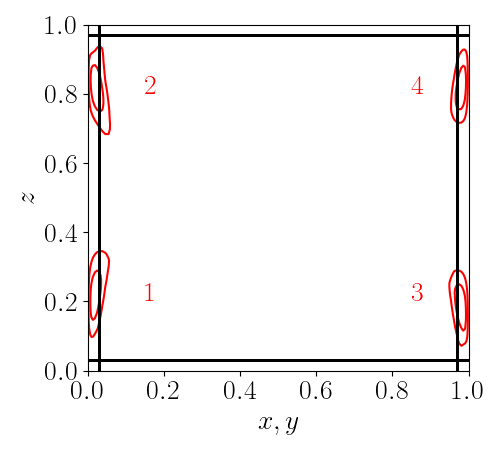} &
         \includegraphics[trim=0 0 0cm 0, clip, height=4cm]{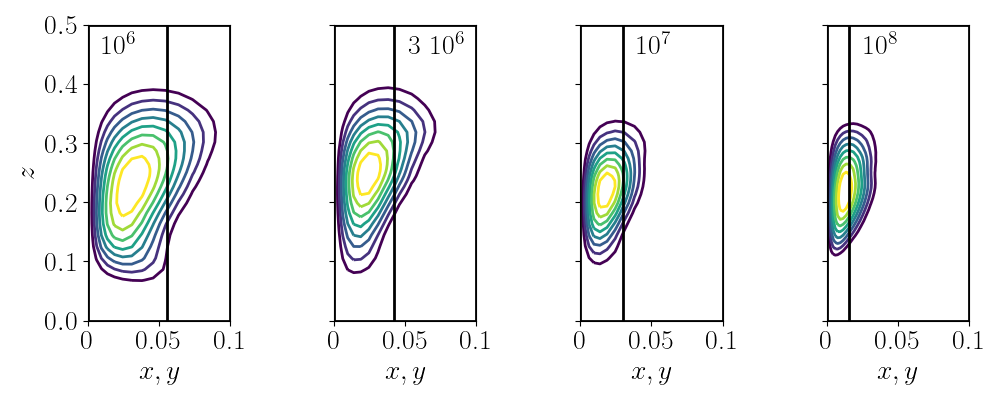}  \\
          \end{tabular}
	  \caption{  left) Dominant motifs at Rayleigh numbers $Ra=10^7$ for  $ N_T=100$. The contour lines correspond to 0.1 $\psi^{max}$ and 0.3 $\psi^{max}$. The motif labels correspond to those of 
            figure \ref{fig:pdtt}; right) Characteristic dominant  motif  at different Rayleigh numbers $ N_T =100$. Isocontours of $\psi_1$ at $0.1 i \psi_1^{max}, i=2, \ldots 9 $. The black lines correspond to the boundary layer thickness.
          }
          \label{fig:topicp}
        \end{figure}

Four of these dominant motifs at $Ra=10^7$ are represented in Figure~\ref{fig:topicp} (left) for $N_T=100$.
As noted above, they consist of elongated structures lying mostly in the boundary layer, and located at a vertical distance of
about 0.25  from the horizontal walls.
Although the positions and sizes of the four identified motifs may slightly vary from one to the other, their features are generally similar 
and a characteristic motif can be obtained from taking the average over all four motifs.
Figure~\ref{fig:topicp} (right) represents this characteristic motif  for the various Rayleigh numbers.
We can see that the dominant motifs are always located mostly within the boundary layer, with a maximum at
a height of about $0.25$.
Their characteristic width $l_y$ was found to decrease as $Ra^{-0.23 \pm 0.04}$, which matches the scaling of the boundary layer thickness.

\begin{figure}
    \centering
	\begin{tabular}{cc}
	  \includegraphics[height=6cm]{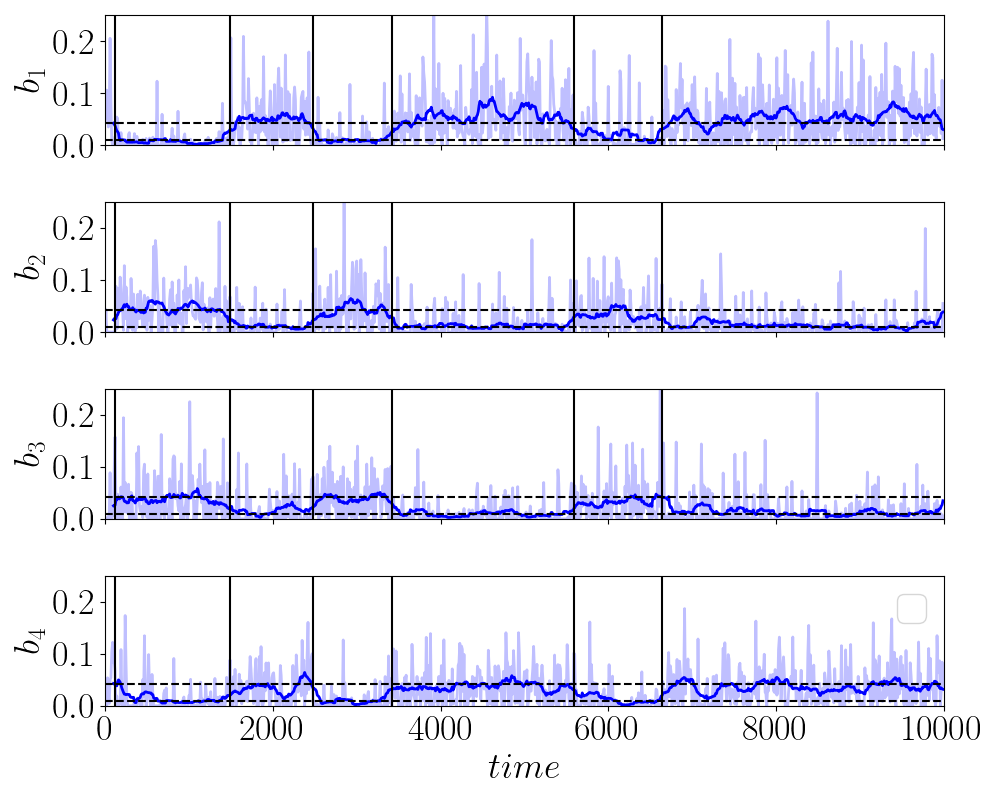} & 
	  \includegraphics[height=6cm]{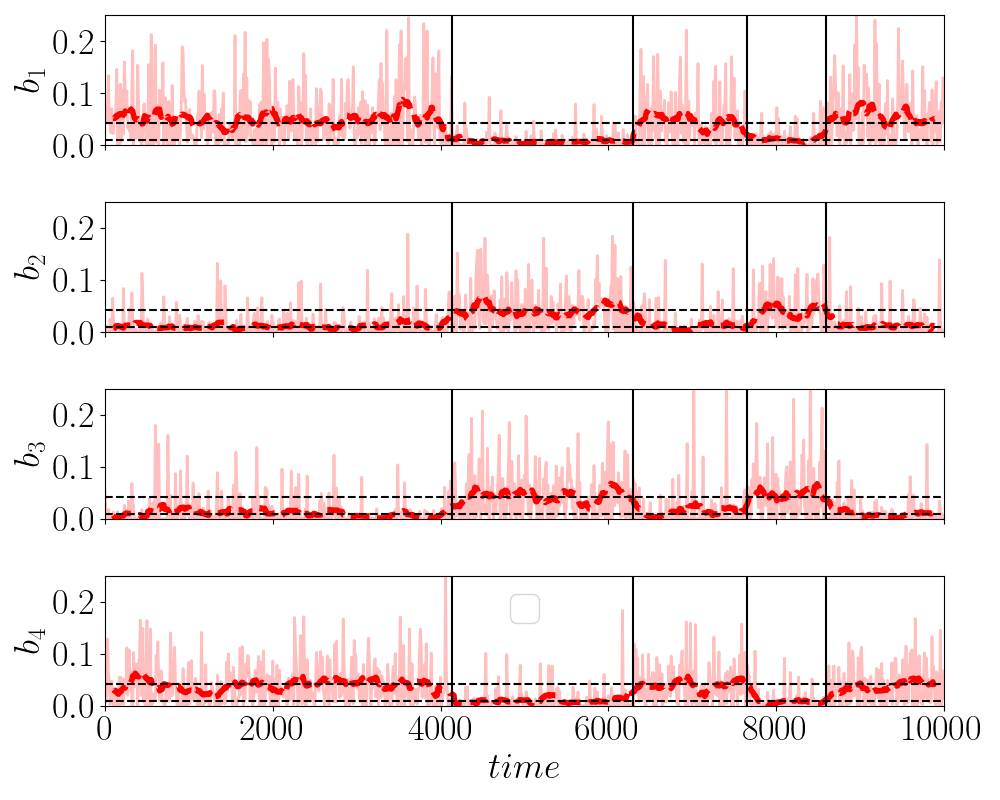 }  \\
	\end{tabular}
	   \caption{ Evolution of the  snapshot-motif distributions $b_n$ for the four dominant motifs
(see  figure \ref{fig:topicp} for labels) at $Ra=10^7$ and for $N_T=100$: left) plane $x=0.5$ right) plane $y=0.5$. The thick line corresponds to a moving average over 200 convective units (4 recirculation times $T_c$). The horizontal dashed lines correspond to the values $b_-=0.017$ and $b_+=0.035$. The vertical lines correspond to the changes in angular momentum.} 
	\label{fig:pdtt}
      \end{figure}

\begin{figure}
	\begin{tabular}{cc}
	  \includegraphics[height=6cm]{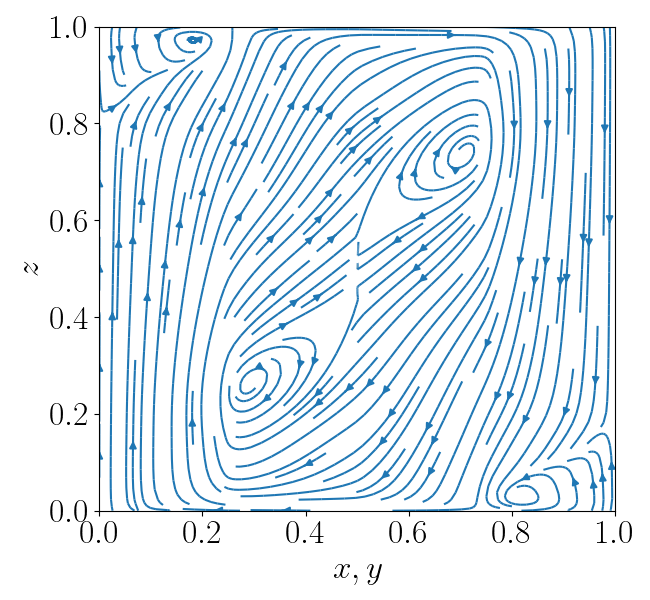} & 
	  \includegraphics[height=6cm]{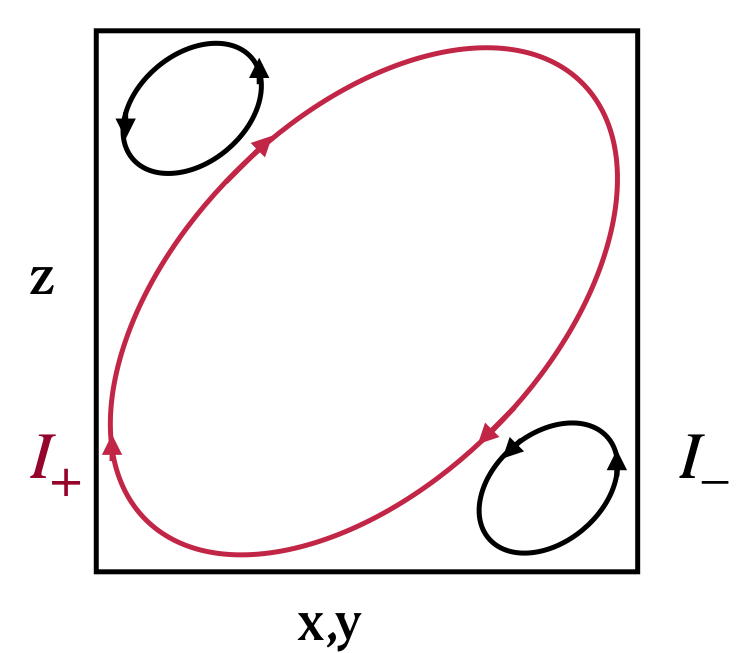 }  \\
	\end{tabular}
	
	  \caption{ Left) Streamlines of the  flow conditionally averaged on the high weight value of $b_1$; 
   Right) Schematics of the cell organization in the vertical mid-plane - the large-scale circulation (in red) corresponds to the $I_+$ state while the corner structure (in black) corresponds to the $I_-$ state.} 
	\label{fig:schemacube}
	\end{figure}

The evolution of the snapshot-motif distribution, or motif weight, is represented in Figure~\ref{fig:pdtt} for $Ra=10^7$.
We can see that the behavior of the motif weight depends on the sign of the global momentum represented in Figure~\ref{fig:Lxy}.
When a moving average of $T_f=200$ time units, corresponding to 4 recirculation times $T_c$, 
was applied, two quasi-stationary states $b_+$ and $b_-$ could be identified in each plane
- they are materialized by the dashed horizontal black lines indicated in Figure~\ref{fig:pdtt}.
The two states appear to correspond to the sign of the angular momentum component i.e. the orientation of the large-scale circulation $I$. 
Streamlines of the flow conditionally averaged on the higher weight
value of $b_1$ are represented in figure ~\ref{fig:schemacube} left). 
They indicate that for the higher characteristic value of the 
weight, $b_+$, the motif is associated with the  large-scale circulation while it is associated with the corner vortex on the opposite side for the lower weight value, $b_-$, as summarized in figure ~\ref{fig:schemacube} right).

This indicates that information about the large-scale reorientation can be extracted from local measurements.  
Two states, $I_+$ and $I_-$, respectively corresponding to  the large-scale circulation and corner vortex can be defined from the weight of the dominant motif $b_1$  using
\begin{equation}
I_+=\{k |   \langle b_1(t_k)\rangle_{T_f} > \langle b_1\rangle \}  \mbox{ and  } I_-=\{k |   \langle b_1(t_k)\rangle_{T_f} < \langle b_1\rangle \}
\end{equation}
where $\langle\cdot\rangle_{T_f}$ represents the moving average over $T_f$.
 The average weights conditioned on $I_+$ and $I_-$ are
 respectively $b_+$ and $b_-$.

\begin{figure}
    \centering
	\begin{tabular}{ccc}
	  $Ra=10^6$ & $Ra=3$ $10^6$   \\  
	 \includegraphics[trim=0 0 0cm 0, clip,height=4cm]{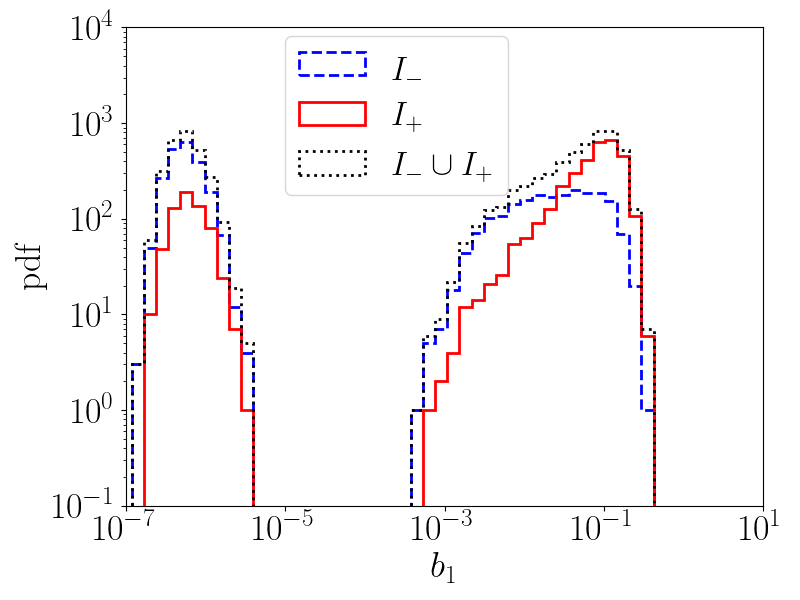} &
	 \includegraphics[trim=0cm 0 0cm 0, clip,height=4cm]{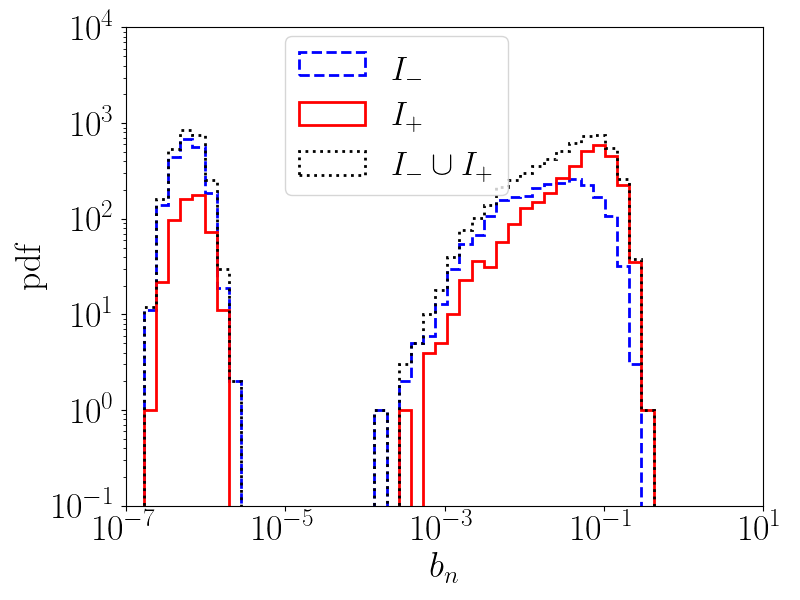}  \\ 
$Ra=10^7$ &  $Ra= 10^8$  \\  
	 \includegraphics[trim=0cm 0 0cm 0, clip,height=4cm]{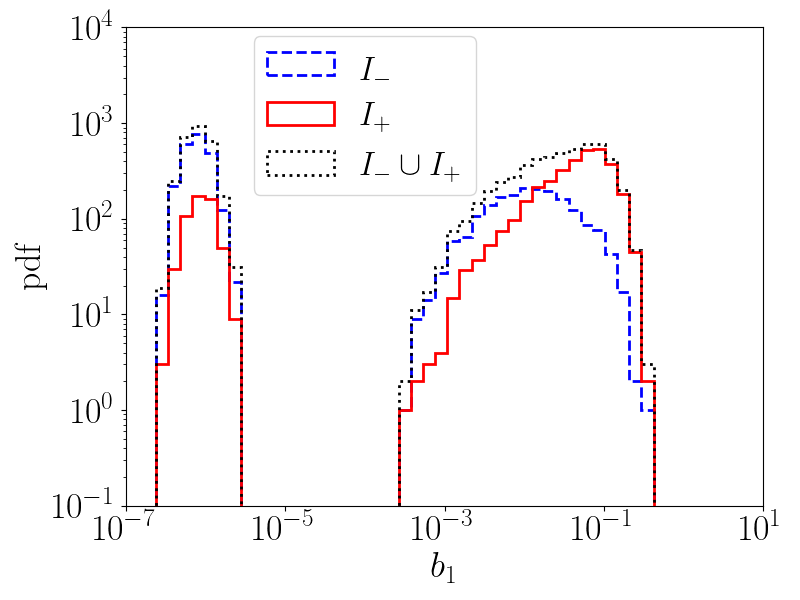}  &
	 \includegraphics[trim=0cm 0 0cm 0, clip,height=4cm]{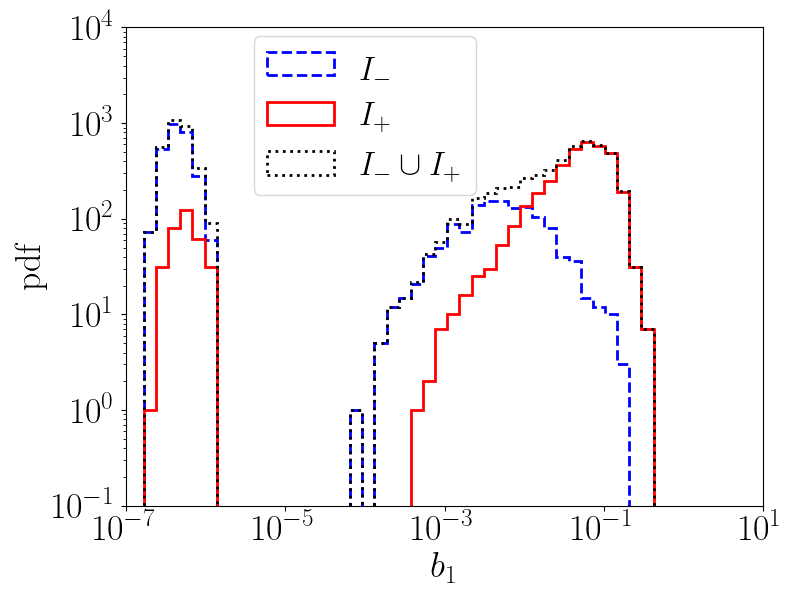}  \\ 
	\end{tabular}
	  \caption{ Distribution of the dominant motif weight $b_1$ for different Rayleigh numbers and
$N_T=100$. }
	\label{fig:pdthist}
	\end{figure}

Figure~\ref{fig:pdthist} displays the histogram of the weight  of the dominant motif $b_1$ (motifs 2 to 4 displayed similar features).
At all Rayleigh numbers, the total distribution is characterized by two distinct lobes, which correspond to the absence and the presence of the motif in the snapshot. 
The relative importance of the lobes therefore provides an indirect measure of the motif intermittency, which can be related to plume emission.
The ratio of motif presence to motif absence was about 0.5-0.6 in the range of Rayleigh numbers considered
 - no significant variation was observed with the Rayleigh number.
 
However, further insights can be obtained by examining the respective contributions of  the  $I_+$ and $I_-$ states to the distribution of $b_1$, which are also represented
in Figure~\ref{fig:pdthist}. 
For all Rayleigh numbers, 
 $I_+$ states contribute more to the higher-value lobe than $I_-$ states, while $I_-$ contributes more to the lower-value lobe.
This shows that the rate of buoyancy production is less intense in the corner rolls than in the large-scale
circulation, or equivalently that  
plumes are emitted at a lower frequency in the corner rolls than in the large-scale circulation.
Moreover, the relative contributions of the $I_+$ and the $I_-$ states vary non-monotonically with the Rayleigh number. 
In the higher-value lobe, 
 the relative contribution of $I_-$ appears to increase relatively to $I_+$ with more high values of $I_-$ at $Ra=3$ $10^6$,
while $I_-$ represents  more low values at $Ra=10^8$.
In the lower-value lobe, the contribution of $I_+$ is least at $Ra=3$ $10^6$ and largest
at $Ra=10^8$.
These observations suggest that both the intensity of the large-scale circulation and that of the corner roll appear to
change with the Rayleigh number, in agreement with the findings of \citet{vishnu20}.

\subsubsection{A model for the reorientation time scale}

A simple model can be made to link these observations with the dynamics of  reorientations. 
The conditionally averaged  weight of the dominant motif in the region close to the wall  $b_\pm$ 
represents the rate of buoyancy production, which can be linked to the emission rate of plumes  
and can be modelled as a Poisson point process.
This means that the time separating two plume ejections $T_{\pm}$ follows
an exponential distribution with mean $1 / b_\pm$, where
$+$ and $-$ respectively characterize the large-scale circulation ($I_+$)
and the corner vortex  ($I_-$) states. $b_\pm$ therefore represents the parameter
of the exponential distribution.
A reorientation can be associated with the event where the corner vortex becomes stronger than the large-scale circulation state, i.e. the time separating two emissions in the corner vortex state becomes smaller than that separating two emissions in the large-scale circulation state. This event can occur independently in
either one of the two horizontal directions $x$ or $y$.

\begin{figure}
	
	\centerline{ \includegraphics[height=7cm]{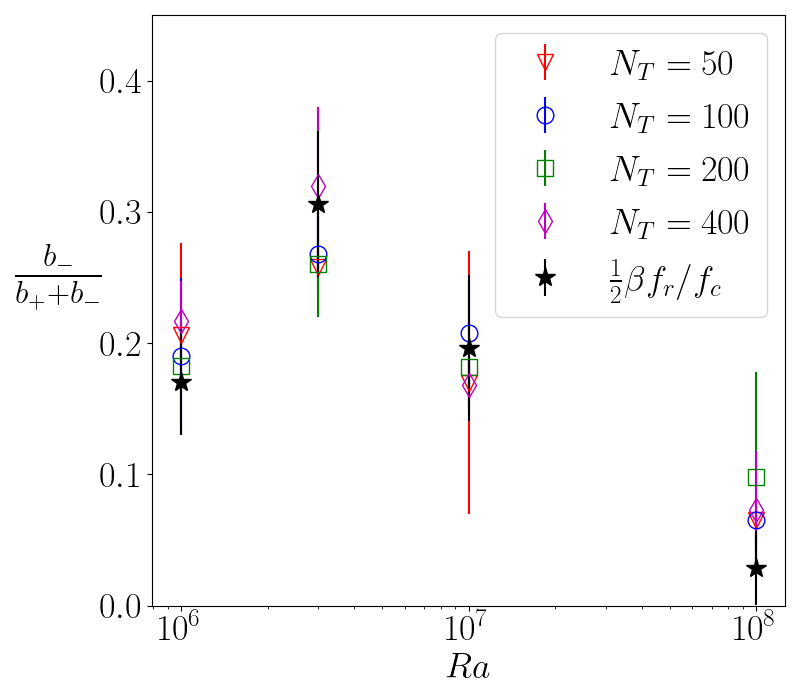} }  

	  \caption{ Probability $p(T_+ > T_-)$ (see text) and comparison with ratio of reorientation  to recirculation time scale  at different Rayleigh numbers -
the rescaling factor is $\beta=5.6$. } 
	\label{fig:pdtra}
	\end{figure}

One can show that the probability $p$ that this event occurs at any given time is given by
\begin{equation}
p= p(T_- > T_+)= \frac{b_-}{b_+ + b_-}
\end{equation}
Owing to the memoryless nature of the exponential distribution, this holds for the time separating
an arbitrary number of emissions, in particular over a characteristic  time $T_s$ sufficiently long to reverse the circulation in that direction. 
$T_s$ should be on the order of the recirculation time $T_c$   so that we have 
$T_s = \beta T_c$ with  $\beta =O (1)$.
If $f_c$ is the recirculation frequency,  
one would then expect the frequency between reorientations $f_r$  to depend on $p$ and $f_c$ following
\begin{equation}
f_r =   2 p  \beta^{-1} f_c, 
\end{equation}
where the factor $2$ comes from the fact that a reorientation
can occur in each direction.
Figure~\ref{fig:pdtra} (right) compares for different Rayleigh numbers the probability $p$  with the  ratio of the frequency between reorientations and the recirculation frequency  estimated in \citep{soucasse21}. 
We see that a very good agreement is obtained between the variations of the average reorientation rate and the measure of the relative intensity of the large-scale circulation and corner vortices.  We note that the largest discrepancy is observed for the highest Rayleigh number, for which the reorientation rate is very low and therefore cannot be determined with good
precision from the DNS.
The value of $\beta$ used in the figure was determined empirically and was found to 
be $5.6$, which makes $T_s$ close to the filtering time scale $T_f = 4 T_c$.  
This suggests that an estimate for the reorientation rate can be obtained by comparing directly the average weight of the motif associated with the large-scale circulation with that of its counterpart 
in the corner structure. This could be of particular interest in cases where
the observation time is  smaller than the expected reorientation time, a situation that is often encountered in  - but not limited to - numerical simulations.



\section{Temperature and velocity motifs}

In this section we try to understand the physics associated with the  lower
reorientation rate observed as the Rayleigh number increases. 
For this we turn to temperature and velocity fluctuations, to which we independently applied LDA.
Although these are not intermittent quantities, and therefore might not be 
considered {\it a priori} appropriate  for LDA application, table \ref{tab:correl} shows that the temperature and kinetic energy 
fields are relatively well reconstructed. 

\begin{table}
\centering
\begin{tabular}{|l|l|c|c|c|c|}
   \hline
    $ < C(q_R, q) > $ &  $N_T$    & $Ra=10^6$ & $Ra=3$ $10^6$ & $Ra=10^7$ & Ra=$10^8$ \\ \hline

 $<C(\theta_R, \theta)>$ & 100  &  0.90 & 0.86  & 0.84 & 0.66 \\ 

 $<C(\theta_R, \theta)>$ & 400  & 0.94 & 0.92  & 0.90 & 0.78 \\ 

 $<C(k_R,k)>$  & 100  &   0.91 & 0.88  & 0.85 & 0.78  \\ 

 $<C(k_R,k)>$  & 400 &  0.94 & 0.92  &  0.89  & 0.82  \\ 

 $<C(\Phi_R, \Phi)>$ & 100 & 0.96   &   0.93   &     0.89  &     0.84 \\

 $<C(\Phi_R, \Phi)>$ & 400 &  0.98  &   0.96   &    0.95  &  0.89  \\ \hline
\end{tabular}
\caption{Average correlation coefficient between the original and the
reconstructed field for the temperature, kinetic energy and heat flux.}
\label{tab:correl}
\end{table}

\subsection{Temperature fluctuations }

\begin{figure}
\begin{tabular}{ccc}
$Ra=10^6$ &  $Ra=10^7$ & $Ra=10^8$ \\ 
\includegraphics[trim=0 0 3.5cm 0, clip, height=4.2cm]{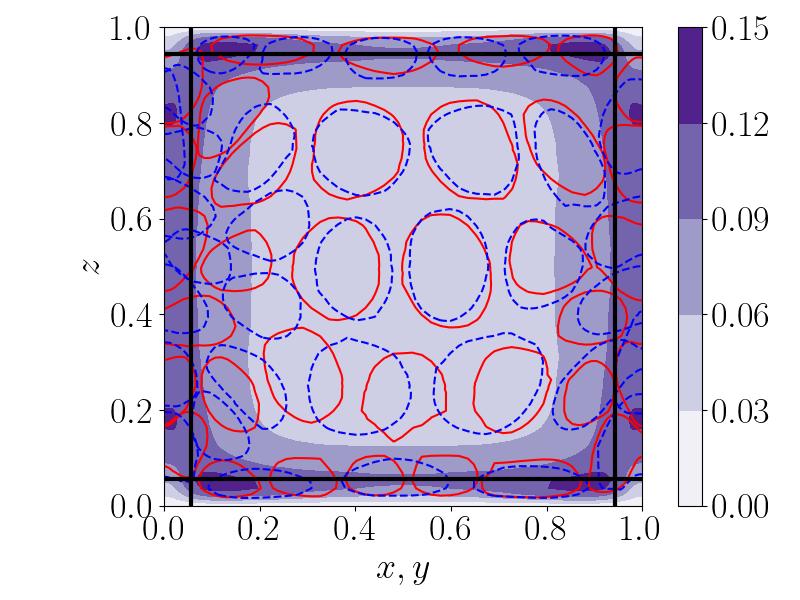}  &
\includegraphics[trim=0 0 3.5cm 0, clip,height=4.2cm]{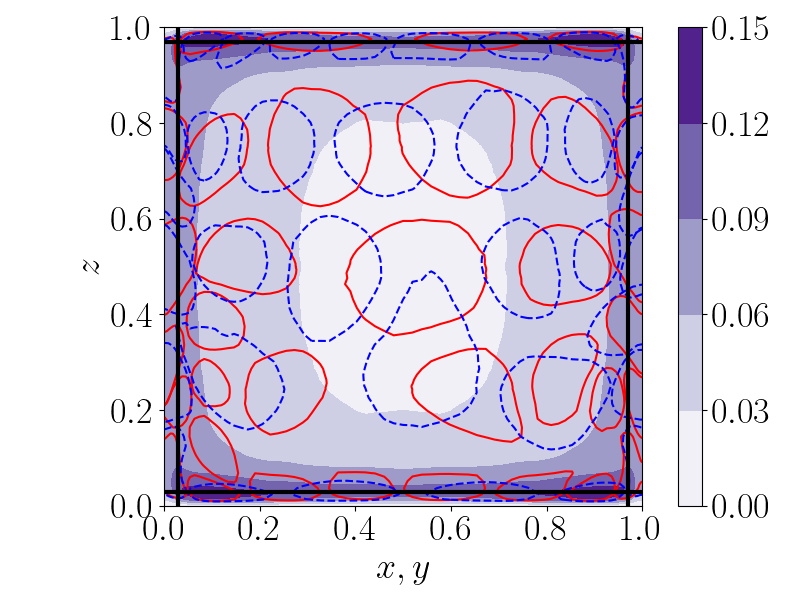}  &
\includegraphics[trim=0 0 0cm 0, clip,height=4.2cm]{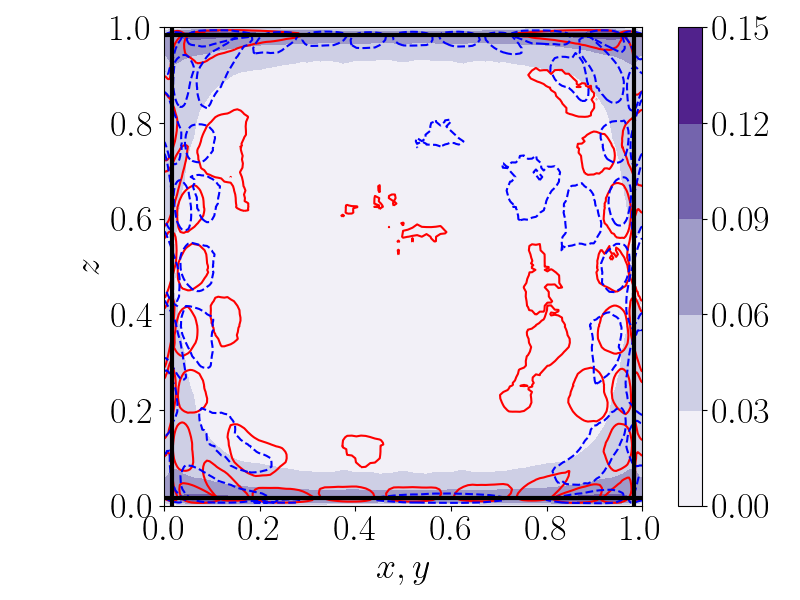}  \\ 
\includegraphics[trim=0 0 3.5cm 0, clip,height=1.9cm]{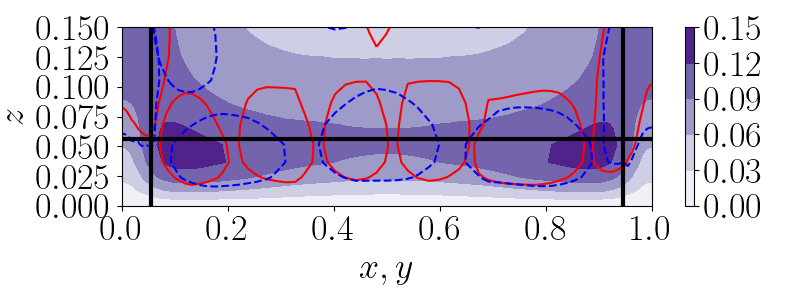}  &
\includegraphics[trim=0 0 3.5cm 0, clip,height=1.9cm]{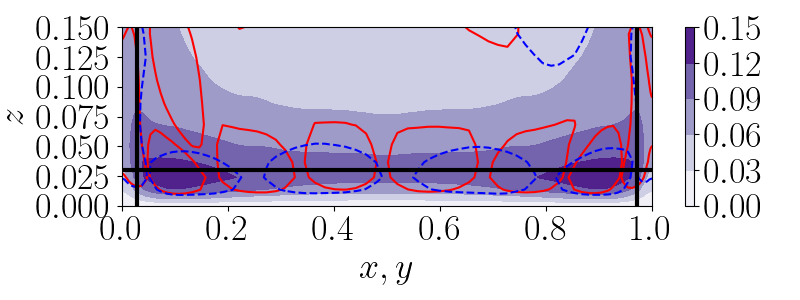}  &
\includegraphics[trim=0 0 0cm 0, clip,height=1.9cm]{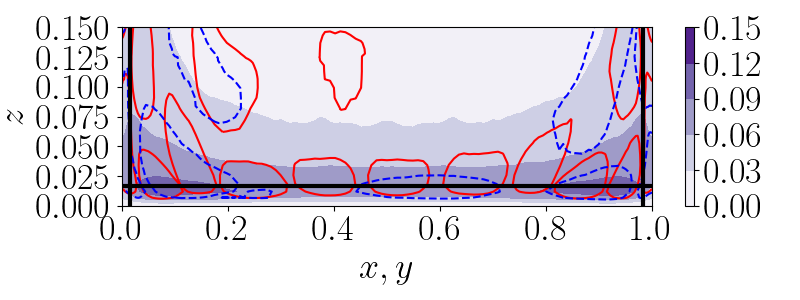}  \\ 
\end{tabular}
\caption{ Top row: Distribution of temperature motifs  in the  cell mid-plane at different Rayleigh numbers; The motifs are materialized by a black line corresponding to a probability contour of $0.606\,\psi_n^{max}$. Contours of the time-averaged variance are represented in the background.
Bottom row: blow-up of the bottom part of the cell. }
\label{fig:temperaturetopic}
\end{figure}

Figure \ref{fig:temperaturetopic} shows the temperature motifs at three different Rayleigh numbers, along with the variance of the fluctuations, for $N_T=100$.
As mentioned above, some symmetry is expected but 
not perfectly enforced, due to the statistical character of the method.
As for heat flux motifs there is a clear difference between the boundary layers and the bulk,
as well as a  strong decrease of motifs in the central part of the cell at $Ra=10^8$.
We can see that temperature fluctuations are also important close to the horizontal walls.
The bottom row of figure \ref{fig:temperaturetopic} shows a close-up of the lower part of the cell.
The maximum of the motif spatial distribution is located at the edge of the boundary layer.
The height of the motifs scale with the boundary layer height in the center of the cell, 
with  negative motifs  shorter and wider than  positive ones in the bottom layer.  
Analogous observations can be made for the top wall, by swapping the role of cold
and hot fluctuations. 

\begin{figure}
\begin{tabular}{cccc}
\includegraphics[trim=0 0 4.cm 0, clip, height=3.5cm]{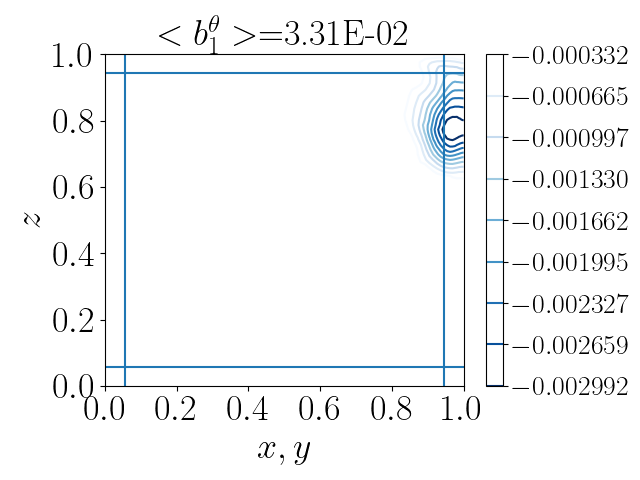}   &
\includegraphics[trim=0 0 4.cm 0, clip, height=3.5cm]{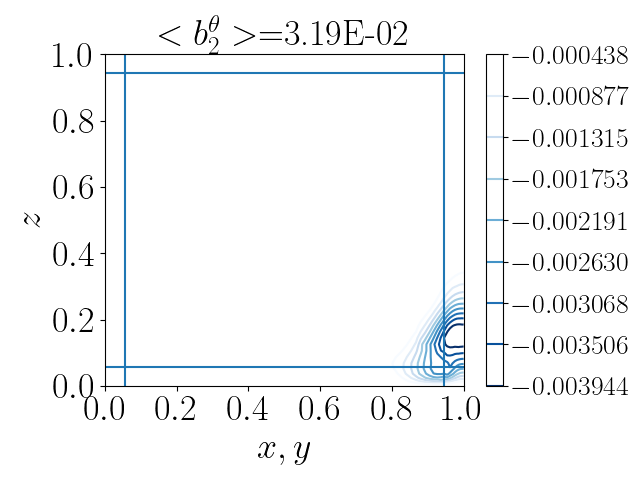}   &
\includegraphics[trim=0 0 4.cm 0, clip,height=3.5cm]{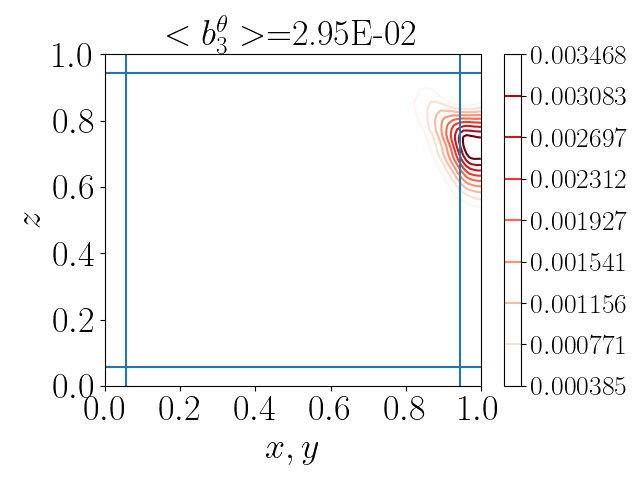}   &
\includegraphics[trim=0 0 4.cm 0, clip,height=3.5cm]{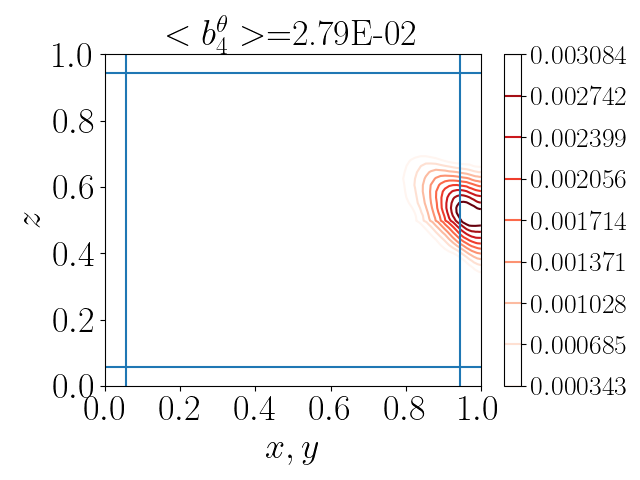}   \\ 
\end{tabular}
\caption{First four dominant temperature motifs at $Ra=10^6$.}
\label{fig:topicra1E6t}
\end{figure}

\begin{figure}
\begin{tabular}{lllll}
\includegraphics[trim=0 0 4.cm 0, clip,height=3.5cm]{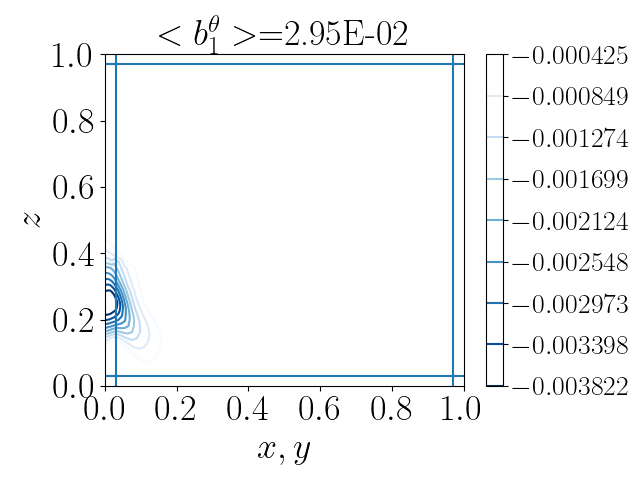}   &
\includegraphics[trim=0 0 4.cm 0, clip,height=3.5cm]{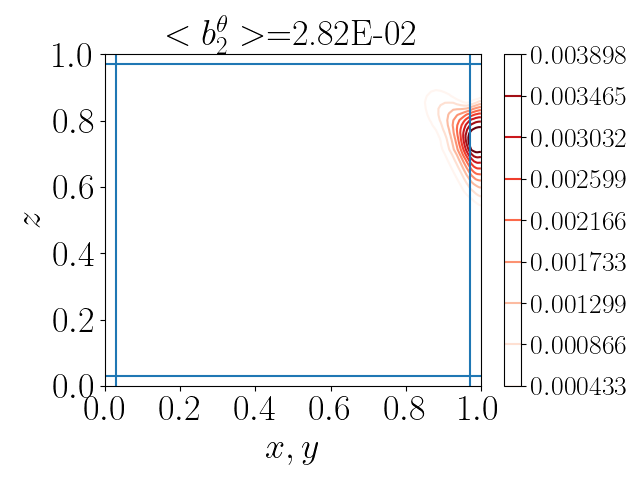}   &
\includegraphics[trim=0 0 4.cm 0, clip,height=3.5cm]{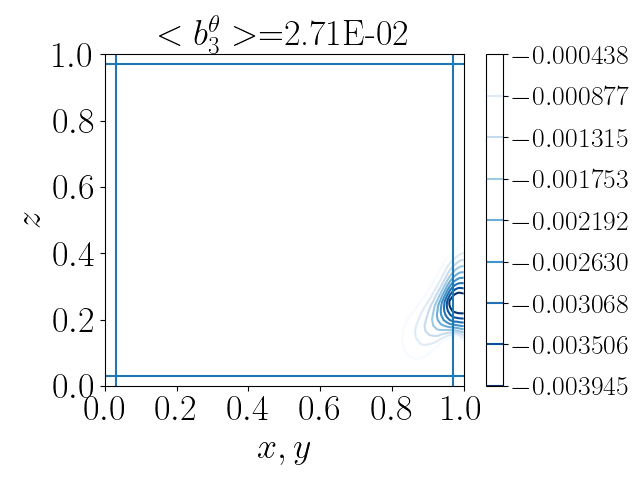}   &
\includegraphics[trim=0 0 4.cm 0, clip,height=3.5cm]{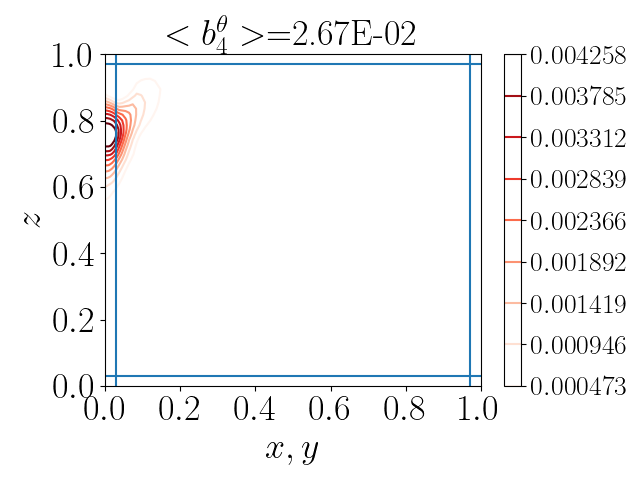}    \\ 
\end{tabular}
\caption{First four dominant temperature motifs at $Ra=$ $10^7$.}
\label{fig:topicra1E7t}
\end{figure}

Figure \ref{fig:topicra1E6t} represents
the first four dominant motifs for the temperature at $Ra=10^6$ (similar observations can be made at $Ra=3$ $10^6$). 
Although the most likely heat flux motifs corresponded to hot  plumes near the bottom wall and cold plumes near the
top wall, this is not the case for the temperature motifs.
For the two lower Rayleigh numbers, temperature motifs are as likely to be found near the bottom wall
than near the top wall.
However, at $Ra=10^7$, figure \ref{fig:topicra1E7t} shows that  
the most likely temperature motifs correspond to hot fluctuations along the bottom 
side walls and cold near the top side wall, corresponding to late-stage plumes arriving at the opposite 
wall.

\begin{figure}
\begin{tabular}{cc}
\includegraphics[height=6cm]{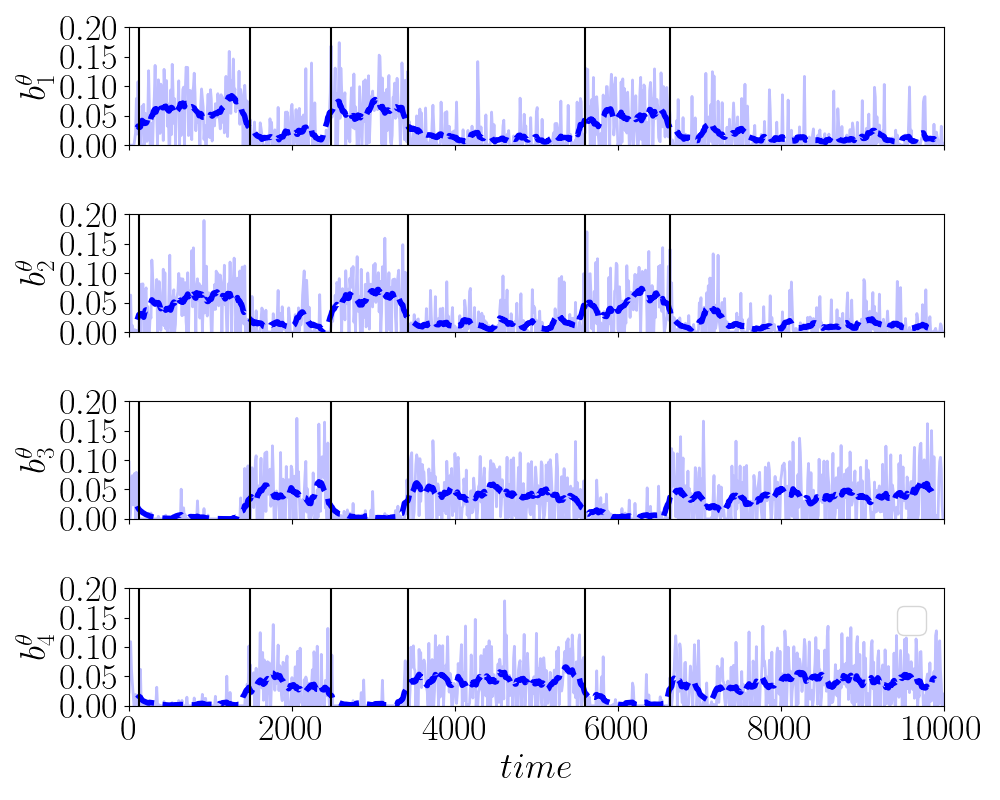}   &
\includegraphics[height=6cm]{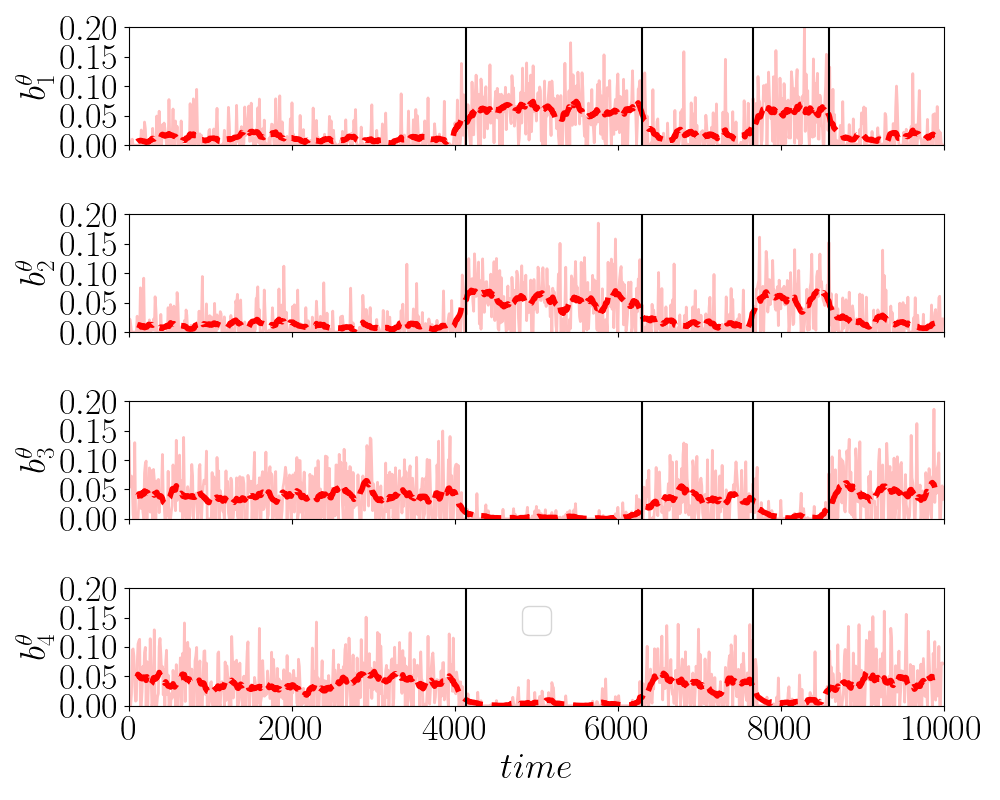} \\
\end{tabular}
\caption{First four dominant temperature motif weights at $Ra=10^7$: {left) plane $x=0.5H$ right) plane $y=0.5H$. The thick line corresponds to a moving average over 200 convective units (4 recirculation times $T_c$).  The vertical lines correspond to the changes in angular momentum.}}
\label{fig:weightra1E7t}
\end{figure}

Figure \ref{fig:weightra1E7t} shows the evolution of the temperature motif weights $b_n^\theta$ on both planes along 
the filtered  representation $\overline{b}_n^\theta$ (i.e. corresponding to an 
average of $T_f$. ).
As observed for the heat flux (figure \ref{fig:pdtt}), the importance of the weights 
depends on the orientation of the large-scale circulation  $I$.
Similar evolutions  were observed at the lower Rayleigh numbers (not shown).

\begin{figure}
\begin{tabular}{lllll}
\includegraphics[trim=0 0 4.cm 0, clip,height=3.5cm]{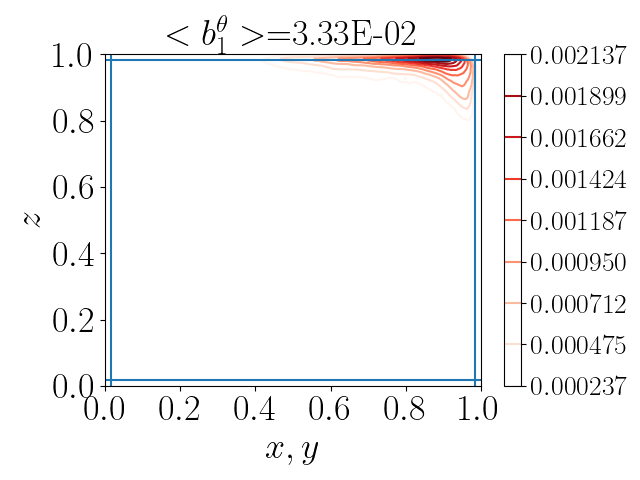}   &
\includegraphics[trim=0 0 4.cm 0, clip,height=3.5cm]{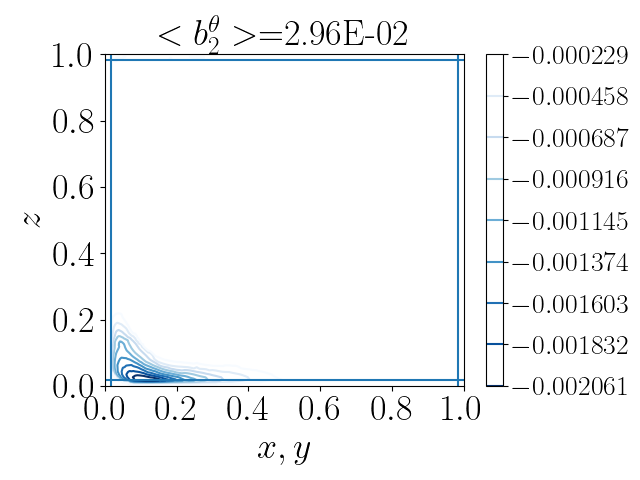}   &
\includegraphics[trim=0 0 4.cm 0, clip,height=3.5cm]{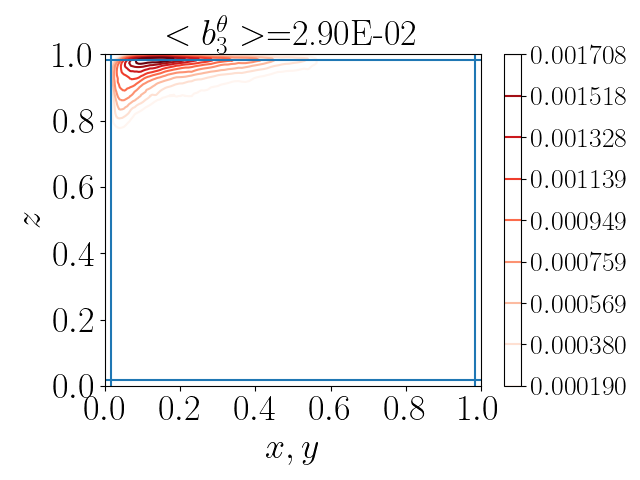}   &
\includegraphics[trim=0 0 4.cm 0, clip,height=3.5cm]{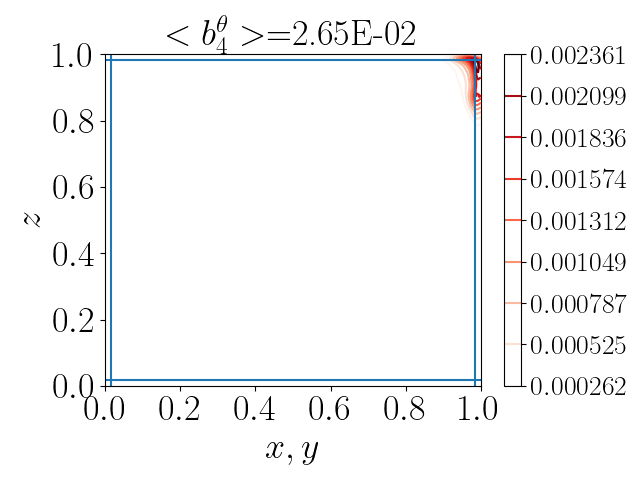}  \\ 
\includegraphics[trim=0 0 4.cm 0, clip,height=3.5cm]{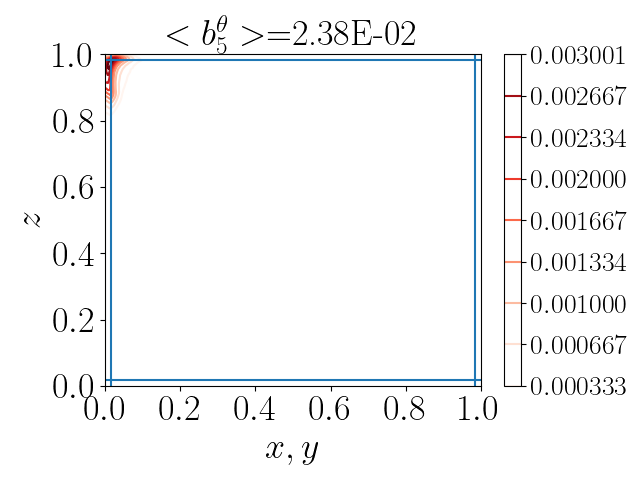}   & 
\includegraphics[trim=0 0 4.cm 0, clip,height=3.5cm]{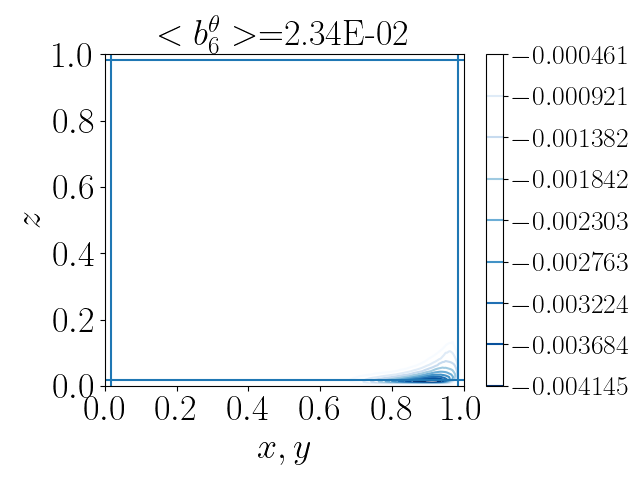}   &
\includegraphics[trim=0 0 4.cm 0, clip,height=3.5cm]{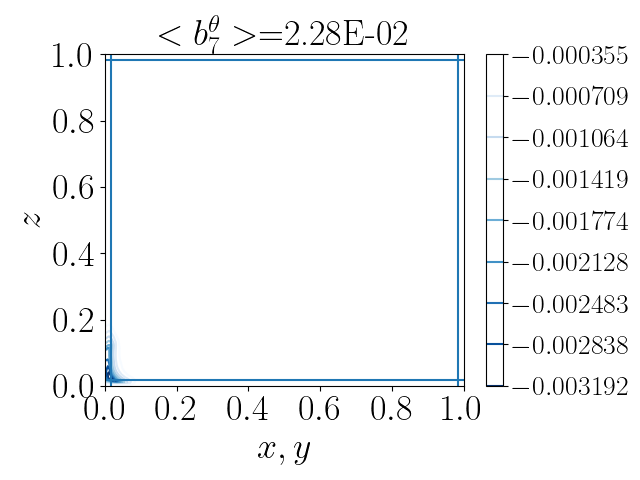}   &
\includegraphics[trim=0 0 4.cm 0, clip,height=3.5cm]{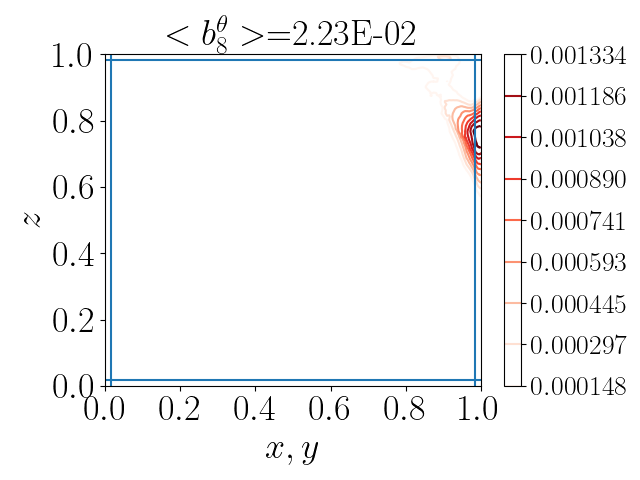}  \\ 
\end{tabular}
\caption{First eight dominant temperature motifs at $Ra=$ $10^8$.}
\label{fig:topicra1E8t}
\end{figure}

\begin{figure}
\begin{tabular}{ll}
\includegraphics[height=6cm]{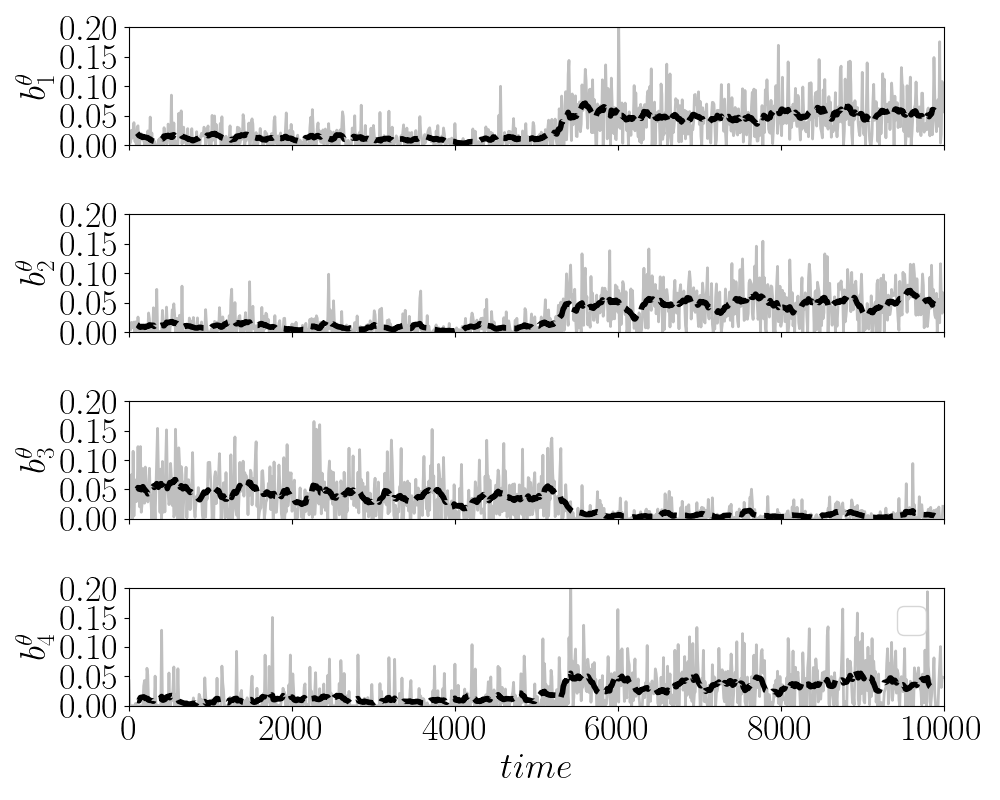}   &
\includegraphics[height=6cm]{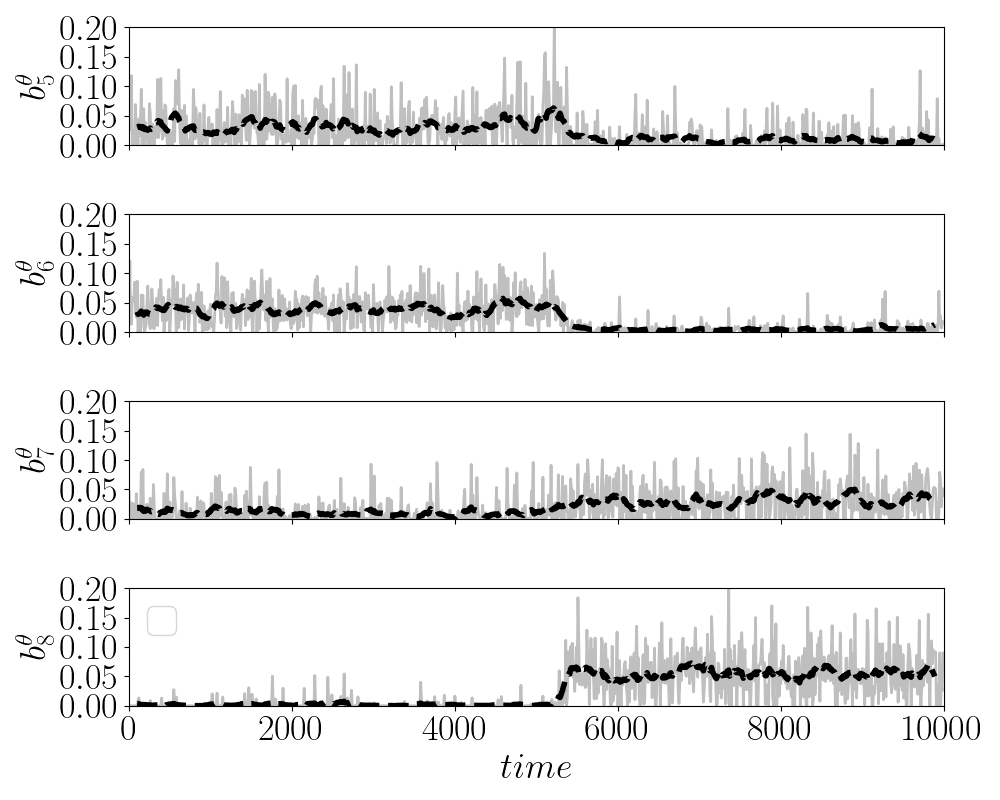}  \\ 
\end{tabular}
\caption{First eight temperature motif  weights  at $Ra=$ $10^8$ {in the plane $x=0.5$}}
\label{fig:pdtavxra1E8t}
\end{figure}

Strong differences can be observed when comparing figures  \ref{fig:topicra1E7t} and  
 \ref{fig:topicra1E8t}. 
At $Ra=10^8$, the most likely temperature motifs are no longer located within 
the vertical boundary layers, but extend from the corner of the cell along the horizontal walls.
The first eight dominant structures consist of two types of corner motifs: 
large,  predominantly horizontal ones, and small, vertical ones located within the boundary layers.  
Motifs near the top (resp. bottom) wall are hot (resp. cold) and therefore correspond to
late-stage plumes.
This is confirmed by the evolution 
of the motif weights shown in figure \ref{fig:pdtavxra1E8t} for the plane $x=0.5$.  
These motifs correspond  to hot fluid being 
brought from the bottom layer by the large-scale circulation next to the top wall and 
into the corner structure, thus decreasing buoyancy effects there. 
These observations are consistent 
with the reduction in intensity of the corner roll and the  significant decrease in the 
reorientation rate observed at this Rayleigh number.
We note that although the small vertical temperature motifs are similar to the heat flux motifs
4 and 6 identified in figure \ref{fig:topicra1E7t} at $Ra=10^7$, they represent 
fluctuations of the opposite sign, and they are well correlated (or anti-correlated) with the orientation $I$ of the large-scale circulation.
This confirms the dominance of the impinging plumes in the corners of the cell.

\subsection{Kinetic energy }

\begin{figure}
\begin{tabular}{cccc}
$Ra=10^6$ & $Ra=3$ $10^6$ & $Ra=10^7$ & $Ra=10^8$ \\ 
\includegraphics[trim=1.5cm 0 3.9cm 0, clip,height=4.1cm]{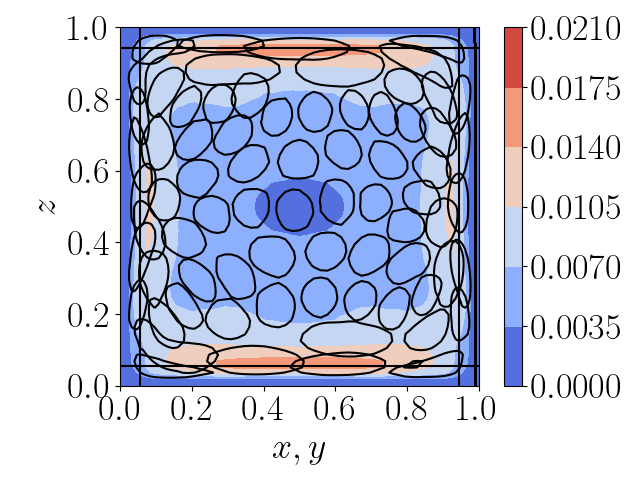}   &
\includegraphics[trim=1.5cm 0 3.9cm 0, clip,height=4.1cm]{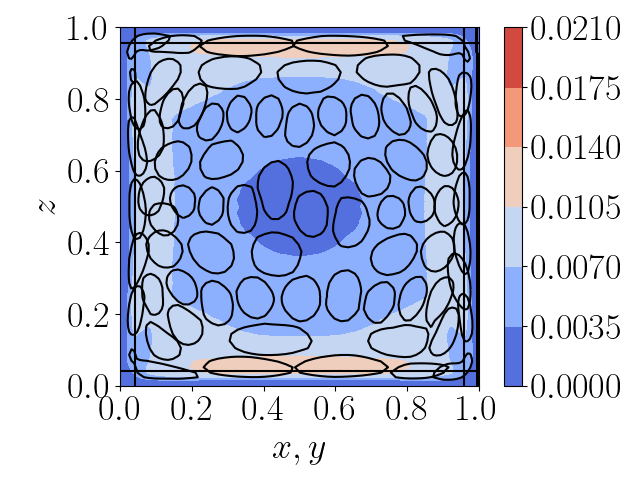}   & 
\includegraphics[trim=1.5cm 0 3.9cm 0, clip,height=4.1cm]{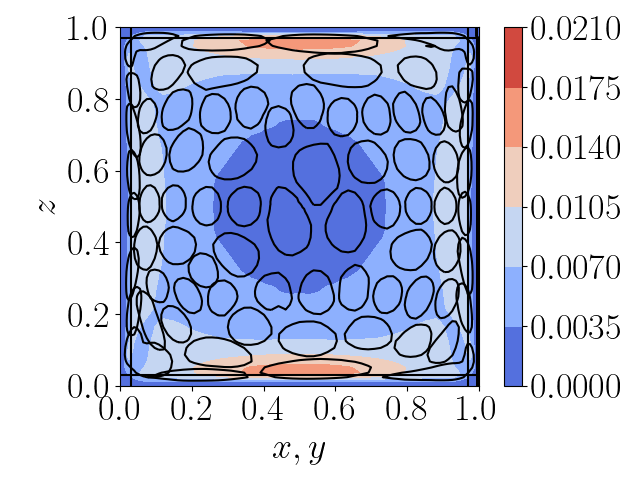}   & 
\includegraphics[trim=1.5cm 0 0 0, clip, height=4.1cm]{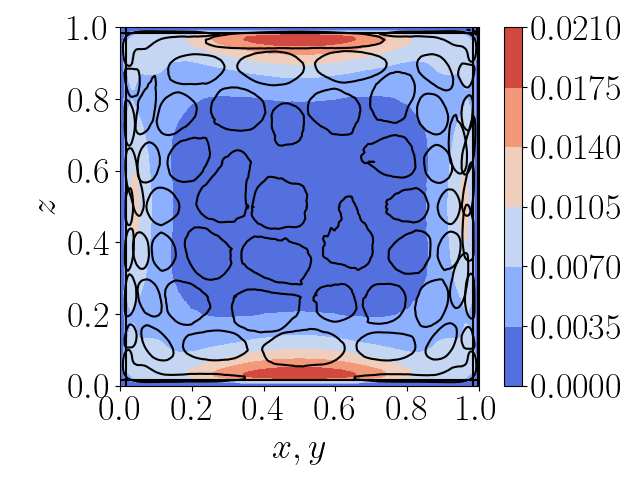}    \\ 
\end{tabular}
\caption{ Spatial distribution of kinetic energy motifs  in the cell mid-plane at different Rayleigh numbers. The motifs are materialized by a black line corresponding to a probability contour of $0.606\,\psi_n^{max}$. Contours of the time-averaged kinetic energy are represented in the background.}
\label{fig:ktopics}
\end{figure}

More details about the structure of the large-scale circulation can be obtained by 
examining kinetic energy motifs.
Figure \ref{fig:ktopics} shows 
the spatial distribution of the velocity  motifs for the different Rayleigh numbers and
$N_T=100$. The spatial distribution of the time-averaged kinetic energy is also represented on the same plot. The size of the core (low-velocity region) appears to increase with the 
Rayleigh number.
The size of the motifs did not appear to change significantly with the Rayleigh number, except
for horizontal corner structures that seem to scale with the boundary layer thickness. 
The kinetic energy motifs have elongated shapes along the walls, with a significantly higher 
extent along the horizontal walls, which shows  
the importance of entrainment in the horizontal boundary layers, in particular in the middle of the cell.
It is lowest at $Ra=3 \:10^6$ and highest at $Ra=10^8$, which varies like the time between reorientations
$T_r$. 
The question is whether this reinforcement of the large-scale circulation can be associated with 
characteristic temperature fluctuations.

\begin{figure}
\begin{tabular}{lllll}
\raisebox{1.8cm}{HBL} &  
\includegraphics[trim=3.1cm 2cm 4.0cm 0cm, clip, height=3.5cm]{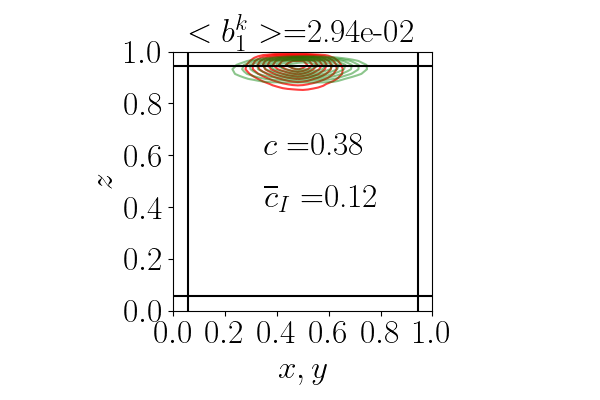}   & 
\includegraphics[trim=4.1cm 2cm 4.0cm 0cm, clip, height=3.5cm]{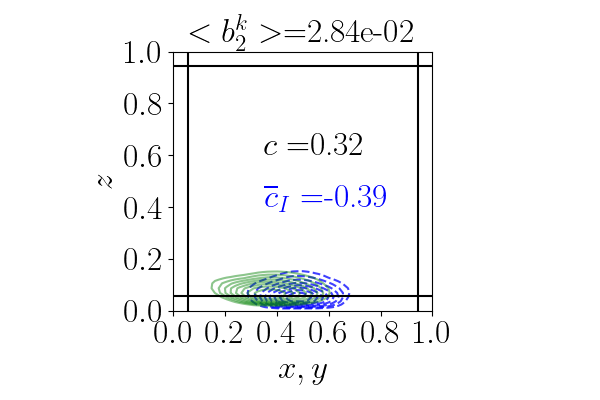}   & 
\includegraphics[trim=4.1cm 2cm 4.0cm 0cm, clip, height=3.5cm]{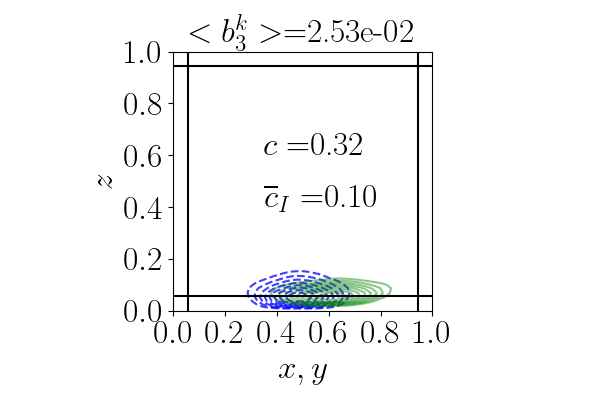}   &  
\raisebox{-0.3cm}{  
\includegraphics[trim=5cm 2cm 4.05cm 0cm, clip, height=5cm]
{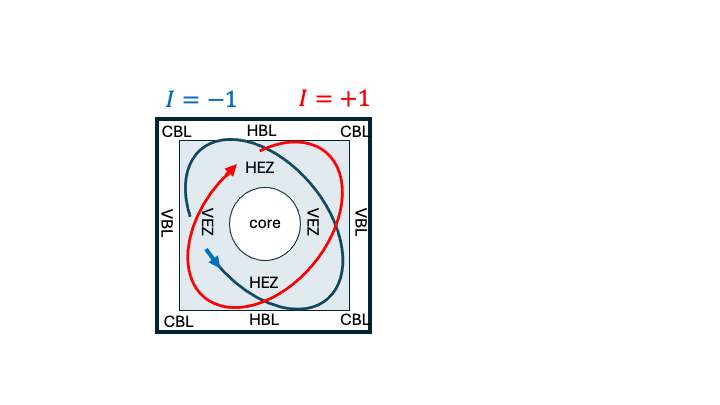}    } 
 \\
\raisebox{1.8cm}{CBL} &  
\includegraphics[trim=3.1cm 2cm 4.05cm 0cm, clip, height=3.5cm]{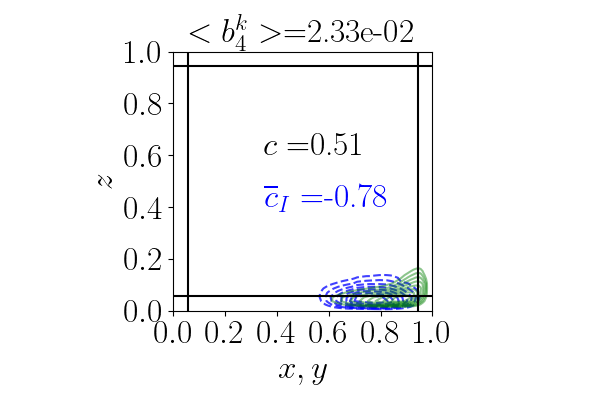}   & 
\includegraphics[trim=4.1cm 2cm 4.05cm 0cm, clip, height=3.5cm]{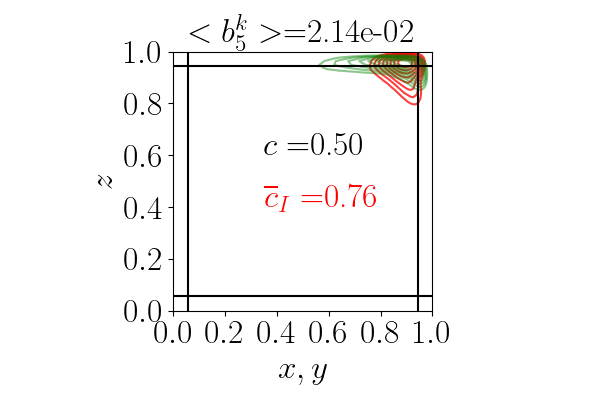}   &  
\includegraphics[trim=4.1cm 2cm 4.05cm 0cm, clip, height=3.5cm]{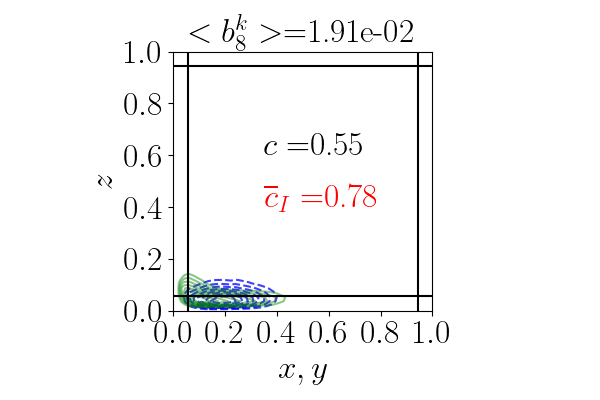} &   \\ 
\raisebox{1.8cm}{VEZ} &  
\includegraphics[trim=3.1cm 2cm 4.05cm 0cm, clip, height=3.5cm]{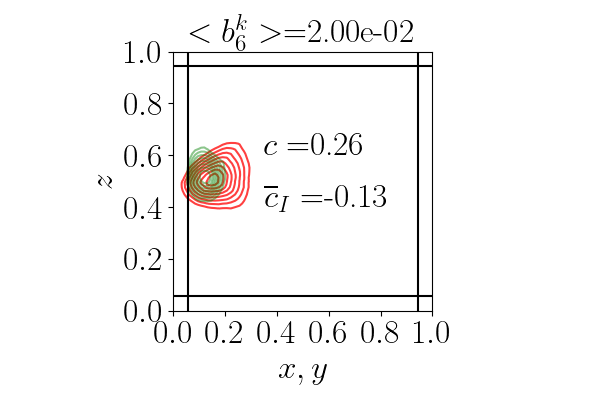}   &  
\includegraphics[trim=4.1cm 2cm 4.0cm 0cm, clip, height=3.5cm]{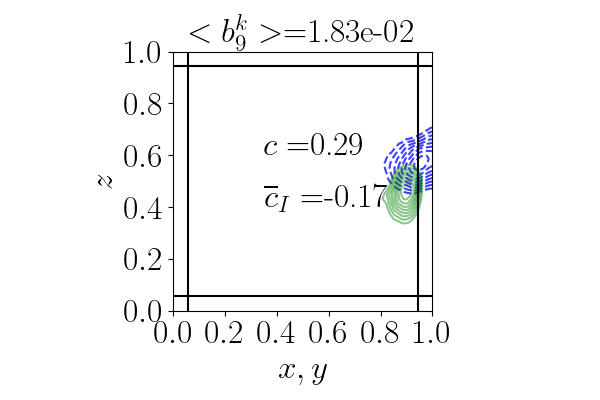}   &  &  \\  
\raisebox{1.8cm}{VBL} &  
\includegraphics[trim=3.1cm 2cm 4.0cm 0cm, clip, height=3.5cm]{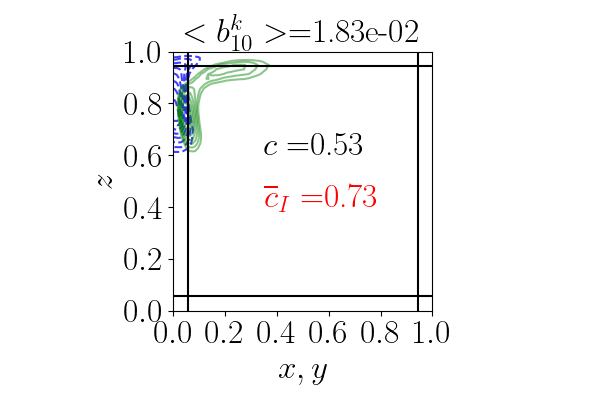}  &  
\includegraphics[trim=4.1cm 2cm 4.0cm 0cm, clip, height=3.5cm]{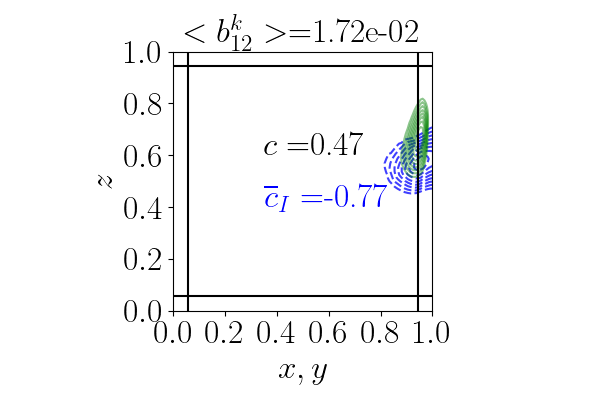}  &  
\includegraphics[trim=4.1cm 2cm 4.0cm 0cm, clip, height=3.5cm]{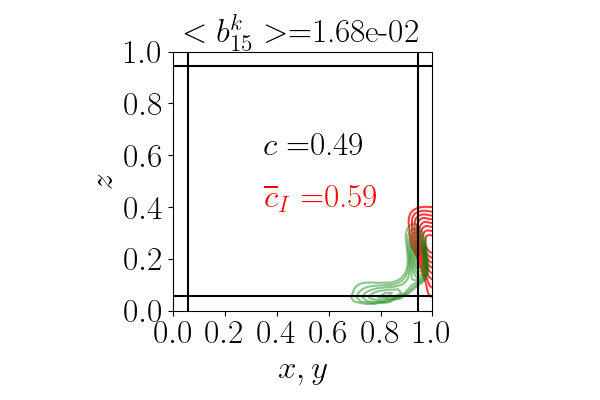}  &   \\
\raisebox{1.8cm}{HEZ} &  
\includegraphics[trim=3.1cm 0cm 4.0cm 0cm, clip, height=4.1cm]{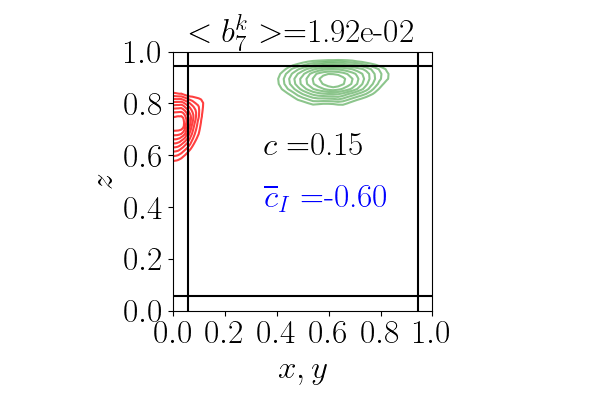}   &  
\includegraphics[trim=4.1cm 0cm 4.0cm 0cm, clip, height=4.1cm]{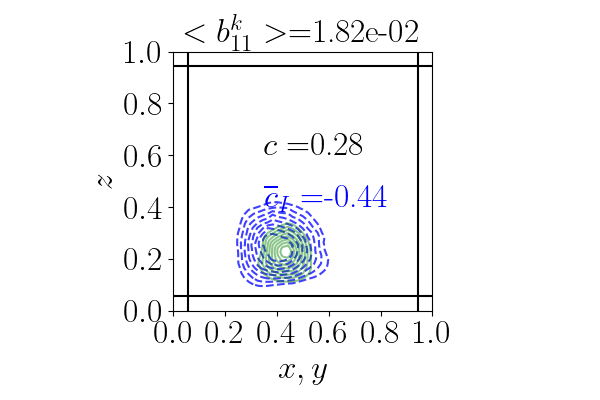}   & 
\includegraphics[trim=4.1cm 0cm 4.0cm 0cm, clip, height=4.1cm]{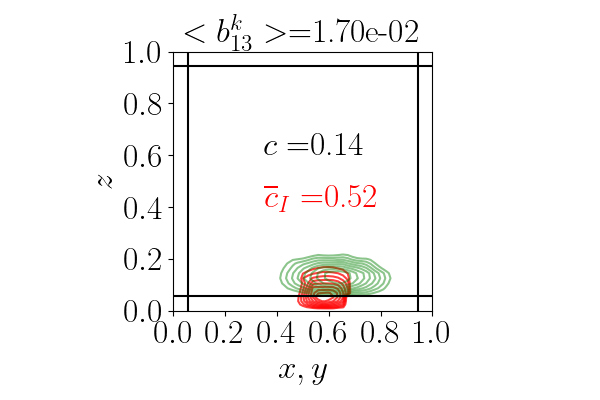}   &  
\includegraphics[trim=4.1cm 0cm 4.0cm 0cm, clip, height=4.1cm]{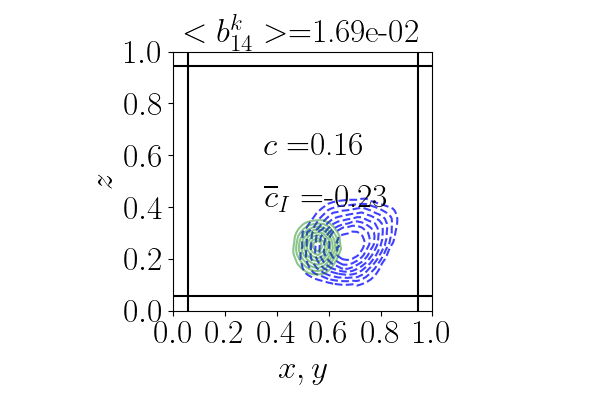}    \\  
\end{tabular}
\caption{ Dominant kinetic energy motifs at $Ra=10^6$ (green lines) ordered by prevalence and location as indicated  at top right. 
Temperature motifs with the highest correlation coefficient $c$ are shown in blue (resp. red) for negative (resp. positive) fluctuations.
Motif contour levels range from $0.2$ to $0.9$ $\psi_n^{max}$ with increments of 0.1 $\psi_n^{max}$.
$\overline{c}_I$ is the correlation coefficient between the heat flux motif weight and 
the LSC indicator $I$. 
Values of $\overline{c}_I$ larger than 0.3
(resp. lower than -0.3) are represented in red (resp. blue). }
\label{fig:comptopicktra1E6}
\end{figure}

\begin{figure}
\begin{tabular}{cccccc}
\raisebox{1.8cm}{HBL} &  
\includegraphics[trim=3.1cm 2cm 4.0cm 0, clip, height=3.85cm]{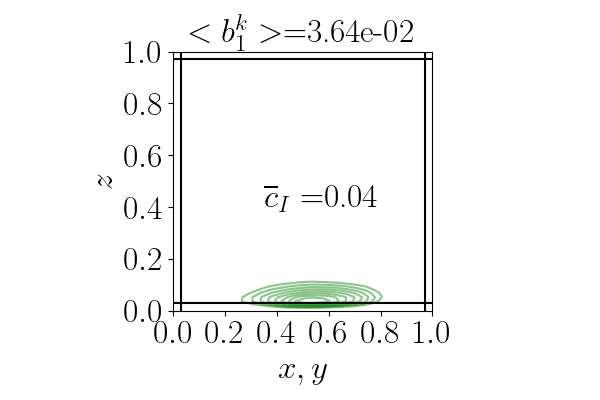}   & 
\includegraphics[trim=4.10cm 2cm 4.0cm 0, clip, height=3.85cm]{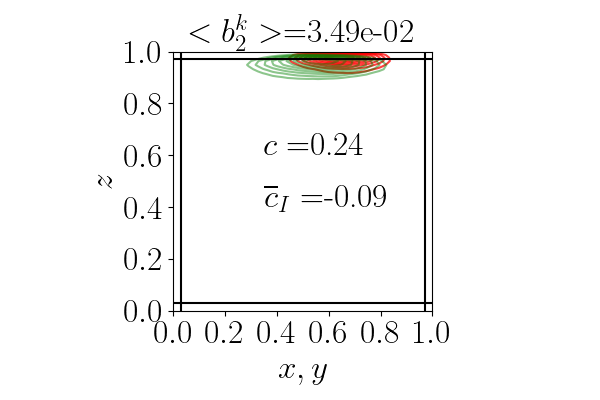}   & 
\includegraphics[trim=4.1cm 2cm 4.0cm 0, clip, height=3.85cm]{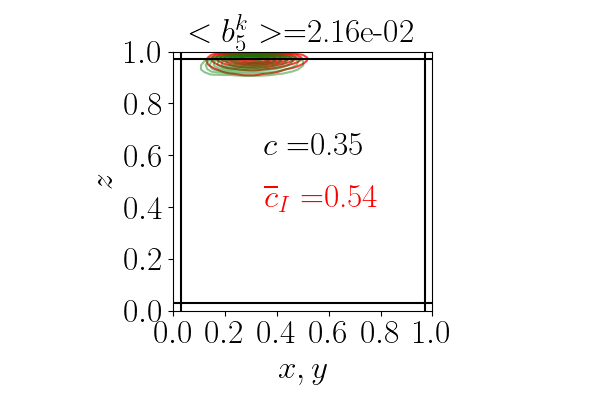}   &  & \\
\raisebox{1.8cm}{CBL} &  
\includegraphics[trim=3.1cm 2cm 4.0cm 0, clip, height=3.85cm]{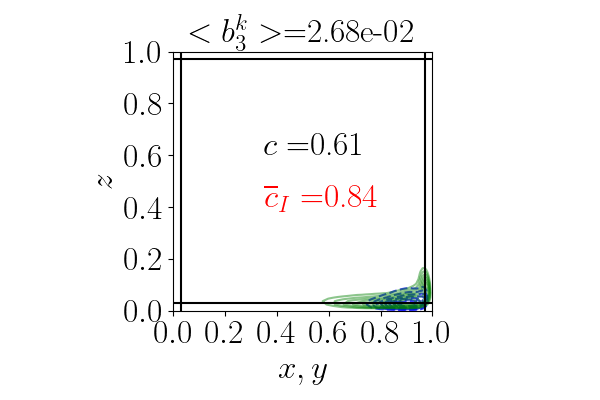}   & 
\includegraphics[trim=4.1cm 2cm 4.0cm 0, clip, height=3.85cm]{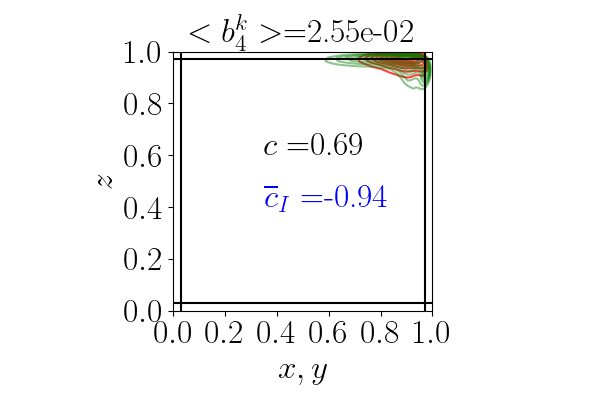}   & 
\includegraphics[trim=4.1cm 2cm 4.0cm 0, clip, height=3.85cm]{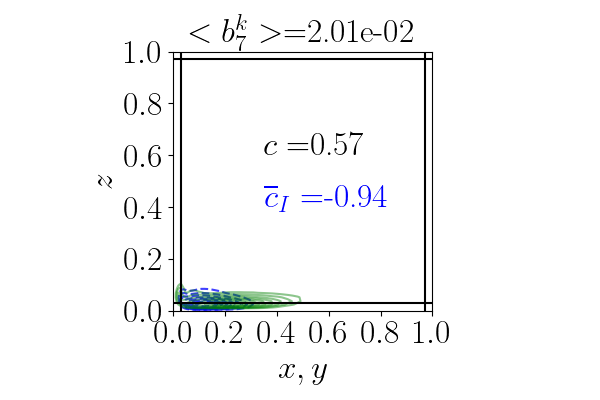}   & 
\includegraphics[trim=4.1cm 2cm 4.0cm 0, clip, height=3.85cm]{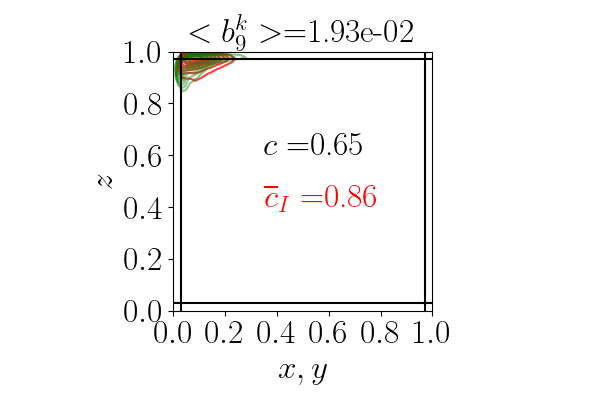}   &  \\
\raisebox{1.8cm}{VEZ} &  
\includegraphics[trim=3.1cm 2cm 4.0cm 0, clip, height=3.85cm]{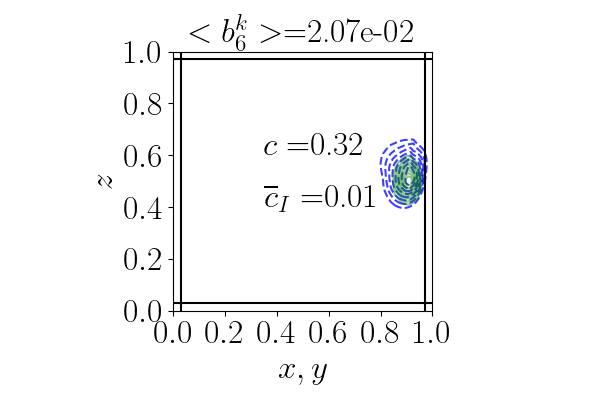}   & 
\includegraphics[trim=4.1cm 2cm 4.0cm 0, clip, height=3.85cm]{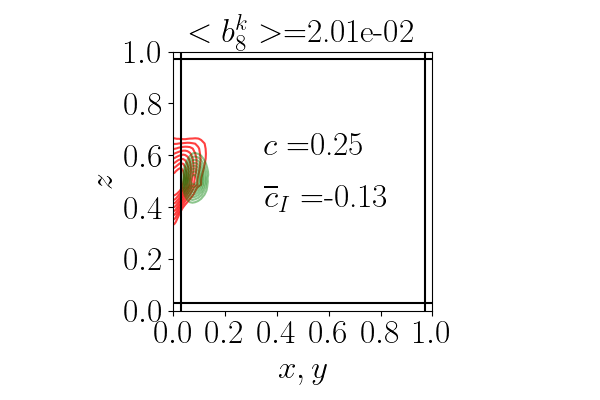}   & 
\includegraphics[trim=4.1cm 2cm 4.0cm 0, clip, height=3.85cm]{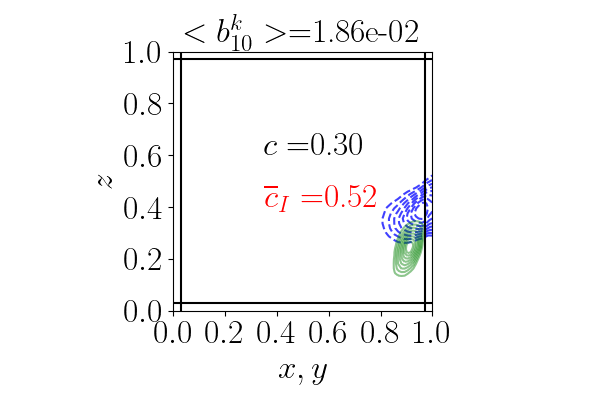}   & 
\includegraphics[trim=4.1cm 2cm 4.0cm 0, clip, height=3.85cm]{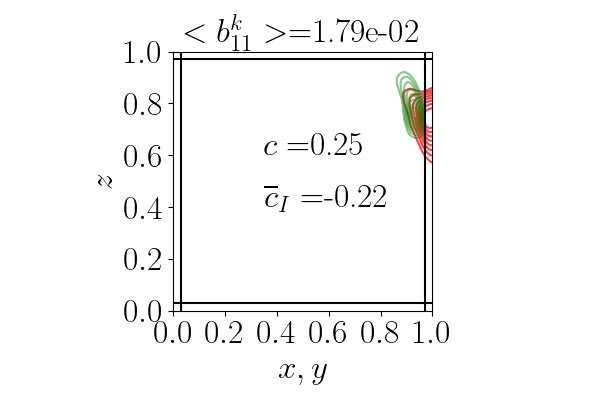}   & \\ 
\raisebox{1.8cm}{VBL} &  
\includegraphics[trim=3.1cm 2cm 4.0cm 0, clip, height=3.85cm]{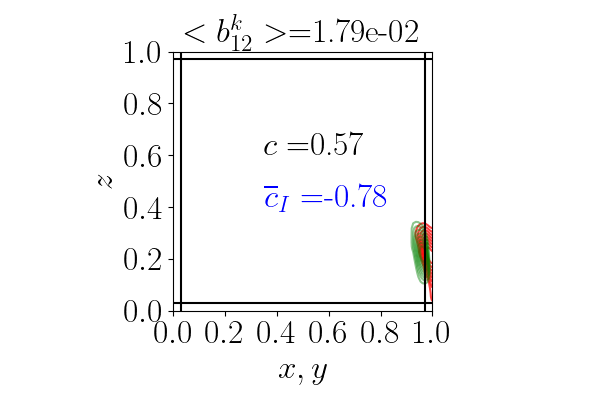} &
\includegraphics[trim=4.1cm 2cm 4.0cm 0, clip, height=3.85cm]{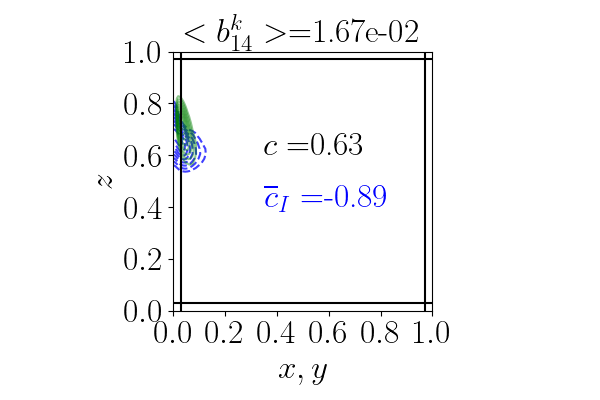} & 
& & \\ 
\raisebox{1.8cm}{HEZ} &  
\includegraphics[trim=3.1cm 0cm 4.0cm 0, clip, height=4.7cm]{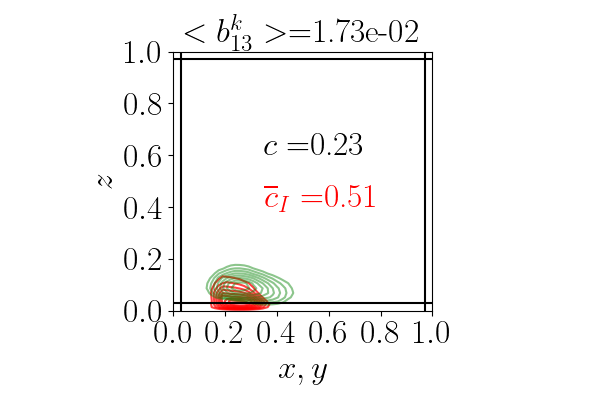}    & 
\includegraphics[trim=4.1cm 0cm 4.0cm 0, clip, height=4.7cm]{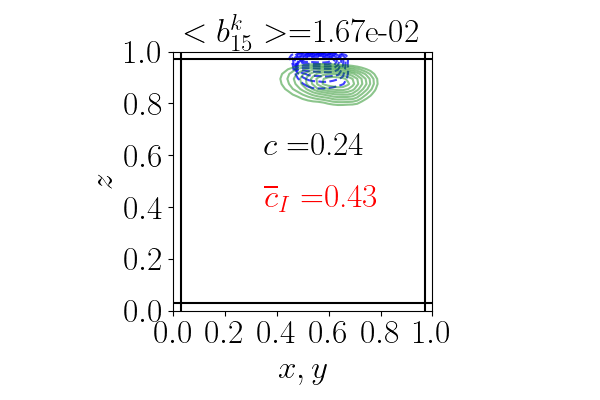}     \\ 
 \\
\end{tabular}
\caption{ Comparison between kinetic energy and temperature motifs at $Ra=$ $10^7$. See legend of figure 
\ref{fig:comptopicktra1E6}. }
\label{fig:comptopicktra1E7}
\end{figure}

\begin{figure}
\begin{tabular}{ccccccc}
\raisebox{1.8cm}{HBL}   &
\includegraphics[trim=3.1cm 2cm 4.0cm 0cm, clip, height=3.85cm]{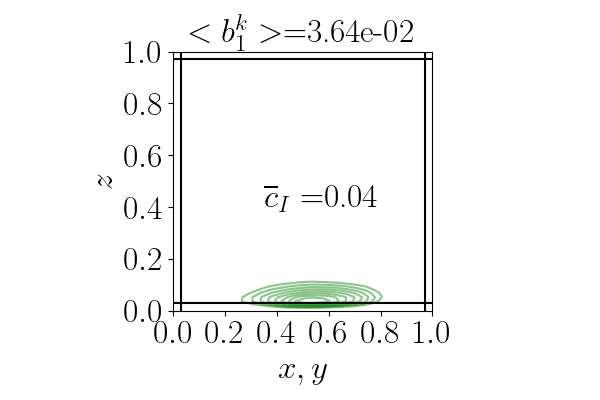}   &
& & & & \\
\raisebox{1.8cm}{HEZ}   &
\includegraphics[trim=3.1cm 2cm 4.0cm 0cm, clip, height=3.85cm]{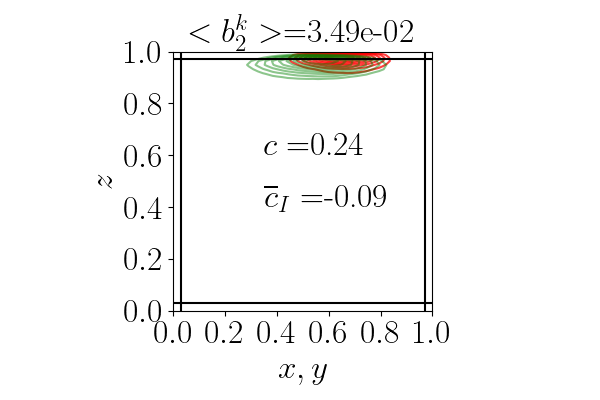}   & 
\includegraphics[trim=4.1cm 2cm 4.0cm 0cm, clip, height=3.85cm]{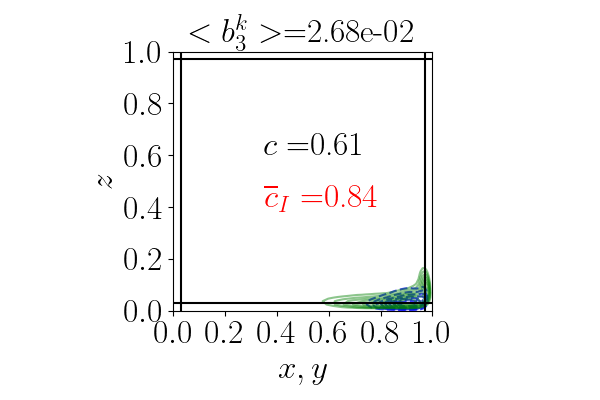}   &
\includegraphics[trim=4.1cm 2cm 4.0cm 0cm, clip, height=3.85cm]{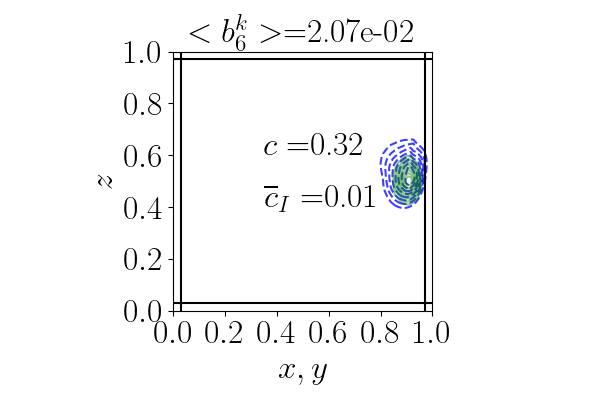}   &  &  
\\ 
\raisebox{1.8cm}{HEZ}   &
\includegraphics[trim=3.1cm 2cm 4.0cm 0cm, clip, height=3.85cm]{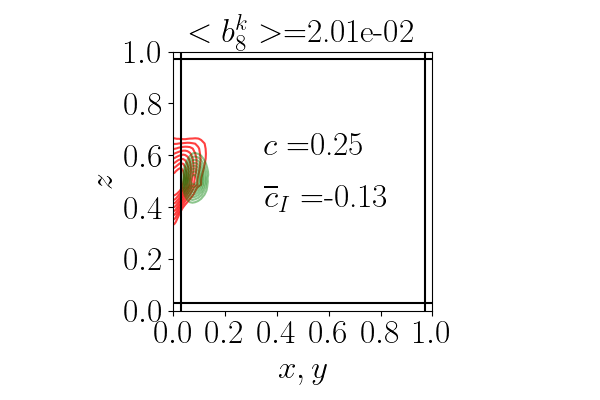}    & 
\includegraphics[trim=4.1cm 2cm 4.0cm 0cm, clip, height=3.85cm]{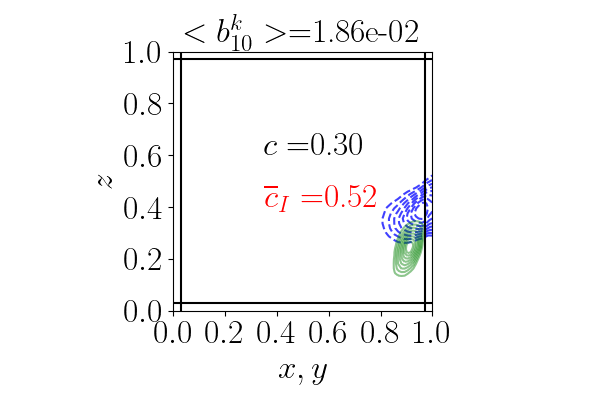}   &   
\includegraphics[trim=4.1cm 2cm 4.0cm 0cm, clip, height=3.85cm]{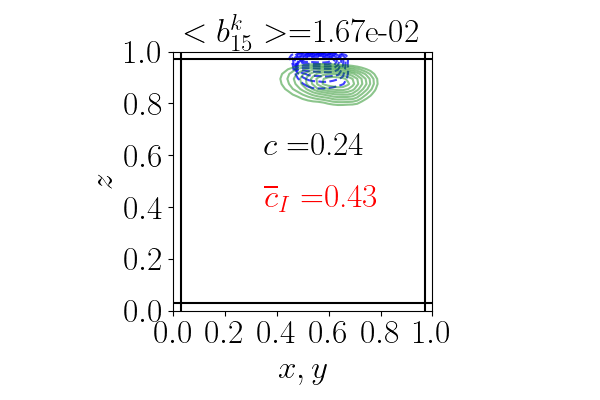}    \\ 
\raisebox{1.8cm}{CBL}   &
\includegraphics[trim=3.1cm 2cm 4.0cm 0cm, clip, height=3.85cm]{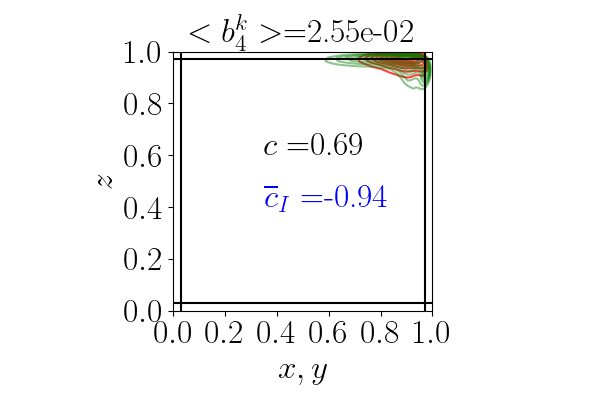}    &
\includegraphics[trim=4.1cm 2cm 4.0cm 0cm, clip, height=3.85cm]{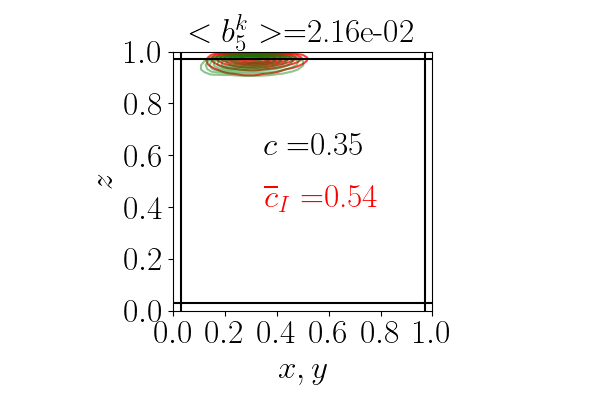}   &
\includegraphics[trim=4.1cm 2cm 4.0cm 0cm, clip, height=3.85cm]{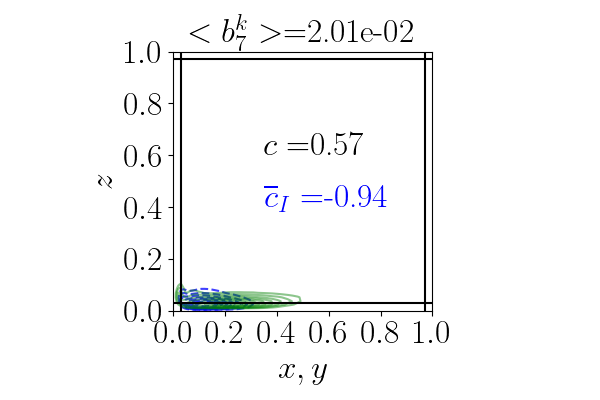}   & 
\includegraphics[trim=4.1cm 2cm 4.0cm 0cm, clip, height=3.85cm]{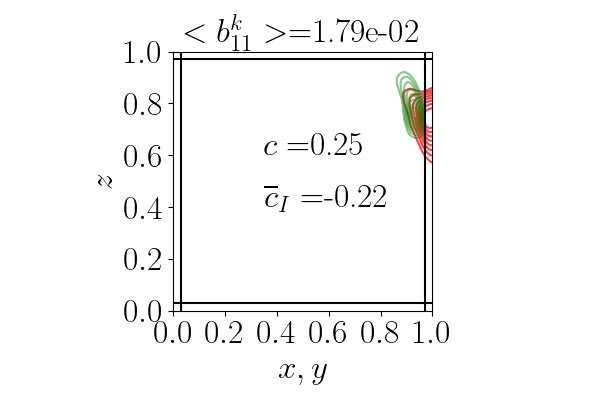}  & & \\ 
\raisebox{1.8cm}{VBL}   &
\includegraphics[trim=3.1cm 0cm 4.0cm 0cm, clip, height=4.7cm]{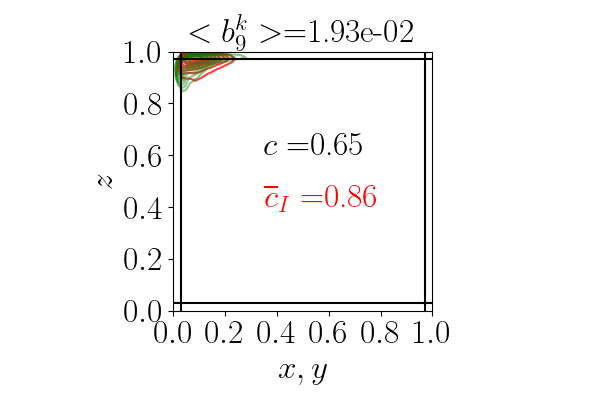} &    
\includegraphics[trim=4.1cm 0cm 4.0cm 0cm, clip, height=4.7cm]{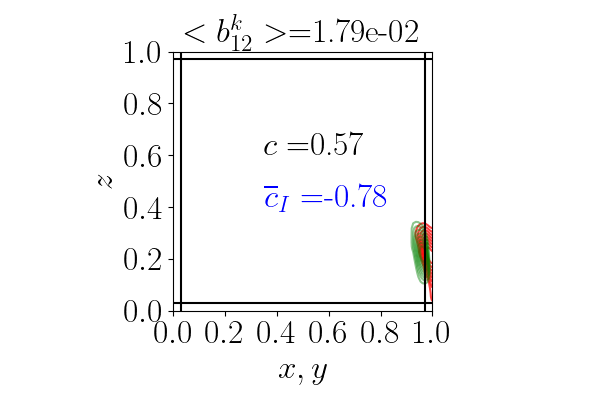} &  
\includegraphics[trim=4.1cm 0cm 4.0cm 0cm, clip, height=4.7cm]{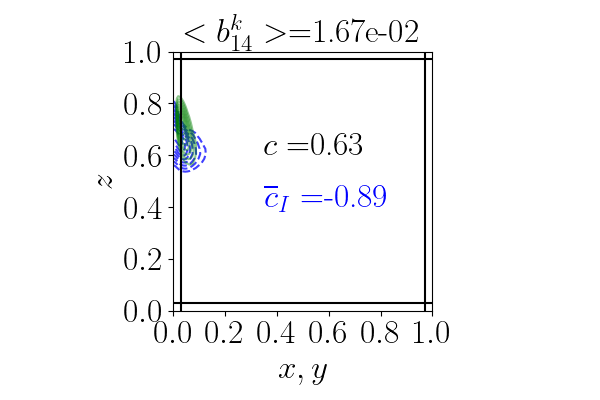} &
\raisebox{1.8cm}{VEZ}   &
\hspace{-1cm}
\includegraphics[trim=3.1cm 0cm 4.05cm 0cm, clip, height=4.7cm]{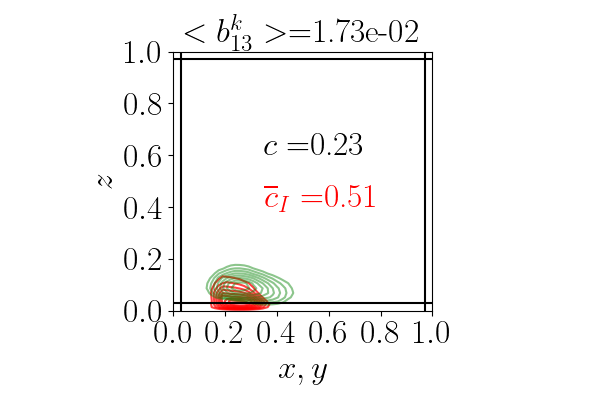} \\ 
\end{tabular}
\caption{ Comparison between kinetic energy and temperature motifs at $Ra=$ $10^8$. See legend of figure 
\ref{fig:comptopicktra1E6}. }
\label{fig:comptopicktra1E8}
\end{figure}

In figures \ref{fig:comptopicktra1E6} to \ref{fig:comptopicktra1E8} the 16 most prevalent kinetic 
energy motifs are represented at  Rayleigh numbers $10^6$, $10^7$ and $10^8$ (the case
$Ra=3$ $10^6$, not shown, was found generally similar to $10^6$ and $10^7$).
The motifs were organized according to the location of their maximum: within 
the horizontal  or vertical boundary layers, which  we will  refer to as respectively HBL or VBL motifs,  
at the corners of the horizontal and the vertical  boundary layer (CBL motifs), and  
outside the boundary layers in the horizontal or vertical entrainment zones, which were 
termed HEZ or VEZ motifs.
The different locations are shown in the top right illustration of figure \ref{fig:comptopicktra1E6}.
For each category the motifs are ordered according to their prevalence, indicated at the top of each plot.
Generally speaking, the prevalence of the motifs increased with the Rayleigh number, which 
is consistent with a strengthening of the large-scale circulation.

For each kinetic energy motif $n$ (represented with green lines), we determined   
the temperature motif $j$ (represented with blue or red lines, depending on its sign) for which  the  correlation coefficient $ C(b_n^k, b_j^\theta)$ is maximal. 
The maximal value $c$ and the temperature motif are represented on each plot, except
in two cases corresponding to HBL motifs, for which the associated temperature motif
had a very low prevalence
$<b_j^\theta>$ and  was considered to be irrelevant. 
In almost all cases, the kinetic energy and temperature motifs are located close to each other in space.
Although the correlation coefficients are typically lower 
than those between the flux and temperature motifs represented in figure \ref{fig:comptopicwttra1E7}, 
several are high enough to associate kinetic energy patterns with specific  temperature fluctuations. 
We also represented on each plot the correlation coefficient $\overline{c}_I$, 
defined as  $\overline{c}_I=C(\overline{b}_n^k, I)$, where  $\overline{b}_n^k$
is the low-pass-filtered kinetic energy motif weight (using $T_f$) and $I$ is the large-scale 
circulation indicator defined in section II B (see also figure \ref{fig:comptopicktra1E6} top right).
High positive (resp. negative) values of $\overline{c}_I$  are indicated in red (resp. blue) for each motif,
and show that the motif can be associated with a specific orientation of  the large-scale circulation.

In all cases, the most frequent motifs consist of centered motifs close to the edge of 
the horizontal boundary layers (HBL).
Evidence of weak correlation (0.3) for some motifs suggested possible association with impinging plumes, however generally low values of $|\overline{c}_I|$  suggest that the weights of the motifs do not depend on the orientation of the large-scale circulation.
In contrast, high values of $c$ and $|\overline{c}_I|$ were found for corner (CBL) motifs,
that were best correlated with impinging plumes.
Corner motifs have a relatively high prevalence, which 
shows that impinging plumes make a significant contribution to the horizontal wind at the edge of  
the boundary layer. 
The correlation coefficient $\overline{c}_I$ increased in absolute value with the Rayleigh number, and was larger than 0.9
at  $Ra=10^8$. In contrast, the maximum correlation coefficient $c$ tended to decrease (but remained significant)
at $Ra=10^8$. 

The next prevalent category of motifs at $Ra=10^6$ and $Ra=10^7$ consisted of
motifs in the vertical entrainment zone (VEZ motifs). They were generally weakly correlated
with temperature motifs of a slightly larger size ($c \sim 0.2-0.3) $ and  were 
 still less correlated with the orientation of the large-scale circulation
 ($\overline{c}_I$ close to zero), which is consistent with their mid-height location.
 Two of the motifs at $Ra=10^7$ (third and fourth motifs) were located closer to a horizontal wall and showed a stronger correlation with $I$. They were found to be correlated with "upstream" 
 temperature fluctuations originating from the opposite wall (arriving plumes). 
 At $Ra=10^8$, only one VEZ motif, with a lower prevalence (compared with the other motifs), was identified. It also corresponded to an arriving plume and was strongly correlated with the orientation of the large-scale circulation.

High values of $|\overline{c}_I|$ were also observed for motifs 
within the vertical boundary layers (VBL), as well as significant values of $c$.
The corresponding temperature motifs were also located within the vertical boundary layers
and consisted of hot (resp. cold) temperature 
fluctuations close to  the bottom (resp. top plate), suggesting that they correspond 
to plumes in the early formation stage (leaving plumes).

At $Ra=10^6$ and $Ra=10^7$, the last category of motifs consisted of motifs
in the horizontal entrainment zone (HEZ).  At the lowest Rayleigh number $Ra=10^6$, 
two of the HEZ motifs (second and fourth motif in the last  row in figures \ref{fig:comptopicktra1E6})
 have a predominantly vertical shape and are associated with large temperature motifs
originating from the opposite (here, top) wall. They are therefore likely to represent coalescing plumes drifting towards the center of the cell as they reach the  opposite wall.
In contrast, all other HEZ motifs at all Rayleigh numbers have a horizontal shape and are associated with 
smaller temperature motifs originating from the closest wall. They are  very well correlated with the orientation of the large-scale circulation.
Significant changes were observed at $Ra=10^8$, with a much larger  
number of HEZ  motifs  and a noticeable increase in their prevalence
- the prevalence of the dominant HEZ motif is twice as large at $Ra=10^8$ than at $Ra=10^7$.

To sum up, a significant difference is observed between $10^7$ and $10^8$. At the highest Rayleigh number, the large-scale circulation is largely reinforced in the horizontal direction due to the formation of new plumes, while stronger impinging plumes remain confined to the corner boundary layers.


	\section{Conclusion}

	We have applied  a  new analysis technique, Latent Dirichlet Allocation,  to characterize the spatio-temporal organization of fluctuations in Rayleigh-Bénard convection.
The method  is based on the inference of  probabilistic latent factors, spatially localized motifs, 
from a collection of instantaneous fields. 
It provides a local yet compact description of the flow 
in terms of  quantitative indicators  such as the  (spatial) size and the (temporal) weight of the motifs.
The technique was applied  to the vertical mid-plane of a Rayleigh-Bénard cubic cell in a range of Rayleigh numbers in $[10^6, 10^8]$.
The method was found to be  robust  with respect to the user-defined parameters.
When applied to the heat flux, it was found to provide good reconstructions of the snapshots 
and  was able to generate new datasets that reproduced key statistics of the original one. 

For all Rayleigh numbers,  dominant heat flux motifs  consisted of  elongated vertical structures located mostly within  the vertical boundary layer, at a height of a quarter of the cell. The width of these motifs scaled with the boundary layer thickness.
These motifs were found to be very well correlated with temperature motifs corresponding to 
plumes in their early formation stage (leaving plumes). 
The motif  weights  were found to depend on the large-scale organization of the flow:
two  states could be identified - one  corresponding to the large-scale circulation  and one to a corner roll structure. The 
two states were characterized by different average weights
which varied non-monotonically with the Rayleigh number.
A simple  model was able to relate the  weights of the dominant heat flux motif  associated with the two states with the average reorientation rate of the large-scale circulation in the cell. This suggests that the model could  be used as a predictor of this rate in cases where few or even no reorientations are observed.

Additional insight about the flow physics was obtained by examining dominant motifs for the temperature
and the kinetic energy.  
While dominant heat flux motifs seemed to be associated with  early-stage  (leaving) plumes, dominant
temperature motifs were associated with later-stage (arriving) plumes.
In contrast with the lower Rayleigh numbers, dominant temperature motifs at $Ra=10^8$ 
were no longer within the vertical boundary layers, but consisted
of plumes  impinging onto the corners of the horizontal boundary layers, which led 
to a reduction of temperature gradients within the corner structure and a decrease in its potential energy. 
This is consistent with the significant drop in the large-scale reorientation rate 
observed at this Rayleigh number.
LDA analysis of the kinetic energy showed that corner impinging 
plumes contributed to the kinetic energy of both the corner structure and  
the large-scale circulation.
The reduction of the reorientation rate at $Ra=10^8$  was also associated 
with a reinforcement of the horizontal wind in the central part of the cell due to 
the formation and entrainment of new plumes. 
The LDA model therefore appears as a promising statistical tool that 
can help track subtle transitions in  the dynamics of turbulent flows. 

 \begin{acknowledgements}
 { This work was granted access to the HPC resources of IDRIS under the allocation 2023- AD012A62062R1  
 made by GENCI. We thank Anouar Soufiani and Philippe Rivière for helpful discussions about the manuscript.   We are also grateful to Jean-Michel Dupays, Rémy Dubois and Camille 
 Parisel for technical support and helpful discussions.}
 \end{acknowledgements}

	\bibliography{savedrecs, biblio_RB}

\pagebreak


\renewcommand{\arraystretch}{2}

\renewcommand{\arraystretch}{1}

\setlength{\tabcolsep}{1pt}

	\end{document}